%% file: DTcombined.tex
\documentclass[11pt, reqno]{amsart}
\input{header}

\title{Doublethink: simultaneous Bayesian-frequentist 
model-averaged hypothesis testing}
\author{
    Helen R. Fryer \textsuperscript{1},
    Nicolas Arning \textsuperscript{1},
    Daniel J. Wilson \textsuperscript{1,2,}*
}
\date{}
\begin{document}
\maketitle

\noindent{\begin{minipage}{\textwidth}

\begin{center}
    9 May 2025
\end{center}

\vspace{5mm}
\begin{itemize}[left=0.3em]
\item[1.] Big Data Institute, Oxford Population Health, University of Oxford

\item[2.] Oxford University Department for Continuing Education

\item[*\phantom{.}] Correspondence: Big Data Institute, Li Ka Shing Centre for Health Information and Discovery, Old Road Campus, Oxford, OX3 7LF, United Kingdom. \includegraphics[height=1em]{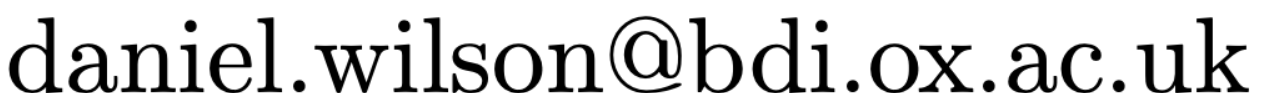} 
\end{itemize}

\vspace{5mm}
ORCIDs: \orcid{0000-0001-9987-8160} (HRF), \orcid{0000-0002-9105-9571} (NA), \orcid{0000-0002-0940-3311} (DJW)

\vspace{5mm}
Running head: `Doublethink: joint Bayesian-frequentist testing'

\vspace{5mm}
This research was funded in whole or in part by the Robertson Foundation, the Wellcome Trust, and the Royal Society (grant no.\ 101237/Z/13/B). For the purpose of Open Access, the author has applied a CC BY public copyright licence to any Author Accepted Manuscript (AAM) version arising from this submission.
\end{minipage}}

\include{DTmaintext}

\setcounter{equation}{0}
\setcounter{section}{0}
\include{DTsupp}

\end{document}

%% file: header.tex
\usepackage[margin=2cm]{geometry}
\usepackage{verbatim}
\usepackage{dirtytalk}
\usepackage{graphicx}
\usepackage{amsthm}
\usepackage{xcolor}
\newcommand{\hl}[1]{\colorbox{yellow}{{#1}}}
\usepackage{amsaddr}
\usepackage{float}
\theoremstyle{definition}
\newtheorem{theorem}{Theorem}
\newtheorem{corollary}{Corollary}
\newtheorem{lemma}{Lemma}
\newtheorem{definition}{Definition}
\newtheorem{example}{Example}

\newtheorem{remark}{Remark}
\newcommand{\appropto}{\mathrel{\vcenter{
  \offinterlineskip\halign{\hfil$##$\cr
    \propto\cr\noalign{\kern2pt}\sim\cr\noalign{\kern-2pt}}}}}
\usepackage{enumerate}
\usepackage{enumitem}

\newcommand{\B}{\mathcal{B}}
\newcommand{\F}{\mathcal{F}}

\newcommand{\BF}{\mathrm{BF}}
\newcommand{\bF}{\mathrm{BF}} 
\newcommand{\PO}{\mathrm{PO}}
\newcommand{\po}{\mathrm{PO}} 
\newcommand{\MLR}{R}
\newcommand{\mlr}{R} 

\newcommand{\bs}{\boldsymbol}

\usepackage{accents}

\newcommand{\upstar}[1]{\accentset{\star}{#1}}

\usepackage{natbib}

\usepackage{mathrsfs}
\usepackage{amsmath}
\usepackage{mathtools}  
\usepackage{xfrac} 
\newcommand{\V}{{\bs v}}
\newcommand{\sizeV}{\nu - |\bs v|}
\newcommand{\W}{\mathcal{W}}

\usepackage{amssymb}
\renewcommand{\complement}{\mathsf{c}}
\newcommand{\bbar}[1]{{#1}}
\usepackage{chngcntr}
\counterwithout{table}{section}
\newcommand{\orcid}[1]{%
  \href{https://orcid.org/#1}{#1}%
}
\usepackage[hidelinks]{hyperref}
\usepackage{chapterbib}  

%% file: DTmaintext.tex
\section*{Abstract} 
Establishing the frequentist properties of Bayesian approaches widens their appeal and offers new understanding. In hypothesis testing, Bayesian model averaging addresses the problem that conclusions are sensitive to variable selection. But Bayesian false discovery rate (FDR) guarantees are sensitive to subjective prior assumptions. Here we show that Bayesian model-averaged hypothesis testing is a closed testing procedure that controls the frequentist familywise error rate (FWER) in the strong sense. To quantify the FWER, we use the theory of regular variation and likelihood asymptotics to derive a chi-squared tail approximation for the model-averaged posterior odds. Convergence is pointwise as the sample size grows and, in a simplified setting subject to a minimum effect size assumption, uniform. The ‘Doublethink’ method computes simultaneous posterior odds and asymptotic $p$-values for model-averaged hypothesis testing. We explore Doublethink through a Mendelian randomization study and simulations, comparing to approaches like LASSO, stepwise regression, the Benjamini-Hochberg procedure, the harmonic mean $p$-value and $e$-values. We consider the limitations of the approach, including finite-sample inflation, and mitigations, like testing groups of correlated variables. We discuss the benefits of Doublethink, including post-hoc variable selection, and its wider implications for the theory and practice of hypothesis testing.

\emph{Keywords}: Bayesian model averaging, closed testing procedure, false discovery rate, familywise error rate, hypothesis tests, post-hoc variable selection

\section{Introduction}

Hypothesis testing is a fundamental scientific approach for identifying variables that affect an outcome of interest \citep{pearson1900criterion, fisher1925statistical, neyman1933ix}.
In observational studies, however, the statistical evidence that one variable directly affects an outcome typically depends on which other variables are included in the model, because of confounding and mediation \citep{vandenbroucke2002history, pearl2016causal}. Uncertainty in model selection can, therefore, strongly influence conclusions and should be accounted for, particularly in big data settings featuring thousands of variables. Bayesian model averaging provides a solution \citep{raftery1995bayesian}, but in many settings Bayes has not become mainstream \citep{hinne2020conceptual}.

The advantages of Bayesian methods, like model-averaging and interpretability, are counterbalanced by various concerns \citep{gelman2008objections}, chiefly the sensitivity to debatable prior assumptions. Whereas Bernstein-von Mises theorem \citep{borwanker1971bernstein} shows that Bayesian and frequentist parameter estimates and confidence regions can converge for large, informative samples, the influence of the prior on hypothesis tests remains uncomfortably direct. The posterior odds of one hypothesis versus another can be doubled by simply doubling the prior odds. Bayes factors are similarly manipulable through the prior on effect sizes \citep{gelman2013philosophy}. This is partly a problem of scale: in large samples, the prior contributes only a constant term to the log posterior odds and gets overwhelmed, and indeed, Bayesian hypothesis testing is consistent by Doob’s theorem \citep{doob1949application, kass1995reference, ohagan1995fractional}. Yet, even in large samples, priors influence testing at the boundary of statistical significance, which can lead different researchers to different conclusions. Likewise, Bayesian false discovery rate (FDR) guarantees are contingent on the prior, which may be disputed.

Establishing the frequentist properties of Bayesian tests can broaden their appeal and afford new insights \citep{bayarri2004interplay}. There is long-standing interest in bridging this Bayesian-frequentist divide \citep{held2018p}.
Formalizing earlier ideas expressed by \citet{good1992bayes}, \citet*{CIS-170509} introduced methods that convert $p$-values to maximum Bayes factors and, by implication, Bayes factors to minimum $p$-values. \citet{zhou2018null} derived the null distribution of Bayes factors in linear regression, enabling direct interconversion of Bayesian and frequentist measures of evidence.
For models in which chi-squared statistics are sufficient, Johnson's \citeyearpar{johnson2005bayes, johnson2008properties} approach (see also \citet{wakefield2009bayes,hu2009bayesian}) enabled one-to-one interconversion between Bayes factors and $p$-values via the maximized likelihood ratio (MLR), similarly to the Bayesian information criterion (BIC; \citet{schwarz1978estimating}), to which it converges.
\citet{vovk2021values} developed conversion of Bayes factors involving a simple null and a possibly model-averaged alternative hypothesis ($e$-values) to maximum $p$-values, together with a closed testing procedure (CTP; \citet{marcus1976closed}) to control the familywise error rate (FWER). FWER is considered standard control in frequentist multiple testing \citep{tukey1953problem}. \citet{goeman2011multiple} developed general $p$-value-based CTPs for FWER control that allow multilevel testing, meaning that thresholds for arbitrary combinations of hypotheses are pre-determined and significant groups of variables can be freely identified post-hoc \citep{meinshausen2008hierarchical, goeman2019simultaneous}.

In this paper, we investigate the frequentist properties of Bayesian model-averaged hypothesis tests involving nested models, with model uncertainty in both the null and alternative hypotheses. Perhaps surprisingly, we show that Bayesian tests are CTPs that control both the Bayesian FDR and the frequentist FWER in the strong sense at some rate. We quantify a large-sample asymptotic false positive rate and FWER, which enables interconversion of Bayesian model-averaged posterior odds and asymptotic $p$-values, on which we base a new method ‘Doublethink’. We investigate the limitations of Doublethink, including finite-sample inflation, and mitigation by grouping correlated variables. We demonstrate the benefits of the approach, including multilevel testing and post-hoc variable selection, through a simulation study benchmarking alternative approaches and a Mendelian randomization study of age-related macular degeneration. Finally, we discuss the broader implications of this work.

\section{Bayesian hypothesis testing is a closed testing procedure that controls the familywise error rate in the strong sense}
\begin{definition}[\bf{Frequentist familywise error rate control}] \label{definitionctp}
\citet*{marcus1976closed} introduced the closed testing procedure (CTP) to control the frequentist familywise error rate (FWER) in the strong sense. Suppose random element $\bs y$ has a probability mass or density function $p({\bs y} ; {\bs x}, \theta)$ that depends on data $\bs x$ and parameters $\theta \in \Theta$ that are partitioned into parameters-of-interest, $\beta$, and nuisance parameters, $\gamma$. 
The aim is to test a set of hypotheses about $\beta$ defined by $\Omega = \{ \omega_{\bs s} \}$, where $\omega_{\bs s} \subset \Theta$ and $\bs s$ is an index. 
$\Omega$ must be closed under intersection, meaning $\omega_{\bs s}, \omega_{\bs s^\prime} \in \Omega$ implies $\omega_{\bs s} \cap \omega_{\bs s^\prime} \in \Omega$. A `local' test rejects the null hypothesis $\theta \in \omega_{\bs s}$ if test function $\psi_{\bs s}({\bs y})$ returns 1, not 0. This controls the false positive rate (FPR) at or below level
\begin{eqnarray}
    \alpha_{{\bs s}} &:=& \sup_{\theta \in \omega_{\bs s}} \alpha_{{\bs s}, \theta} \nonumber
\end{eqnarray}
where
\begin{eqnarray}
    \alpha_{{\bs s}, \theta} &:=& \Pr\left(\psi_{\bs s}({\bs y}) = 1 ; {\bs x}, \theta\right), \qquad \theta \in \omega_{\bs s} .
\end{eqnarray}
By definition, a CTP rejects $\theta \in \omega_{\bs s}$ when all intersection hypotheses involving $\omega_{\bs s}$ are rejected by local tests. That is, when the function
\begin{eqnarray}
    \phi_{\bs s}({\bs y}) &=& \min_{\omega_{\bs r} \subseteq \omega_{\bs s} } \psi_{\bs r}({\bs y})  \label{CTPtestfunction} 
\end{eqnarray}
equals 1. A familywise error occurs when one or more null hypotheses are falsely rejected. A CTP controls the FWER in the strong sense (see Appendix \ref{background_on_ctps}) at level
\begin{eqnarray}
\mathrm{FWER} &:=& \underset{\omega_{{\bs s}} \in \Omega \ \theta \in \omega_{{\bs s}}}{\max \ \sup} \ \Pr\left( \phi_{\bs s}({\bs y}) = 1 ; {\bs x}, \theta \right) \nonumber\\
    &\leq& \max_{\omega_{\bs s} \in \Omega} \ \alpha_{\bs s}.
\end{eqnarray}
\end{definition}

\begin{definition}[\bf{Bayesian false discovery rate control}] \label{definitionbayestest}
A Bayesian test rejects the null hypothesis that $\theta \in \omega_{\bs s}$ when the posterior odds
\begin{eqnarray} \label{posterioroddsctp}
    \po_{\omega_{\bs s}^\complement:\omega_{\bs s}} &=& \frac{
        \int_{\omega_{\bs s}^\complement} p({\bs y};{\bs x}, \theta)\ \mathrm{d}\Pi(\theta)
    }{
        \int_{\omega_{\bs s}} p({\bs y};{\bs x}, \theta)\ \mathrm{d}\Pi(\theta)    
    },
\end{eqnarray}
exceed some threshold $\tau$, or equivalently, when the test function $\psi_{\bs s}(\bs y) = \mathbb{I}(\po_{\omega_{\bs s}^\complement:\omega_{\bs s}}\geq\tau)$ equals 1. Here, $\mathbb{I}$ is an indicator function, $\omega_{\bs s}^\complement = \Theta \setminus \omega_{\bs s}$, and $\Pi(\theta)$ is the distribution function of the prior on $\theta \in \Theta$. A Riemann-Stieltjes integral is employed since the prior may not be continuous. The test controls the Bayesian false discovery rate (FDR) locally \citep{efron2001empirical} at level
\begin{eqnarray}
 \mathrm{FDR}_{\bs s} 
    &:=& \Pr\left(\theta \in \omega_{\bs s} \mid {\bs y},\, {\bs x} \right) \, \psi_{\bs s}({\bs y}) \nonumber\\
    &=& \dfrac{1}{1+\po_{\omega_{\bs s}^\complement:\omega_{\bs s}}} \, \mathbb{I}\left(\po_{\omega_{\bs s}^\complement:\omega_{\bs s}}\geq\tau\right)  \nonumber\\
    &\leq& \frac{1}{1+\tau}
\end{eqnarray}
and, for the set of hypotheses defined by $\Omega = \{ \omega_{\bs s} \}$, globally at level
\begin{eqnarray}
    \mathrm{FDR} &:=& \frac{\sum_{\bs s} \mathrm{FDR}_{\bs s}}
    {\max\left\{ 1, \sum_{\bs s} \psi_{\bs s}(\bs y) \right\}} \ \leq\ \dfrac{1}{1+\tau}.
\end{eqnarray}
\end{definition}

\begin{theorem}[\bf{Bayesian hypothesis tests simultaneously control the Bayesian FDR and the frequentist FWER}] \label{theorem_bayesCTP}
Bayesian hypothesis tests are a type of CTP known as a shortcut CTP. That is, $\phi_{\bs s}({\bs y})=\psi_{\bs s}({\bs y})=1$ (rejection of $\theta\in\omega_{\bs s}$) automatically implies $\phi_{\bs r}({\bs y})=\psi_{\bs r}({\bs y})=1$  for all intersection hypotheses $\omega_{\bs r} \subset \omega_{\bs s}$. Therefore, they simultaneously control the frequentist FWER in the strong sense at or below level $\max \{ \alpha_{\bs s} : \omega_{\bs s}\in \Omega\}$ and the Bayesian FDR at or below level $1/(1+\tau)$.

\begin{proof}\label{Lemma 1 section proof}
By definition, $\omega_{\bs r} \subset \omega_{\bs s}$ and $\omega_{\bs r}^\complement \supset \omega_{\bs s}^\complement$, so ${
        \int_{\omega_{\bs r}} p({\bs y};{\bs x}, \theta)\ \mathrm{d}\Pi(\theta)
    } \leq {
        \int_{\omega_{\bs s}} p({\bs y};{\bs x}, \theta)\ \mathrm{d}\Pi(\theta)
    }$ and ${
        \int_{\omega_{\bs r}^\complement} p({\bs y};{\bs x}, \theta)\ \mathrm{d}\Pi(\theta)
    } \geq {
        \int_{\omega_{\bs s}^\complement} p({\bs y};{\bs x}, \theta)\ \mathrm{d}\Pi(\theta)
    }$. Therefore
\begin{eqnarray}
\po_{\omega_{\bs r}^\complement:\omega_{\bs r}} = \frac{{\int_{\omega_{\bs r}^\complement} p({\bs y};{\bs x}, \theta)\ \mathrm{d}\Pi(\theta)}}{ {\int_{\omega_{\bs r}} p({\bs y};{\bs x}, \theta)\ \mathrm{d}\Pi(\theta)}} &\geq& \frac{{\int_{\omega_{\bs s}^\complement} p({\bs y};{\bs x}, \theta)\ \mathrm{d}\Pi(\theta)}}{ {\int_{\omega_{\bs s}} p({\bs y};{\bs x}, \theta)\ \mathrm{d}\Pi(\theta)}}
 = \po_{\omega_{\bs s}^\complement:\omega_{\bs s}} \nonumber
\end{eqnarray}
so
\begin{eqnarray}
\psi_{\bs s}(\bs y)=1 \iff \po_{\omega_{\bs s}^\complement:\omega_{\bs s}}\geq \tau \implies \po_{\omega_{\bs r}^\complement:\omega_{\bs r}}\geq \tau \iff \psi_{\bs r}({\bs y})=1 .
\end{eqnarray}
Control of the strong-sense FWER and Bayesian FDR follows by Definitions \ref{definitionctp} and \ref{definitionbayestest}

\end{proof}
\end{theorem}

The rest of this paper focuses on quantifying the level at which a Bayesian test controls the FWER, asymptotically for large samples, in a general regression setting.

\section{Frequentist false positive rate of a Bayesian model-averaged regression converges pointwise as the sample size grows}

We apply Theorem \ref{theorem_bayesCTP} to construct a joint Bayesian-frequentist approach to model-averaged hypothesis testing in a regression setting.

\begin{definition}[\bf{Regression problem}] \label{iidregressionprob} \label{define_regression_problem}
We consider a general regression with $n$ observed outcomes, ${\bs y} = (y_1,\dots,y_n)^T$, $\nu$ regression coefficients, $\beta = (\beta_1,\dots,\beta_\nu)^T$, for candidate explanatory variables $x_{ij}, i=1\dots n, j=1\dots \nu$, and $\zeta$ nuisance parameters, $\gamma = (\gamma_1,\dots,\gamma_\zeta)^T$. Together, we write $\theta = \binom{\beta}{\gamma}$. The aim is to identify which regression coefficients are non-zero. We index hypotheses by a binary vector ${\bs s} \in \mathcal{S} = \{0, 1\}^\nu$ such that $\omega_{\bs s} = \{\theta:\, \beta_j = 0 \;\, \forall \ s_j = 0\} \subseteq \Theta = \mathbb{R}^{\nu+\zeta}$. The null hypotheses are $\Omega = \mathcal{S} \setminus {\bs 1}$. We define models in terms of variable selection. Unlike the null hypotheses, the models are disjoint with respect to $\beta$, such that model $\bs s$ has parameter space $\Theta_{\bs s} = \{ \theta \, : \, \beta_j = 0\ \forall\ s_j = 0\ ;\ \beta_j \neq 0\ \forall\ s_j = 1 \} \subseteq \Theta$. We say that hypothesis $\omega_{\bs r}$ is nested in $\omega_{\bs s}$ if $\omega_{\bs r} \subset \omega_{\bs s}$.
\end{definition}

In general, the finite sample properties of frequentist methods, including FWER, are difficult to obtain. Instead, it is standard to use approximations based on the asymptotic distribution of methods as the sample size, $n$, tends to infinity. For nested hypotheses, the workhorse is the likelihood ratio test (LRT).

\begin{definition}[\bf{Likelihood ratio test; LRT}] \label{define_lrt}
\citet{neyman1928use} introduced the maximized likelihood ratio (MLR) for comparing hypotheses $\omega_{\bs s}$ and $\omega_{\bs t}$,
\begin{eqnarray}
    \mlr_{\omega_{\bs t}:\omega_{\bs s}} &:=&
        \frac{
            \sup_{\theta \in {\omega_{\bs t}}} p({\bs y};{\bs x}, \theta)
        }{
            \sup_{\theta \in \omega_{\bs s}} p({\bs y};{\bs x}, \theta)
        } .
\end{eqnarray}
\citet{wilks1938large} and \citet{wald1943tests} showed that for nested hypotheses, $\omega_{\bs s} \subset \omega_{\bs t}$, differing in $|\bs  t|-|\bs s|$ dimensions, the distribution of the MLR converges such that the deviance,
\begin{eqnarray}
2 \, \log \mlr_{\omega_{\bs t}:\omega_{\bs s}} &\overset{d}{\mathop{\to }}&   \chi^2_{|\bs t|-|\bs s|} \label{define_lrt_d2logR_given_theta}, \qquad n\to\infty,
\end{eqnarray}
given $\theta \in \omega_{\bs s}$, where $\chi^2_k$ represents a chi-squared distribution with $k$ degrees of freedom,
assuming the $n$ outcomes are independent realizations, each with likelihood that satisfies standard regularity conditions (Appendix \ref{regCon}), under the `local alternatives' assumption \citep{davidson1970limiting}. The FPR of the test $\psi_{{\bs s}}({\bs y}) = \mathbb{I}(2 \, \log \mlr_{\omega_{\bs t}:\omega_{\bs s}} \geq x_\mathrm{crit})$ converges in general pointwise (in the sense of \citet{leeb2008sparse}) to
\begin{eqnarray}
    \alpha_{{\bs s}} &\overset{\mathrm{pw}}{\sim}& \Pr\left(\chi^2_{\bs |t|-|\bs s|} \geq x_\mathrm{crit}\right), \qquad n\rightarrow\infty,
\end{eqnarray}
meaning that
\begin{eqnarray}
    \underset{\theta\in\omega_{\bs s} \ n\rightarrow\infty}{\sup \, \ \lim} \ \frac{\alpha_{{\bs s},\theta}}{\Pr\left(\chi^2_{|\bs t|-| \bs s|} \geq x_\mathrm{crit}\right)} &=& 1.
\end{eqnarray}
As shorthand, we will write $\mlr_{{\bs t}:{\bs s}} := \mlr_{\omega_{\bs t}:\omega_{\bs s}}$, noting that $\mlr_{\omega_{\bs t}:\omega_{\bs s}} = \mlr_{\Theta_{\bs t}:\Theta_{\bs s}}$ by our use of point null hypotheses. We will also write $\mlr_{{\bs s}} := \mlr_{{\bs s}:{\bs 0}}$, where $\omega_{\bs 0}$ is the `grand null' hypothesis and $\bs 0=(0,\dots,0)^T$.
\end{definition}

One method that addresses the regression problem while controlling the FWER, approximately for large $n$, is given below.

\begin{example}[\bf{FWER control of the regression problem}] \label{example_fwer_regression}
The LRT \citep{wilks1938large, wald1943tests} can be combined with Bonferroni correction \citep{bonferroni1936teoria} to define a leave-one-out test that drops variables relative to the `grand alternative' hypothesis $\omega_{\bs 1}$, where $\bs{1}=(1, \dots, 1)^T$.
\begin{eqnarray}
    p_{{\bs 1}:{\bs t}} &=& \Pr\left(\chi^2_1 \geq 2\, \log \MLR_{{\bs 1}:{\bs t}} \right), \qquad {\bs t} \in \mathcal{S},\ |\bs t|=\nu-1 \\
    \phi_{\bs s}(\bs y) &=& 
    \mathbb{I}\left(\min_{\genfrac{}{}{0pt}{}{\omega_{\bs s} \subset \omega_{\bs t}}{|\bs t|=\nu-1}} p_{{\bs 1}:{\bs t}} \leq \frac{\alpha}{\nu} \right).
\end{eqnarray}
This controls the FWER at level $\alpha$, to the order of the large $n$ approximation. There is a large literature on combined tests for $p$-values that are more powerful than Bonferroni correction under certain conditions \citep{loughin2004systematic}, for example \citet{simes1986improved}, \citet{hommel1988stagewise}, \citet{meijer2019hommel}, \citet{wilson2019harmonic}; and  \citet{liu2020cauchy}.
\end{example}

Despite its familiarity, it can be hard to solve the regression problem while controlling the FWER and maintaining power, and Example \ref{example_fwer_regression} has various problems. When the candidate explanatory variables, $\bs x$, are correlated, parameter estimates can be noisy and $p$-values deflated (conservative), which reduces power. Under collinearity of $\bs x$, it may not be possible to fit the grand alternative model at all, in which case the method does not work.

Instead, it is common to select a narrower hypothesis, $\bs s : \omega_{\bs s} \subset \omega_{\bs 1}$, for example using expert opinion, univariable associations, machine learning, or statistical optimization \citep{miller2002subset,porwal2022comparing}, then perform $|\bs s|$ leave-one-out likelihood ratio tests and, perhaps, $(\nu-|\bs s|)$ add-one-in tests. Sparse models ($|\bs s|\ll\nu$) are usually preferred on principle, and to mitigate variance and deflation due to correlated variables (see e.g.\ \citet{tibshirani1996regression}). Often, adjustment is made for multiple testing, as per Example 1, but uncertainty in model selection -- which can affect conclusions -- is rarely accounted for, partly because the effects of selecting $\bs s$ on the distribution of test statistics is not known, in general. 

Bayesian model averaging presents a solution to these challenges and, by Theorem \ref{theorem_bayesCTP}, controls the FWER. To be practical, we need to quantify the level at which the FWER is controlled. As a first step we pursue an asymptotic approximation to the FPR by extending work by \citet{schwarz1978estimating} and \citet{johnson2005bayes, johnson2008properties}.

\begin{definition}[\bf{Posterior odds via the Bayesian information criterion; BIC}]
\citet{schwarz1978estimating} developed the BIC for model selection as an approximate Bayes factor in large samples. It implies the posterior odds of model $\bs t$ versus model $\bs s$, (not necessarily nested) are related to the MLR, $\mlr_{\bs t:\bs s}$, as
\begin{eqnarray} 
   \po_{\Theta_{\bs t}:\Theta_{\bs s}} &=& 
    \frac{
        \int_{\Theta_{\bs t}} p({\bs y};{\bs x}, \theta)\ \mathrm{d}\Pi(\theta)
    }{
        \int_{\Theta_{\bs s}} p({\bs y};{\bs x}, \theta)\ \mathrm{d}\Pi(\theta)    
    }\nonumber
\\&\approx& \label{define_po_bic}\mu_{\Theta_{\bs t}:\Theta_{\bs s}}\ n^{-(|\bs t|-|\bs s|)/2}\ \mlr_{\bs t:\bs s},
\end{eqnarray}
where $\mu_{\Theta_{\bs t}:\Theta_{\bs s}} = \int_{\Theta_{\bs t}} \mathrm{d}\Pi(\theta)/ \int_{\Theta_{\bs s}} \mathrm{d}\Pi(\theta)$ are the prior odds. The BIC has error of the order $O_p(1)$ when used to approximate the log posterior odds or log Bayes factor \citep{kass1995reference}. As shorthand we will write $\po_{{\bs t}:{\bs s}} := \po_{\Theta_{\bs t}:\Theta_{\bs s}}$ and $\po_{{\bs s}} := \po_{{\bs s}:{\bs 0}}$, and likewise for $\mu_{\bs t:\bs s}$.
\end{definition}

\begin{definition}[\bf{Joint Bayesian-frequentist test: Johnson model}]\label{define_johnson}
\citet{johnson2005bayes, johnson2008properties} developed simultaneous Bayesian-frequentist inference that defines the posterior odds of model $\bs t$ versus model $\bs s$ (not necessarily nested) via the MLR as
\begin{eqnarray} \label{define_johnson_eq}
    \po_{\bs t:\bs s} &\approx& \mu_{\bs t:\bs s}\ {\xi_n}^{(|\bs t|-|\bs s|)/2}\ {\mlr_{\bs t:\bs s}}^{1-\xi_n},\quad \xi_n=\tfrac{h}{n+h},
\end{eqnarray}
where $h$ is the prior precision parameter. This closely resembles Equation \ref{define_po_bic}, to which it converges as $n\rightarrow\infty$, but is proper in the sense that $\mathbb{E}_\Pi[\po_{\bs t:\bs s}\mid\theta\in\Theta_{\bs s}] = \mu_{\bs t:\bs s}$. Broadly, the model mirrors the assumptions of the LRT (Definition \ref{define_lrt}), including large $n$ and local alternatives. It can be derived under the following conjugate prior:
\begin{eqnarray}
\theta_{\F_{\bs s}} \,\big|\, \theta \in{\omega}_{\bs s} &\overset{d}{=}& \mathcal{N}_{|\bs s|+\zeta}\left( \bs{0}, h^{-1} \left[ \mathcal{I}(\bs 0)_{\F_{\bs s},\F_{\bs s}}  \right]^{-1}  \right), \label{define_johnson_ConjugatePrior} \label{ConjugatePrior}
\end{eqnarray}
where $\F_{\bs s}$ is the index set of unconstrained parameters in ${\omega}_{\bs s}$, $\mathcal{N}_k(\bs m, \bs \Sigma)$ represents a multivariate Normal distribution on $k$ dimensions with mean vector $\bs m$ and variance matrix $\bs \Sigma$, and $\mathcal{I}(\theta)_{\F_{\bs s},\F_{\bs s}}$ is the per-observation Fisher information matrix for model $\bs s$ evaluated at $\theta$ (defined in SI Equation \ref{fisher}). The prior resembles Zellner's \citeyearpar{zellner1986assessing} $g$-prior with $g=n/h$; a unit information prior \citep{kass1995reference, liang2008mixtures} arises when $h=1$. 
The test $\psi_{\bs s}(\bs y) = \mathbb{I}(\po_{\bs t:\bs s}\geq\tau)$ controls the local Bayesian FDR at or below level $1/(1+\tau)$ (Definition \ref{definitionbayestest}), and when $\omega_{\bs s} \subset \omega_{\bs t}$, the FPR converges in general pointwise to
\begin{eqnarray}
    \alpha_{\bs s} &\overset{\mathrm{pw}}{\sim}& \Pr\left(\chi^2_{\bs t: \bs s} \geq 2 \log \frac{\tau}{\mu_{\bs t:\bs s}\ {\xi_n}^{(|\bs t|-|\bs s|)/2}} \right), \quad n\rightarrow\infty,
\end{eqnarray}
by the LRT (Definition \ref{define_lrt}), meaning that
\begin{eqnarray}
    \underset{\theta\in\omega_{\bs s} \ n\rightarrow\infty}{\sup \, \ \lim} \ \frac{\alpha_{\bs s,\theta}}{\Pr\left(\chi^2_{\bs t: \bs s} \geq 2 \log \frac{\tau}{\mu_{\bs t:\bs s}\ {\xi_n}^{(|\bs t|-|\bs s|)/2}}\right)} &=& 1.
\end{eqnarray}
Johnson's model expresses an asymptotic relationship not only between classical and Bayesian hypothesis tests, but point estimates and variances too (Table \ref{maintable}).
\end{definition}

{\begin{table}[t]
\scriptsize
\renewcommand{\arraystretch}{2.5}
\begin{tabular}{|l|l|l|l|l|l|}
\hline
\multicolumn{3}{|c|}{Classical quantity}&\multicolumn{3}{c|}{Bayesian counterpart}\\ 
\hline
Quantity & Expression & SI Eqn. & Quantity & Expression & SI Eqn.\\
\hline
Maximum likelihood estimate & $\hat\theta^{\bs s}$ & \ref{classmean} & Posterior mean & $\frac{n}{n+h} \hat\theta^{\bs s}$ & \ref{postmean} \\
Estimator variance & $[{{{J}}}(\hat\theta^{\bs s})_{\F_{\bs s},\F_{\bs s}}]^{-1}$ & \ref{classvar} & Posterior variance & $\frac{n}{n+h} [ {{{J}}}(\hat\theta^{\bs s})_{\F_{\bs s},\F_{\bs s}}]^{-1}$ &  \ref{postvar}\\

Maximized likelihood ratio & $\mlr_{\bs s}$ & \ref{Rsdef} & Bayes factor & $\left(\frac{h}{n+h}\right)^{{|\bs s|}/2}\,{\mlr_{\bs s}}^{n/(n+h)}$ & \ref{MLR0} \\
\hline
\end{tabular}
\caption{Connections between Bayesian and classical inference in Johnson's \citeyearpar{johnson2005bayes, johnson2008properties} model. {\smaller ${{{J}}}(\hat\theta^{\bs s})_{\F_{\bs s},\F_{\bs s}}$ is the observed information matrix for hypothesis $\bs s$ at the MLE.}}\label{maintable} 
\end{table}
}

Next, we define a model-averaged extension of Johnson's \citeyearpar{johnson2005bayes,johnson2008properties} model, including a prior on models. Notably, the posterior odds in Definition \ref{define_johnson} would be pre-determined if the prior on $\theta$ were continuous since that would imply $\int_{\Theta_{\bs s}} \mathrm{d}\Pi(\theta)=\mathbb{I}(\bs s=\bs 1)$. Therefore pursuing Bayesian hypothesis testing with point null  hypotheses implies a discontinuous prior.

\begin{definition}[\bf{Bayesian model-averaged hypothesis testing: Doublethink model}]
\label{define_bmatest}
We assume that the prior odds of model $\bs t$ versus model $\bs s$ take the form
\begin{eqnarray}\label{doubletpriorodds}
{\mu}_{\bs t:\bs s} &=& \mu^{|\bs t|-|\bs s|}
\end{eqnarray}
meaning the prior probability that $\beta_j\neq 0$ is independent for each variable and equal to $\mu/(1+\mu)$, and we approximate, by the Johnson model (Definition \ref{define_johnson}), that
\begin{eqnarray}
    \po_{\bs t:\bs s} &:=& \mu^{|\bs t|-|\bs s|}\ {\xi_n}^{(|\bs t|-|\bs s|)/2}\ {\mlr_{\bs t:\bs s}}^{1-\xi_n},\quad \xi_n=\tfrac{h}{n+h}.
\end{eqnarray}
We can write the model-averaged posterior odds against null hypothesis $\omega_{\bs v} = \{\theta:\ \beta_j=0\ \forall\ v_j=0\}$ as
\begin{eqnarray} \label{define_bmatest_eq} \label{modelavpostodd} \label{MPO1}
    \po_{\mathcal{A}_\V:\mathcal{O}_\V} \ :=\ 
    \po_{\omega_{\bs v}^\complement:\omega_{\bs v}} &=& 
    \frac{
        \int_{\omega_{\bs v}^\complement} p({\bs y};{\bs x}, \theta)\ \mathrm{d}\Pi(\theta)
    }{
        \int_{\omega_{\bs v}} p({\bs y};{\bs x}, \theta)\ \mathrm{d}\Pi(\theta)    
    } \nonumber\\
    &=&
    \frac{
        \sum_{\bs s\in\mathcal{A}_\V} \int_{\Theta_{\bs s}} p({\bs y};{\bs x}, \theta)\ \mathrm{d}\Pi(\theta)
    }{
        \sum_{\bs s\in\mathcal{O}_\V} \int_{\Theta_{\bs s}} p({\bs y};{\bs x}, \theta)\ \mathrm{d}\Pi(\theta)    
    } \ =\ 
    \dfrac{\sum_{\bs s \in \mathcal{A}_\V} {\po}_{\bs s}}{\sum_{\bs s \in \mathcal{O}_\V} {\po}_{\bs s}},
\end{eqnarray}
where we define the set of models compatible with null hypothesis $\omega_{\bs v}$ as $\mathcal{O}_\V = \left\{ {\bs s} \in \mathcal{S} : \Theta_{\bs s} \subseteq \omega_{\bs v} \right\}$, and the models consistent with the alternative hypothesis $\omega_{\bs v}^\complement$ as $\mathcal{A}_\V = \mathcal{S} \setminus \mathcal{O}_\V$. The test of the model-averaged posterior odds, $\PO_{\mathcal{A}_\V:\mathcal{O}_\V}$,
\begin{eqnarray}
    \psi_{\bs v}(\bs y) &=& \mathbb{I}\left(\PO_{\mathcal{A}_\V:\mathcal{O}_\V} \geq \tau\right)
\end{eqnarray}
is a shortcut CTP because $\psi_{\bs v}(\bs y)=1$ implies $\psi_{\bs u}(\bs y)=1$ for all $\omega_{\bs v} \supseteq \omega_{\bs u}$. It controls the Bayesian FDR locally and globally at level $1/(1+\tau)$ (by Definition \ref{definitionbayestest}) and (by Definition \ref{definitionctp} and Theorem \ref{theorem_bayesCTP}) the FWER at level
\begin{eqnarray}
    \underset{\omega_{\bs v} \in \Omega \ \theta\in \omega_{\bs v}}{\max \ \sup} \
    \Pr\left( \psi_{\bs v}(\bs y) = 1; {\bs x}; \theta \right).
\end{eqnarray}
\end{definition}

We note that whereas the Johnson model is derived specific assumptions, its close connection to the BIC implies that the model-averaged posterior odds in Definition \ref{define_bmatest} may serve as a useful large-sample approximation in a wide variety of settings. It remains to quantify the rate at which the FPRs and FWER are controlled. First, we present an intermediate result for large samples.

\begin{lemma}[\bf{Convergence of the scaled model-averaged posterior odds to a sum of maximized likelihood ratios}]
\label{lemma_scaled_posterior_odds}
Under the Doublethink model (Definition \ref{define_bmatest}), the scaled model-averaged posterior odds against null hypothesis $\omega_{\bs v}$ converge in distribution, for large samples, to a sum of maximized likelihood ratios involving one degree-of-freedom tests relative to the true model $\tilde{\bs s}$, when $\theta\in\Theta_{\tilde{\bs s}}$ and $\Theta_{\tilde{\bs s}} \subseteq \omega_{\bs v}$:
\begin{eqnarray}
\dfrac{\po_{\mathcal{A}_\V:\mathcal{O}_\V}}{\mu \sqrt{\xi_n}} &\overset{d}{\rightarrow}& \sum_{j\, :\, v_j=0} \mlr_{\tilde{\bs s}+\bs e_j:\tilde{\bs s}}, \qquad n\rightarrow\infty,
\end{eqnarray}
meaning
\begin{eqnarray}
\lim_{n\rightarrow\infty} \Pr\left( \dfrac{\po_{\mathcal{A}_\V:\mathcal{O}_\V}}{\mu \sqrt{\xi_n}} \geq x;\; \theta \right) &=& \Pr\left( \sum_{j\, :\, v_j=0} \mlr_{\tilde{\bs s}+\bs e_j:\tilde{\bs s}} \geq x;\; \theta \right), \qquad \forall\ x,
\label{lemma_scaled_posterior_odds_eqn}
\end{eqnarray}
where $\bs e_j$ is a unit vector, so $\{\bs e_j\}_i = \mathbb{I}(i=j)$.

\begin{proof}
By Slutsky \citeyearpar{slutsky1925stochastische}, Wilks \citeyearpar{wilks1938large}, Doob \citeyearpar{doob1949application} and Schwartz \citeyearpar{schwartz1965onbayes}, assuming that all MLRs in which the true null model $\tilde{\bs s}$ is nested are bounded in probability, and assuming $\{ \mlr_{\tilde{\bs s}+\bs e_j:\tilde{\bs s}} : v_j = 0 \}$ converge jointly in distribution. See Appendix \ref{proof_lemma_scaled_posterior_odds}.
\end{proof}
\end{lemma}

Next we derive an analytic result for the FPR by characterizing this heavy tailed sum of MLRs using asymptotic likelihood theory (see e.g.\ \citet{cox1974theoretical}) and the theory of regular variation (see e.g.\ \citet{mikosch1999regular}).

\begin{theorem}[\bf{Frequentist false positive rate of the Bayesian model-averaged hypothesis test in large samples}]
\label{Corr20} \label{theorem_fpr}
The frequentist FPR of the Bayesian model-averaged hypothesis test in Definition \ref{define_bmatest}, which rejects the null hypothesis $\omega_{\bs v} = \{ \theta : \beta_j = 0 \ \forall \ v_j=0 \}$, when ${\PO}_{\mathcal{A}_\V:\mathcal{O}_\V}\geq\tau$, converges asymptotically, for large sample size, to
\begin{eqnarray} \label{theorem_fpr_eq1}
    \alpha_{\bs v} &\overset{\mathrm{pw}}{\sim}&  \Pr\left( \chi^2_1
    \geq 2 \log \dfrac{\tau}{(\sizeV) \, \mu \sqrt{\xi_n}} \right), 
    \quad n\rightarrow\infty, \label{alphav}
\end{eqnarray}
where pw indicates that convergence is pointwise with respect to $\theta$, meaning that
\begin{eqnarray}
    \underset{\theta \in \omega_{\bs v} \ n\rightarrow\infty}{\sup \ \lim} \
    \dfrac{\Pr\left( \po_{\mathcal{A}_\V:\mathcal{O}_\V} \geq \tau;\; \theta \right)}{\Pr\left( \chi^2_1
    \geq 2 \log \frac{\tau}{(\sizeV) \, \mu \, \sqrt{\xi_n}} \right)} &=& 1.
\end{eqnarray}
\begin{proof}
By Fisher \citeyearpar{fisher1925statistical}, Karamata \citeyearpar{karamata1933mode},
Wilks \citeyearpar{wilks1938large}, Nagaev \citeyearpar{nagaev1965limit}, \citet{davis1996limit}, and Lemma \ref{lemma_scaled_posterior_odds}. See Appendix \ref{proof_theorem_fpr}.
\end{proof}
\end{theorem}

\begin{corollary}[\bf{The model-averaged deviance has a limiting chi-squared distribution in the tail}]\label{corollary_mlr_limiting_dist}
Define the model-averaged MLR as
\begin{eqnarray}
    \mlr_{\mathcal{A}_\V:\mathcal{O}_\V} &:=& \frac{\po_{\mathcal{A}_\V:\mathcal{O}_\V}}{\left(\mu\,\sqrt\xi_n + 1\right)^{\sizeV}-1} \ \sim\ \frac{\po_{\mathcal{A}_\V:\mathcal{O}_\V}}{(\sizeV)\, \mu\, \sqrt\xi_n},\quad n\rightarrow\infty,
\end{eqnarray}
The model-averaged deviance converges pointwise in the tail to a chi-squared distribution with one degree of freedom:
\begin{eqnarray}
    2\, \log \mlr_{\mathcal{A}_\V:\mathcal{O}_\V} \overset{d}{\rightarrow} \chi^2_1, \qquad n\rightarrow\infty
\end{eqnarray}
in the sense that
\begin{eqnarray}
    \underset{\theta \in \omega_{\bs v} \ n\rightarrow\infty}{\sup \ \, \lim} \ 
    \frac{\Pr\left( 2 \log \mlr_{\mathcal{A}_\V:\mathcal{O}_\V} \geq x_n;\; \theta \right)}{\Pr\left( \chi^2_1
    \geq x_n \right)} &=& 1
\end{eqnarray}
for $x_n$ increasing in $n$.
\end{corollary}

\begin{corollary}[\bf{The model-averaged posterior odds can be transformed into an asymptotic $p$-value}]\label{unadjustedSIp-values} \label{corollary_po2punadj}
The Bayesian model-averaged test in Definition \ref{define_bmatest}, which rejects the null hypothesis $\omega_{\bs v} = \{ \theta : \beta_j = 0 \ \forall \ v_j=0 \}$, is equivalent to a frequentist test that rejects the null when the asymptotic $p$-value $p_{\mathcal{A}_\V:\mathcal{O}_\V}$ is less than or equal to the asymptotic false positive rate $\alpha_{\bs v}$ (Equation \ref{theorem_fpr_eq1}), where
\begin{eqnarray}
\label{Th1MainEq}
p_{\mathcal{A}_\V:\mathcal{O}_\V}&\sim& 
\Pr\left( \chi^2_1 \geq 2\log \dfrac{\PO_{\mathcal{A}_\V:\mathcal{O}_\V}}{ (\sizeV) \, \mu \sqrt\xi_n } \right), \qquad n\to\infty.
\end{eqnarray}
This result extends Johnson's \citeyearpar{johnson2005bayes,johnson2008properties} interconversion of MLRs and $p$-values for nested hypothesis tests to the model-averaged setting.
\end{corollary}


In the next section we derive the asymptotic FWER and adjusted $p$-values that account for multiple testing.

\section{Strong-sense familywise error rate of a Bayesian model-averaged regression converges pointwise as the sample size grows}

\newcommand{\VV}{\V}
\begin{theorem}[\bf{Familywise error rate of the Bayesian model-averaged hypothesis test in large samples}]\label{TH2FWER} \label{theorem_fwer}
The strong-sense frequentist FWER of the Bayesian model-averaged hypothesis test in Definition \ref{define_bmatest}, which rejects all null hypotheses $\omega_{\bs v} = \{ \theta : \beta_j = 0 \ \forall \ v_j=0 \}$, $\omega_{\bs v} \in \Omega$, for which ${\PO}_{\mathcal{A}_{\V}:\mathcal{O}_{\V}}\geq\tau$, can be bounded asymptotically, for large sample size:
\begin{eqnarray} \label{theorem_fwer_eq1}
    \alpha &\overset{\mathrm{pw}}{\lesssim}& \Pr\left( \chi^2_1
    \geq 2 \log \dfrac{\tau}{\nu \, \mu \, \sqrt{\xi_n}} \right), 
    \quad n\rightarrow\infty \label{alpha_fwer}
\end{eqnarray}
where pw indicates that convergence to the bound is pointwise with respect to $\theta$, meaning that
\begin{eqnarray}
    \underset{\omega_{\bs v} \in \Omega \ \theta \in \omega_{\bs v} \ n\rightarrow\infty}{\max \ \, \sup \ \, \lim} \
    \frac{
        \Pr\left( \po_{\mathcal{A}_{\V}:\mathcal{O}_{\V}} \geq \tau;\; \theta \right)
    }{
        \Pr\left( \chi^2_1
    \geq 2 \log \tfrac{\tau}{\nu \, \mu \, \sqrt{\xi_n}} \right)
    } &\leq& 1.
\end{eqnarray}
\begin{proof} By Definition \ref{definitionctp}, Theorem \ref{theorem_bayesCTP}, Definition \ref{define_bmatest} and Theorem \ref{theorem_fpr}. See Appendix \ref{proofT3}.
\end{proof}
\end{theorem}

\begin{corollary}[\bf{The model-averaged posterior odds can be transformed into an asymptotic adjusted $p$-value; Doublethink}] \label{Corr7} \label{corollary_po2padj}
The Bayesian model-averaged test in Definition \ref{define_bmatest}, which rejects all null hypotheses $\omega_{\bs v} = \{ \theta : \beta_j = 0 \ \forall \ v_j=0 \}$, $\omega_{\bs v} \in \Omega$, for which ${\PO}_{\mathcal{A}_{\V}:\mathcal{O}_{\V}}\geq\tau$, is equivalent to a frequentist test that rejects the same nulls when their asymptotic adjusted $p$-values $p^{\star}_{\mathcal{A}_{\V}:\mathcal{O}_{\V}}$ are less than or equal to the asymptotic FWER $\alpha$ (Equation \ref{theorem_fwer_eq1}), where
\begin{eqnarray}
\label{Pstardef}
p^{\star}_{\mathcal{A}_{\V}:\mathcal{O}_{\V}}  &:=& \Pr\left( \chi^2_1 \geq 2\log \dfrac{\PO_{\mathcal{A}_{\V}:\mathcal{O}_{\V}}}{\nu \, \mu \sqrt\xi_n } \right), \qquad n\rightarrow\infty.
\end{eqnarray}
We call this approach, which (by Theorem \ref{theorem_bayesCTP}) simultaneously controls the Bayesian FDR, locally and globally, at level $1/(1+\tau)$, Doublethink.
\end{corollary}

\begin{corollary}[\bf{Scaling of bounds on the Bayesian FDR and the asymptotic FWER}] \label{corollary_scaling_fdr_fwer}
The asymptotic bound on the FWER, $\alpha$, scales approximately linearly with (a) the bound on the Bayesian FDR, $1/(1+\tau)$, (b) the number of variables $\nu$, (c) the prior odds that $\beta_j\neq0$, $\mu$, (d) the square root of the prior precision, $\sqrt{h}$, and (e) the inverse square root of the sample size, $1/\sqrt{n}$:
\begin{eqnarray}
    \alpha &\sim& \frac{\nu\,\mu\,\sqrt{h}}{\tau}\,\sqrt{\frac{n_0}{n}}\, \Pr\left(\chi^2_1 \geq 2 \log \sqrt{n_0} \right), \quad n\rightarrow\infty,\quad n_0 = O(n).
\end{eqnarray}
The bounds are not necessarily tight.
\begin{proof}
    By Theorem \ref{theorem_fwer}, noting that the function $G(x) = \Pr\left(\chi^2_1 \geq 2 \log x\right)$ is regularly varying in $x$ with index $\lambda = 1$ which, by the definition of regular variation \citep{karamata1933mode}, satisfies
    \begin{eqnarray}
        \lim_{x\rightarrow\infty} \frac{G(c\, x)}{G(x)} = c^{-\lambda}, \qquad \forall\ c>0,\ \lambda\neq 0.
    \end{eqnarray}
\end{proof}
\end{corollary}

\section{Inflation in a simplified two-variable model}

\begin{definition}[\bf{Two-variable model with simplifying assumptions}]\label{define_twovariablemodel}
    To study convergence we consider a normal linear model with $\nu=2$ regression coefficients (the variables-of-interest, $\beta$), an intercept (the nuisance parameter $\gamma$), a known variance $\sigma^2$, and likelihood proportional to $p({\bs y} ; {\bs x}, \theta) = \prod_{i=1}^n  f_{\mathcal{N}(0, 1)}\left((y_i - \gamma - \beta_1\, x_{i1} - \beta_2\, x_{i2} )/\sigma\right)$. We assume the variables are standardized with means $\mathbb{E}[x_{\cdot 1}]=\mathbb{E}[x_{\cdot 2}]=0$, variances $\mathbb{E}[x_{\cdot 1}^2]=\mathbb{E}[x_{\cdot 2}^2]=1$, and correlation coefficient $\mathbb{E}[x_{\cdot 1}\, x_{\cdot 2}]=\rho$. 
    
    There are three null hypotheses of interest, $\omega_{\bs v} = \{ \theta : \beta_j = 0 \ \forall \ v_j=0 \}$, with $\bs v=(0, 1)^T$,  $\bs v=(1, 0)^T$ or $\bs v=(0, 0)^T$. 
\end{definition}

\begin{theorem}[\bf{Uniform convergence in a simplified two-variable model}]\label{theorem_uniformconvergence_twovariablemodel}
In the simplified two-variable model (Definition \ref{define_twovariablemodel}), testing the constrained null hypothesis $\omega_{\bs v}^{*}$ subject to a minimum effect size assumption $\omega_{\bs v}^{*}=\{\theta \in \omega_{\bs v}: |\beta_j|\notin (0, \beta_\mathrm{min})\ \forall\ j=1\dots\nu\}$, the FPR
\begin{eqnarray}
\alpha^{*}_{\bs v} &:=& \sup_{\theta \in \omega^*_{\bs v}}\ \Pr\left( \po_{\mathcal{A}_\V:\mathcal{O}_\V} \geq \tau;\; \theta \right) 
\end{eqnarray}
converges asymptotically, for large sample size, to
\begin{eqnarray}    
    {\alpha^*_{\bs v}} &\overset{\mathrm{u}}{\lesssim}& {\Pr\left( \chi^2_1
    \geq 2 \log \dfrac{\tau}{(\sizeV) \, \mu  \sqrt{\xi_n}} \right)}, \qquad n\to\infty,
\end{eqnarray}
where u indicates that convergence is uniform  (in the sense of \citet{leeb2008sparse}, versus pairwise in Theorem \ref{theorem_fpr}) with respect to $\theta$, meaning that
\begin{eqnarray} \label{theorem_uniformconvergence_twovariablemodel_eq1}
    \underset{n\rightarrow\infty \ \theta \in \omega^*_{\bs v}}{\lim\ \; \sup} \ 
    \dfrac{ \Pr\left( \po_{\mathcal{A}_\V:\mathcal{O}_\V} \geq \tau;\; \theta \right) }{\Pr\left( \chi^2_1
    \geq 2 \log \tfrac{\tau}{(\sizeV) \, \mu  \sqrt{\xi_n}} \right)} &\leq& 1.
\end{eqnarray}

\begin{proof}
    See Appendix \ref{proof_theorem_uniformconvergence_twovariablemodel}. 
\end{proof}
\end{theorem}

\begin{corollary} [\bf{Inflation in the simplified two-variable model}] \label{corollary_inflationbounds}
In the simplified two-variable model (Definition \ref{define_twovariablemodel}), with fixed $n$ and $\bs v=(0, 1)^T$, the FPR can be inflated, with inflation factor $\iota$, such that $\Pr\left( \po_{\mathcal{A}_\V:\mathcal{O}_\V} \geq \tau \ ; \theta \right) = \Pr\left( \chi^2_1 \geq 2 \log \frac{\tau}{\mu \sqrt{\xi_n} \, \iota} \right) \sim \iota\, \Pr\left( \chi^2_1 \geq 2 \log \frac{\tau}{\mu \sqrt{\xi_n}} \right)$. 
    The inflation factor has range $\iota \in \left(\frac{\mu \sqrt{\xi_n}}{1 + \mu \sqrt{\xi_n}}, \frac{1 + \mu \sqrt{\xi_n}}{\mu \sqrt{\xi_n}} \right)$.
At its worst, $\Pr\left( \po_{\mathcal{A}_\V:\mathcal{O}_\V} \geq \tau \ ; \theta \right) \sim \Pr\left( \chi^2_1 \geq 2 \log \tau \right)$. See SI Equations \ref{proof_th4_stochastic_bound}--\ref{proof_th4_inflation_factor}.
\end{corollary}

\begin{figure}
    \centering
    \includegraphics[width=0.9\linewidth]{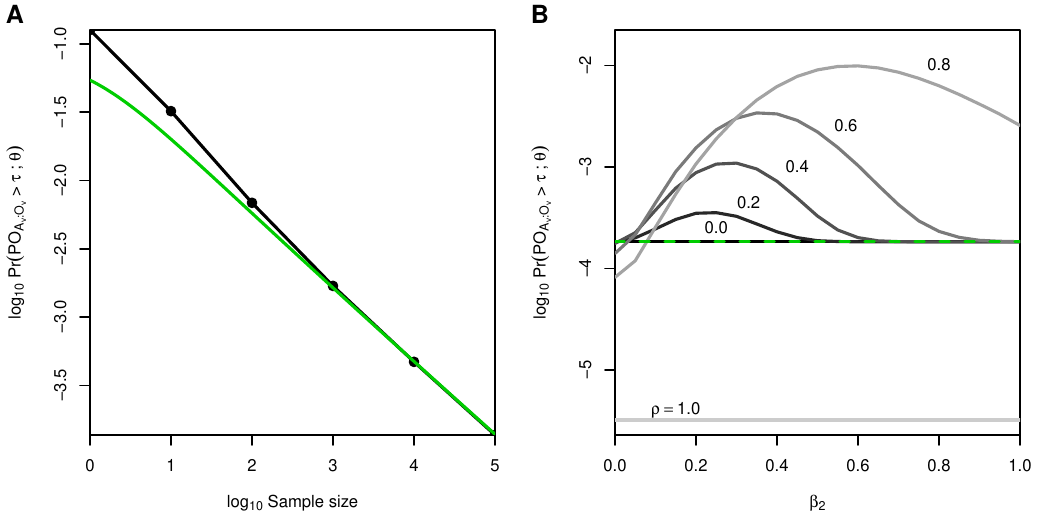}
    \caption{\smaller Inflation in the simplified two-variable model testing $\beta_1=\beta_2=0$ (\textbf{A}) and $\beta_1=0$ (\textbf{B}). \textbf{A}: FPR as a function of sample size: simulations (black line) and Theorem \ref{theorem_fpr} (green line). Assuming $\mu=1$, $h=1$, $\tau=9$, $\rho=0$. \textbf{B}: FPR as a function of $\beta_2$ for $\rho\in[0.0, 1.0]$ (shaded grey lines, labelled by $\rho$) and Theorem \ref{theorem_fpr} (green dashed line). Assuming $n=145$, $\mu=0.1$, $h=1$, $\sigma=1$, $\tau=9$. Each panel is based on 10 million simulations.}
    \label{fig:simulations_null_inflation}
\end{figure}

There are two sources of inflation in the simplified two-variable model, illustrated by Figure \ref{fig:simulations_null_inflation}. The first source (Figure \ref{fig:simulations_null_inflation}A) are higher-order tests involving more than one-degree-of-freedom (here, $\PO_{\bs 1:\bs 0}$), whose contribution is assumed negligible by Theorem \ref{theorem_fpr}. As the sample size increases, this source of inflation recedes and the FPR tends to its asymptotic level.

The second source (Figure \ref{fig:simulations_null_inflation}B) is correlation between variables that are being tested, even if they genuinely have zero effect ($x_{\cdot 1}$ here), and variables not being tested, that have non-zero effect ($x_{\cdot 2}$ here). Inflation peaks at an effect size $\beta_2$ that depends on the value of $\rho$. For the smallest effect sizes, deflation is evident for higher $|\rho|$. For the largest effect sizes, inflation abates to its asymptotic level. Worst-case inflation increases with $|\rho|$, from none when $\rho=0$, to the worst case as $|\rho|\rightarrow 1$. However, at the point $|\rho|=1$, deflation is observed regardless of $\beta_2$. Negative $\rho$ and $\beta_2$ are not shown because the behaviour is symmetric.

This second source of inflation does not abate as the sample size increases; instead, worst-case inflation increases and manifests at ever-smaller values of $\beta_2$, inversely proportional to $\sqrt{n}$. At its worst, the FPR, instead of scaling inversely with $\sqrt{n}$, as per Corollary \ref{corollary_scaling_fdr_fwer}, remains on the order of the FDR, by Corollary \ref{corollary_inflationbounds}.

 Whereas it is impossible to control the minimum effect size in real data analysis, 
 it is possible to mitigate the inflating effect of highly correlated variables, as we will discuss later (Definition \ref{define_fwer_rho}). Another countermeasure would be to increase either $\mu$ or $\xi_n$ via $h$; however, this would diminish the advantages of Bayesian model-averaging in terms of sparsity, variance reduction and over-fitting. 

\section{Application to Mendelian randomization study of age-related macular degeneration}

To trial Doublethink in a real data analysis, we applied it to a Mendelian randomization study of age-related macular degeneration (AMD).

\begin{example}[\bf{Mendelian randomization study of age-related macular degeneration}]
    \citet{zuber2020selecting} used two-sample multivariable Mendelian randomization \citep{sanderson2019examination} to study the causal effects of $\nu = 49$ biomarkers on AMD, and introduced a Bayesian model-averaging approach, MR-BMA. The aim was to identify which biomarkers had a non-zero effect on risk of AMD. In their model, the parameters-of-interest, $\beta$, were the direct causal effects of each biomarker on AMD. The outcome data $\bf y$ were $z$-scores for the direct effect of $n=145$ genetic variants on AMD, estimated by a genome-wide association study (GWAS) \citep{fritsche2016large}. For each genetic variant, the variables $x_{i \cdot}$ were $z$-scores of its direct effects on the 49 biomarkers, estimated by another GWAS \citep{kettunen2016genome}. In computing the $z$-scores, a common standard error was used across all genetic variants, applying the standard error used in the computation of $y_i$ to compute the $z$-scores in $x_{i \cdot}$. There was one nuisance parameter, the error variance, $\gamma$, expected theoretically to equal one. This yielded the normal linear regression $p({\bf y}; {\bf x}, \theta)=\prod_{i=1}^n f_Z\left((y_i - \beta_1\, x_{i 1} - \dots - \beta_\nu\, x_{i \nu})/\sqrt{\gamma} \right)$, where $f_Z$ represents a standard normal density function.
\end{example}

In their Bayesian model-averaged analysis of AMD, the biomarkers for which \citet{zuber2020selecting} found the strongest evidence of a causal effect on AMD were XL.HDL.C and L.HDL.C (extra-large and large high-density lipoprotein cholesterol), with posterior probabilities 0.70 and 0.23, corresponding to Bayes factors of 21 and 2.7, respectively. 
Considering that neither would be considered `strong' evidence according to Jeffreys' \citeyearpar{jeffreys1939theory} scale, and the prior influences Bayes factors anyway, this begs the questions: (a) How would a frequentist evaluate this level of evidence? (b) Would the evidence against the null hypothesis of no effect have been stronger by grouping XL.HDL.C and L.HDL.C together?

In partial answer to the first question, \citet{julian2023phenome} implemented a permutation procedure, which for this dataset yields unadjusted $p$-values
below 0.005 for both XL.HDL.C and L.HDL.C (200 permutations, taking 80 hours). However, such procedures generally control the FWER only in the weak sense, i.e.\ when the grand null is true. The $e$-value approach of \citet{vovk2021values} suggests maximum $p$-values (computed as inverse Bayes factors) of $0.048$ and $0.37$ respectively, unadjusted for multiple testing, although its requirement for no nuisance parameters in the null hypothesis is not met here. 
The leave-one-out tests of Example \ref{example_fwer_regression} yielded unadjusted $p$-values of $0.90$ and $0.16$, probably a dilution of the signal due to many correlated variables.

Since the number of biomarkers was too large to fit all $2^{49} \approx 10^{15}$ models, and since many biomarkers were highly correlated, we selected a subset of $15$ variables, for which we exhaustively analysed all $2^{15}=32~768$ models. We selected the 15 variables in a way that attempted to preserve most of the important variability, and the correlation structure, in the data:
    (1) We ranked the biomarkers by their two-sample univariable Mendelian randomization $p$-values for a causal association with AMD.
    (2) From the most to the least significant, we introduced 15 biomarkers iteratively.
    (3) We omitted biomarkers with an absolute correlation exceeding $|\rho|=0.8$ with two or more biomarkers that we had already introduced.
The 15 remaining biomarkers ranged from the first to the 24\textsuperscript{th} most significantly associated with AMD. Three pairs showed correlation coefficients above 0.80: XL.HDL.C and L.HDL.C ($\rho=0.82$); ApoB and IDL.TG ($\rho=0.91$); S.HDL.TG and S.VLDL.TG ($\rho=0.92$). The other nine biomarkers we included, in decreasing order of univariable association, 
were ApoA1, LDL.D, Ace, XL.HDL.TG, VLDL.D, M.HDL.C, His, Ala and Gln.

An important caveat is that the analysis of only a subset of the candidate variables affects the interpretation of those results. The next theorem clarifies the valid interpretation of such results.

\begin{theorem}[\bf{Testing based on a sub-analysis}]\label{theorem_testing_a_subset}
    Suppose we conduct a sub-analysis in which we exclude a subset of variables $\{j \in \{1, \dots, \nu\} : {v}^-_j=0\}$, in a Bayesian model-averaged setting (Definition \ref{define_bmatest}). Let $\Theta^- = \omega_{\V^-}$ define a reduced parameter space and let $\mathcal{S}^- = \{ {\bs s} \in \mathcal{S}: \Theta_{\bs s} \subseteq \Theta^- \}$ define a reduced model space. Let $\mathcal{O}^-_{\V}\; =\mathcal{O}_{\V}\,\cap \,\mathcal{S}^-$ and $\mathcal{A}^-_{\V}=\mathcal{A}_\V \cap\mathcal{S}^-$ be the sets of models compatible with any null and corresponding alternative hypothesis, respectively. In the sub-analysis, the model-averaged posterior odds against the null hypothesis $\omega_{\bs v} \subseteq \Theta^-$ are
    \begin{align}
        \po_{\mathcal{A}^-_\V:\mathcal{O}^-_\V} &= \frac{\sum_{{\bs s} \in \mathcal{A}^-_\V} \po_{\bs s}}{\sum_{{\bs s} \in \mathcal{O}^-_\V} \po_{\bs s}}.
    \end{align}
   Now $\po_{\mathcal{A}^-_\V:\mathcal{O}^-_\V}$ is a lower bound on the posterior odds against the intersection null hypothesis $\omega_{\V \odot \V^\prime} := \omega_\V \ \cap\  \omega_{\V^\prime}$, where $\odot$ denotes elementwise multiplication, i.e.\ 
    \begin{align}
        \po_{\mathcal{A}^-_\V:\mathcal{O}^-_\V} &\leq \po_{\mathcal{A}_{\V \odot \V^\prime}:\mathcal{O}_{\V \odot \V^\prime}}
\intertext{but it is not a bound on the posterior odds against the null hypothesis $\omega_\V$, i.e.\ }
        \po_{\mathcal{A}^-_\V:\mathcal{O}^-_\V} &\nleq \po_{\mathcal{A}_\V:\mathcal{O}_\V} .
    \end{align}
    Therefore rejection of a null hypothesis $\omega_\V$ in a sub-analysis only implies rejection of the intersection null hypothesis $\omega_{\V \odot \V^\prime}$ in the full analysis and does not imply rejection of $\omega_{\V}$ in the full analysis.
\begin{proof}
    $\mathcal{A}^-_\V \subseteq \mathcal{A}_{\V \odot \V^\prime}$ and $\mathcal{O}^-_\V = \mathcal{O}_{\V \odot \V^\prime}$ whereas $\mathcal{A}^-_\V \subseteq \mathcal{A}_\V$ and $\mathcal{O}^-_\V \subseteq \mathcal{O}_\V$.
\end{proof}
\end{theorem}


We analysed the 15 biomarkers with Doublethink, assuming a prior odds of variable inclusion of $\mu=0.1$ and a prior precision parameter $h=1$. Table \ref{tab:mr_doublethink} reports the five most significant associations between biomarkers and AMD risk. The top two variables were the same as \citet{zuber2020selecting} found, XL.HDL.C and L.HDL.C, with posterior odds of 1.75 and 0.58, corresponding to Bayes factors of 17 and 5.8, and unadjusted asymptotic $p$-values (by Corollary \ref{corollary_po2punadj}) of 0.001 and 0.004. 

However, these tests should be properly interpreted as intersection null hypotheses with the 34 excluded variables, according to Theorem 5. 
After adjusting for $\nu=49$ biomarkers (by Corollary \ref{corollary_po2padj} or Bonferroni correction), neither was significant, even at $\alpha=0.05$. This begs the second question, is there evidence that any of the biomarkers had non-zero causal effects on AMD risk?

\begin{table}
    \centering \smaller
    \begin{tabular}{|l|cccc|}
    \hline
         Biomarker&  Posterior odds&  Asymptotic $p$-value&  Point estimate& Standard error
\\
         &  &  (unadjusted)&  (posterior mean)& (root posterior variance)
\\ \hline
         XL.HDL.C&  1.75&  0.001&  0.33& 0.26
\\
         L.HDL.C&  0.58&  0.004&  0.14& 0.20
\\
         Gln&  0.08&  0.033&  -0.03& 0.11
\\
         IDL.TG&  0.06&  0.048&  -0.01& 0.05
\\
         S.VLDL.TG&  0.03&  0.091&  -0.01& 0.04\\
    \hline         
    \end{tabular}
    \caption{\smaller Two-sample multivariable Mendelian randomization in a Doublethink sub-analysis of 15 biomarkers in \citet{zuber2020selecting} for direct effects on AMD risk. Top 5 variables shown.}
    \label{tab:mr_doublethink}
\end{table}

In fact there was strong evidence that at least one of the biomarkers directly affected AMD risk. The posterior odds against the null hypothesis that none of the 15 biomarkers affected AMD risk were 3~688, corresponding to 
an unadjusted asymptotic $p$-value (by Corollary \ref{corollary_po2punadj}) of $6.0\times10^{-6}$. 
After adjusting for $\nu=49$ variables, the asymptotic $p$-value (by Corollary \ref{corollary_po2padj}) was $2.0\times10^{-5}$.

The smallest group of significant variables comprised XL.HDL.C and L.HDL.C. The posterior odds against the null hypothesis that neither affected AMD risk was 89, corresponding to 
an unadjusted asymptotic $p$-value of $3.4\times10^{-5}$. After sub-analysis adjustment (for 36 variables; Theorem \ref{theorem_testing_a_subset}), 
the asymptotic $p$-value, adjusted for $\nu=49$ variables, was $1.0\times10^{-3}$. This shows that the combined test was able to establish significance (in a Bayesian sense at an FDR $<0.012$ and in a frequentist sense at an asymptotic FWER $<0.0011$), whereas the individual tests were not.

Not only is power increased by testing an intersection null hypothesis that no members of a group have an effect, but consideration of FWER control in the presence of correlation suggests that highly correlated variables should always be treated as a group, as the next section considers.

Note that the Doublethink combined test (Corollary \ref{corollary_po2padj}) was significant even when $p$-value-based combined tests were not. We applied four methods that control the FWER via unadjusted $p$-values: Bonferroni \citeyearpar{bonferroni1936teoria} correction, two multilevel procedures (\citet{hommel1988stagewise} implemented by \citet{meijer2019hommel}, and \citet{wilson2019tradeoffs}) based on Simes' \citeyearpar{simes1986improved} test, and the harmonic mean $p$-value (HMP) procedure \citep{wilson2019harmonic}. We also applied the \citet{benjamini1995controlling} procedure, which controls a frequentist FDR. We furnished these procedures with unadjusted $p$-values from Corollary \ref{corollary_po2punadj}, which test the null hypothesis of no effect for each of the 15 variables individually (or rather, the intersection of each with the 34 excluded variables, as per Theorem \ref{theorem_testing_a_subset}). None of these tests rejected the null hypothesis that neither XL.HDL.C nor L.HDL (nor the 34 excluded variables) influence AMD risk, at $\alpha=0.05$. None of them even rejected the grand null hypothesis that none of the biomarkers affected AMD risk. 
The much-reduced significance among these $p$-value-based combination tests compared to Doublethink likely reflects the information bottleneck when reducing the data to individual $p$-values before combining them.

\section{Inflation between highly correlated variables: simulation approach}

In human GWAS, genetic variants that are physically linked tend to be correlated, so that signals of association often manifest as clusters of significant genetic variants.
When several linked genetic variants are significant, the assumption is that most are not causally associated. So we reject the intersection hypothesis that none of the clustered variants are associated, but we do not reject all the elementary hypotheses, as a straightforward interpretation of statistical significance would suggest, because that would probably produce many false positives among the clustered variants. In this section we formalize the idea and show through simulation that the way statistical significance is interpreted has a direct bearing on FWER control.

\begin{definition}[\bf{Frequentist control of FWER$_\rho$}] \label{define_fwer_rho}
    In the regression problem (Definition \ref{define_regression_problem}), where the binary vector ${\bs s} \in \mathcal{S} = \{0, 1\}^\nu$ indexes the null hypotheses such that $\beta_j=0\ \forall\ s_j=0$, where $\beta_j$ is the coefficient for variable $x_{\cdot j}$, define the correlation coefficient between variables $j$ and $k$ to be $\rho_{j,k}$. Define the maximum absolute correlation between variables whose regression coefficients are constrained versus unconstrained under null hypothesis $\bs s$ to be
    \begin{eqnarray}
        \rho^\mathrm{max}_{\bs s} &=& \begin{cases}
        0 & \bs s\in\{\bs 0,\bs 1\}\\\max_{j, k \, :\, s_j \neq s_k} |\rho_{j,k}|&\textrm{otherwise}\end{cases}
    \end{eqnarray}
    The idea is to group the most correlated variables indivisibly, in order to exclude null hypotheses with the highest $\rho^\mathrm{max}_{\bs s}$ from the set of all tested hypotheses, $\Omega$, because they are most susceptible to inflation. We define $\Omega_\rho = \{ \omega_{\bs s} \in \Omega :  \rho^\mathrm{max}_{\bs s} \leq \rho \}$. By definition, $\Omega_\rho$ remains closed under intersection. We define 
    \begin{eqnarray}
    \mathrm{FWER}_\rho &:=& \underset{\omega_{{\bs s}} \in \Omega_\rho \ \theta \in \omega_{{\bs s}}}{\max \ \sup} \ \Pr\left( \phi_{\bs s}({\bs y}) = 1 ; {\bs x}, \theta \right) \nonumber\\
    &\leq& \max_{\omega_{\bs s} \in \Omega_\rho} \ \alpha_{\bs s}.
    \end{eqnarray}
    This implies the regular FWER (Definition \ref{definitionctp}) equals FWER$_1$. Since $\mathrm{FWER}_\rho \leq \mathrm{FWER}_1$, it is a less stringent error rate to control, and control of $\mathrm{FWER}_1$ implies control of $\mathrm{FWER}_\rho$. 
\end{definition}

\begin{definition}[\bf{Asymptotically equivalent test statistic}] \label{define_asymptotically_equivalent_test}
    Lemma \ref{lemma_scaled_posterior_odds} implies that the model-averaged posterior odds against the null hypothesis $\omega_{\bs v}$, denoted $\po_{\mathcal{A}_\V:\mathcal{O}_\V}$, are asymptotically equivalent to the posterior odds of alternatives differing from the true model $\tilde{\bs s}$, $\theta\in\Theta_{\tilde{\bs s}}$ by one degree-of-freedom:
    \begin{eqnarray} \label{define_asymptotically_equivalent_test_eq1}
        \po_{\mathcal{A}_\V:\mathcal{O}_\V} &\sim& \sum_{j\,:\,v_j=0} \po_{
        \tilde{\bs s}+{\bs e}_j:\tilde{\bs s}} \ =:\ \tilde{\po}_{\mathcal{A}_\V:\mathcal{O}_\V}
    \end{eqnarray}
    as $n\rightarrow\infty$, assuming that $\tilde{\bs s}\in\mathcal{O}_\V$, where $\{{\bs e}_j\}_k = \mathbb{I}(j=k)$.
    This allows us to partition the error of Doublethink into (a) the approximation to the distribution of $\tilde{\po}_{\mathcal{A}_\V:\mathcal{O}_\V}$ and (b) the rate of convergence to the asymptotically equivalent test statistic.
\end{definition}

\begin{remark}[\bf{The $\alpha \leq 0.025$ threshold in Doublethink}] \label{remark_p0.025_threshold}
    Simulations based on the asymptotically equivalent test statistic (Definition \ref{define_asymptotically_equivalent_test}) indicate that the FWER should be set no larger than $\alpha = 0.025$ in Doublethink; equivalently, $p$-values that exceed 0.025 should be reported as not significant, or 1. This is because the chi-squared approximation (Theorem \ref{theorem_fpr}; Corollary \ref{corollary_mlr_limiting_dist}) can be anti-conservative for $\alpha\geq0.0259846$, for some values of $(\sizeV)$, even under independence, even when applied to the asymptotically equivalent test statistic. See Appendix \ref{si_section_p0.02} and SI Figure \ref{si_fig_p0.025_threshold}.
\end{remark}

\begin{definition}[\bf{Bayes FWER}] \label{definition_bayes_fwer}
    For the purpose of simulation it is convenient to define the FWER as an expectation over a (possibly degenerate) distribution of parameter values, rather than in a minimax sense (Definition \ref{definitionctp}). We define the Bayes FWER in a Bayes risk sense (see, e.g.\ \citet{berger1985bayesian}) as
    \begin{eqnarray}
        \mathrm{BFWER} 
        &:=& \int_{\Theta} \Pr\left( 
        \po_{\mathcal{A}_{\V(\theta)}:\mathcal{O}_{\V(\theta)}} \geq \tau ; {\bs x}, \theta \right) \mathrm{d}\Pi(\theta) ,
    \end{eqnarray}
    where the notation $\V(\theta)$, $v_j(\theta) = \mathbb{I}(\beta_j \neq 0)$, emphasizes that the intersection of the true null hypotheses depends on the parameter $\theta$. By analogy, we can define BFWER$_\rho$ with reference to Definition \ref{define_fwer_rho}, and the asymptotically equivalent Bayes FWER,
    \begin{eqnarray}
        \mathrm{AFWER}
        &:=& \int_{\Theta} \Pr\left( 
        \tilde{\po}_{\mathcal{A}_{\V(\theta)}:\mathcal{O}_{\V(\theta)}} \geq \tau ; {\bs x}, \theta \right) \mathrm{d}\Pi(\theta),
    \end{eqnarray}
    with the caveat that the asymptotically equivalent test (Definition \ref{define_asymptotically_equivalent_test}), does not actually define a CTP, and therefore AFWER is not a familywise error rate, for finite $n$.
\end{definition}

\begin{definition}[\bf{Worst-case Bayes FWER in Doublethink}] \label{define_evalue_doublethink}
    The $e$-value approach of \citet{vovk2021values} yields a worst-case inflation for BFWER in Doublethink when $\Pi$ is the `true' distribution of parameter values, which provides a benchmark for simulation performance.
    \begin{eqnarray}
        \mathrm{BFWER}  &=& \Pr\left( 
            \po_{\mathcal{A}_{\V(\theta)}:\mathcal{O}_{\V(\theta)}} \geq \tau ; {\bs x}, \Pi \right) \nonumber\\
            &\leq& \mathbb{E}\left[ \po_{\mathcal{A}_{\V(\theta)}:\mathcal{O}_{\V(\theta)}} ; {\bs x}, \Pi \right] / \tau \qquad \textrm{(by Markov's inequality)} \nonumber \\
            &=& \mathbb{E}\left[ \mu^{\nu-|\V(\theta)|} ; \Pi \right] / \tau \nonumber \\
            &=& \left[\left(\ 1 + \frac{\mu}{1+\mu}\right)^\nu - 1\right] / \tau \ \ \sim\ \ \frac{\nu\,\mu}{\tau}\ \textrm{when}\ \mu\ll 1. \label{define_evalue_doublethink_eq1}
    \end{eqnarray}
\end{definition}

To investigate inflation in Doublethink in a real data setting, we performed simulations based on the AMD Mendelian randomization data \citep{zuber2020selecting}. We simulated $\beta$ for $\nu=15$ variables from the Doublethink prior, assuming $\mu=0.1$ and $h=1$, fixing $\gamma=1$, and selecting the variables as before. To assess performance, we calculated BFWER and AFWER (Definition \ref{definition_bayes_fwer}).

\begin{figure}
    \centering
    \includegraphics[width=0.9\linewidth]{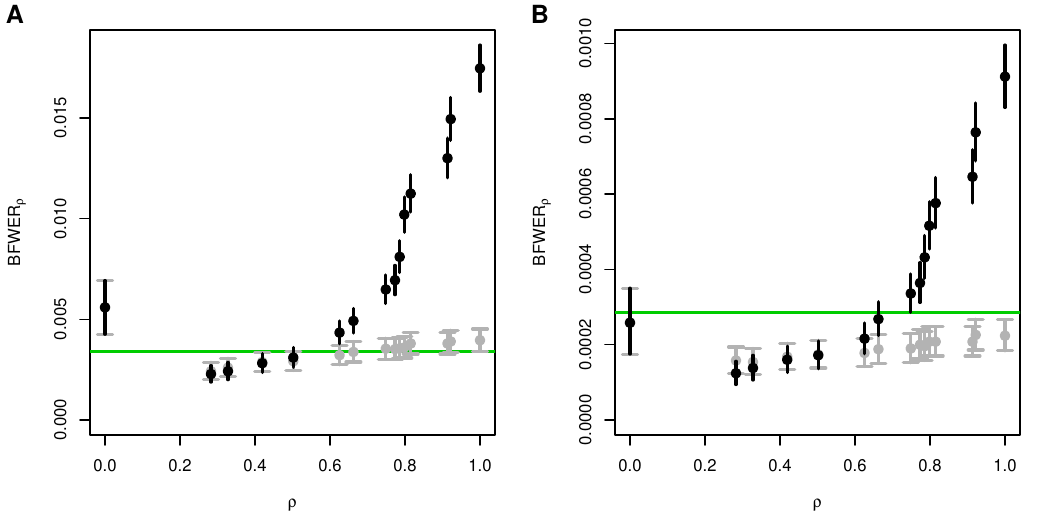}
    \caption{\smaller Inflation in simulations based on the AMD Mendelian randomization example with $\nu=15$ variables and $n=145$ (\textbf{A}) or $n=14 \ 500$ (\textbf{B}). BFWER as a function of $\rho$ (black points), versus AFWER (grey points) and Theorem 3 (green line). Assuming $\mu=0.1$, $h=1$, $\tau=9$.
    Error bars based on $50~000$ (\textbf{A}) and $500~000$ (\textbf{B}) simulations.
    }
    \label{fig:simulations_amd_example}
\end{figure}

Figure \ref{fig:simulations_amd_example}A shows that when $n=145$, BFWER$_\rho$ (black points) exhibited circa five-fold inflation at $\rho=1$, less than the 87-fold worst case inflation determined by the $e$-value approach (Definition \ref{define_evalue_doublethink}). The inflation had subsided by $\rho=0.5$, there was slight deflation around $\rho=0.3$, and 1.6-fold inflation at $\rho=0$. This pattern can be understood by comparing to the asymptotically equivalent error rate, AFWER$_\rho$ (grey points). Near $\rho=1$, BFWER$_\rho$ far exceeded AFWER$_\rho$, indicating that the model-averaged posterior odds were far from converging to the asymptotically equivalent test statistic. This reflects the difficulty in reliably attributing the causal effect between pairs of highly correlated variables. Near $\rho=0.5$, BFWER$_\rho$ had converged to AFWER$_\rho$. AFWER$_\rho$ was slightly conservative for intermediate values of $\rho$, by the conservative nature of the CTP. Near $\rho=0$ and $\rho=1$, there was inflation in AFWER$_\rho$ caused by small-sample error in the distribution of the asymptotically equivalent test statistic.

When $n=14~500$ (Figure \ref{fig:simulations_amd_example}B), the pattern was similar but worst-case deflation was reduced (3.2-fold) and BFWER$_\rho$ (black points) converged to AFWER$_\rho$ (grey points) at larger values of $\rho$, reflecting the improved ability to correctly identify the true model with larger sample size. Theorem 3 (green line) was no longer an anti-conservative approximation to AFWER$_\rho$ at any value of $\rho$, reflecting better performance of the asymptotic approximation at a larger sample size.

With either sample size, BFWER$_\rho$ was less inflated for smaller $\rho$ because the set of true null variables was less likely to split groups of highly correlated variables. For $\rho=1$, the set of true null variables
$\{j : \beta_j = 0 \}$ often included variables that were highly correlated with other variables outside the set. This caused inflation of the posterior odds. When $\rho=0.5$, elementary hypotheses concerning highly correlated variables were removed from $\Omega$, the set of null hypotheses under investigation. Although $\Omega$ still contained groups of the same hypotheses, these groups were indivisible. Therefore the set of true null hypotheses (which can be smaller than under $\rho=1$, but not larger) no longer split highly correlated variables, so inflated variables were not left out of the set of true null hypotheses. When $\rho=0$, there was a single group of all variables, the null hypothesis which corresponds to the grand null. Controlling FWER$_0$ is therefore equivalent to a single combined test with weak-sense control.

These results show that not only it is beneficial, in terms of power, to group correlated variables, but also advisable in terms of controlling the FWER. This formalizes existing practice in terms of interpreting statistically significant signals, as per the GWAS example. Interestingly, decisions concerning choice of $\rho$ for FWER$_\rho$ can be deferred until post-analysis, in the same way that decisions concerning choice of $\tau$ for the FDR and $\alpha$ for the FWER can be deferred. This means the analysis can be executed across all elementary hypotheses, and the level $\rho$ in FWER$_\rho$ control determined when the results come to be interpreted.

Comparing these results to the sub-analysis of the AMD Mendelian randomization data in the previous section, it suggests that any attempt to attribute causal effects within the three pairs of highly correlated variables (XL.HDL.C and L.HDL.C, ApoB and IDL.TG, S.HDL.TG and S.VLDL.TG; $\rho>0.8$) would be prone to inflated familywise error, although the Bayesian FDR would still be properly controlled, contingent on the prior. 
Therefore, not only do we conclude that XL.HDL.C and L.HDL.C form a significant group in the sub-analysis, but from a frequentist perspective it is advisable to exercise restraint in attempting to attribute the signal to XL.HDL.C versus L.HDL.C.






\section{Comparison to related approaches: simulations}


We used simulations to compare the performance of Doublethink to related approaches: LASSO \citep{tibshirani1996regression}, elastic net \citep{zou2005regularization}, ridge regression \citep{hoerl1970ridge}, backward elimination \citep{venables2002modern} using the BIC \citep{schwarz1978estimating}, and MR-BMA \citep{zuber2020selecting}. To these we added as benchmarks the grand null model, the grand alternative model, and the `oracle' or true model \citep{fan2001variable}, as if it were known.

To compare methods for hypothesis testing, we deployed some approaches (LASSO, elastic net, MR-BMA and Doublethink) both natively, and as the first step in a model selection procedure. In the model selection version (taking the maximum \emph{a posteriori} model for the Bayesian methods), we fitted the data without regularization and performed $\nu$ leave-one-out or add-one-in hypothesis tests. To investigate the effect of prior assumptions in Doublethink, we fitted the model at combinations of $\mu = 0.05$, $0.1$ and $0.2$, and $h = 0.25$, $1$ and $4$.

For testing intersection null hypotheses, which only Doublethink supported natively (Corollary \ref{corollary_po2padj}), we employed $p$-value combination approaches: the Bonferroni \citeyearpar{bonferroni1936teoria} procedure, a multilevel Simes' procedure \citep{wilson2019tradeoffs}, Hommel's \citeyearpar{hommel1988stagewise} procedure, which is also based on Simes' \citeyearpar{simes1986improved} test, implemented by \citet{meijer2019hommel}, and the HMP procedure \citep{wilson2019harmonic}. For comparison, we applied the same $p$-value combination procedures to Doublethink's unadjusted $p$-values (Corollary \ref{corollary_po2punadj}), substituting any $p$-values above 0.02 with 1 (following Remark \ref{remark_p0.025_threshold}).

Simulations were based on the AMD Mendelian randomization data \citep{zuber2020selecting}, with $n=145$. We simulated $\beta$ for $\nu=15$ variables from the Doublethink prior, assuming $\mu=0.1$ and $h=1$, fixing $\gamma=1$, and selecting 15 of the 49 biomarker variables as before. Again, we computed the Bayes FWER (Definition \ref{definition_bayes_fwer}) as an average over the simulations.

We compared the following metrics: computation time, out-of-sample prediction error (measured as the root mean squared error, or L2 norm), estimator error (L2 norm), coverage of the standard error, Bayes FWER for type I (false positive) errors, and rates of strikeout (wrongly calling every test) for type II (false negative) errors.
We calculated the FWER overall and for tests of individual variables only (marginal), pairs of variables only (pairwise), and the global test that all variables are null (headline). The target FWER was $\alpha=0.01$. In total we performed $10~000$ simulations.

\emph{Computation time.} On average, most methods took under 0.01 seconds to run in R \citep{Rcore2024}, the exceptions being LASSO (0.08s), elastic net (0.86s), MR-BMA (5.5s) and Doublethink (49s). The 5.5s run-time of MR-BMA, which is efficiently implemented, reflects its exhaustive search of all $32~768$ models. The 49s run-time of Doublethink reflects its exhaustive search, native support for testing intersection null hypotheses, and lack of optimization.

\emph{Prediction error}
\begin{figure}
    \centering
    \includegraphics[width=0.9\linewidth]{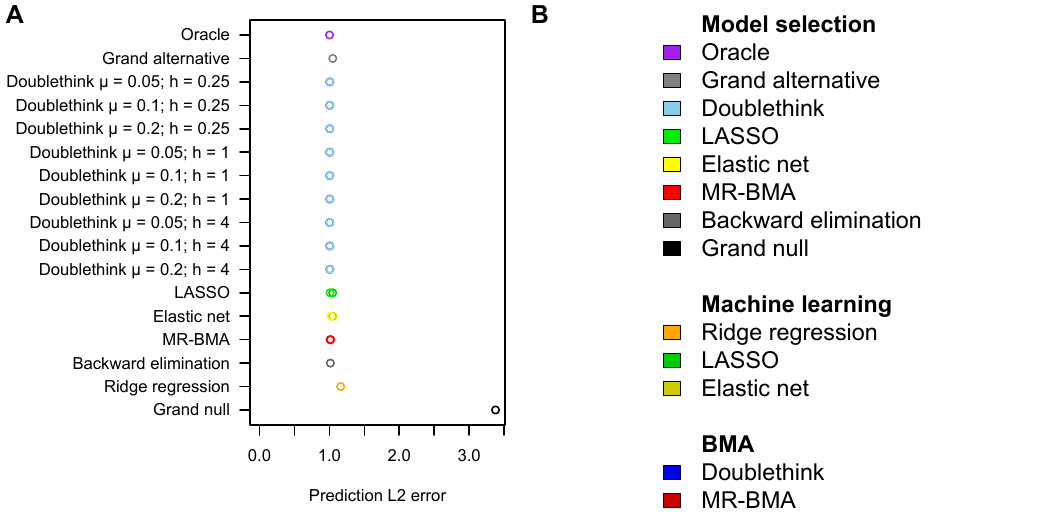}
    \caption{\smaller Out-of-sample prediction error in $10~000$ simulations from the Doublethink prior with $\mu=0.01$ and $h=1$. The key applies to all figures in this section.}
    \label{fig:sims-prediction-error}
\end{figure}
We computed out-of-sample prediction error (root mean squared error, i.e.\ L2 norm) by simulating pairs of datasets for each simulated parameter vector. The first was used for training the model, the second for testing prediction error. Most methods achieved an error below 1.05, the exceptions being ridge regression (1.16) and the grand null model (3.4) (Figure \ref{fig:sims-prediction-error}). The latter reflects the worst case because no coefficients were fit.

\emph{Estimator error.}
\begin{figure}
    \centering
    \includegraphics[width=0.9\linewidth]{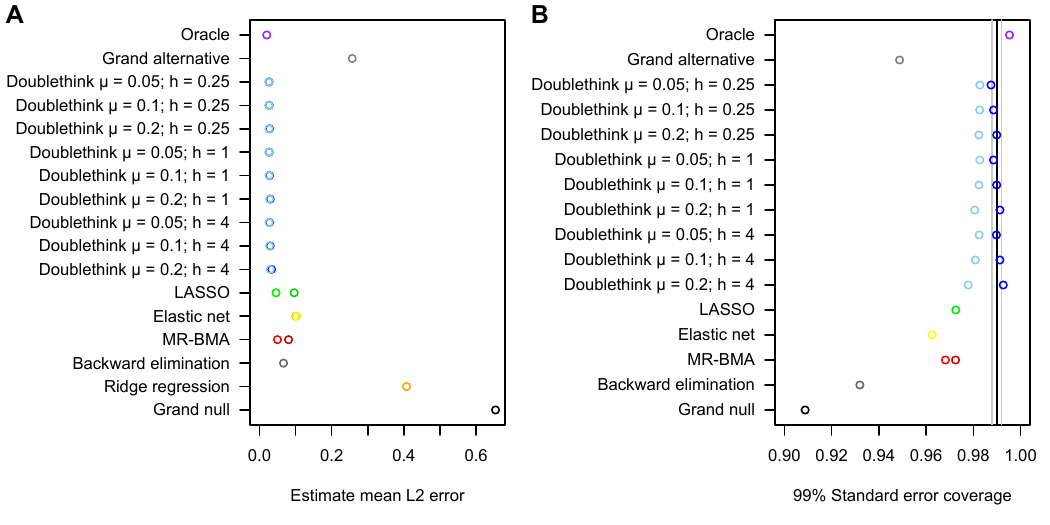}
    \caption{\smaller Estimator error (\textbf{A}) and standard error coverage (\textbf{B}) in $10~000$ simulations from the Doublethink prior with $\mu=0.01$ and $h=1$. Expected coverage (black line) is shown, with allowance for Monte Carlo error (grey lines; 95\% confidence interval).}
    \label{fig:sims-parameter-estimation}
\end{figure}
Methods differed substantially in their estimator error (Figure \ref{fig:sims-parameter-estimation}A). The oracle model was close to optimal (root mean squared error 0.020). Doublethink's Bayesian model-averaged estimates and model selection-based estimates performed closely (below 0.033), for all assumed values of $\mu$ and $h$. Surprisingly, the model selection-based estimates out-performed the native estimates for LASSO and MR-BMA (error below 0.050). Backward elimination (0.067) and elastic net (0.100) followed. Poor estimates were obtained under the grand alternative model (0.257) and ridge regression (0.407), presumably due to over-fitting caused by lack of sparsity. The grand null model, in which all parameters were returned as zero, under-fitted, returning the worst case estimator error (0.654).

\emph{Standard error coverage.}
We defined coverage to be the proportion of simulations in which a 99\% confidence interval, constructed from the standard error assuming a Normal sampling distribution, included the true parameter value (Figure \ref{fig:sims-parameter-estimation}B). The oracle model was conservative (99.5\% coverage), with Doublethink model averaging the best calibrated of the methods (98.7--99.3\% coverage), followed by Doublethink model selection (97.8\%--98.3\%). LASSO, MR-BMA and elastic net model selection were less well calibrated (97.3\%, 96.8\% and 96.3\% respectively), while MR-BMA model averaging performed similarly (97.2\%). Three methods produced the most anti-conservative standard errors: the grand alternative model (94.9\%), backward elimination (93.2\%) and the grand null model (90.9\% coverage). 

\emph{Marginal Bayes FWER.}
\begin{figure}
    \centering
    \includegraphics[width=0.9\linewidth]{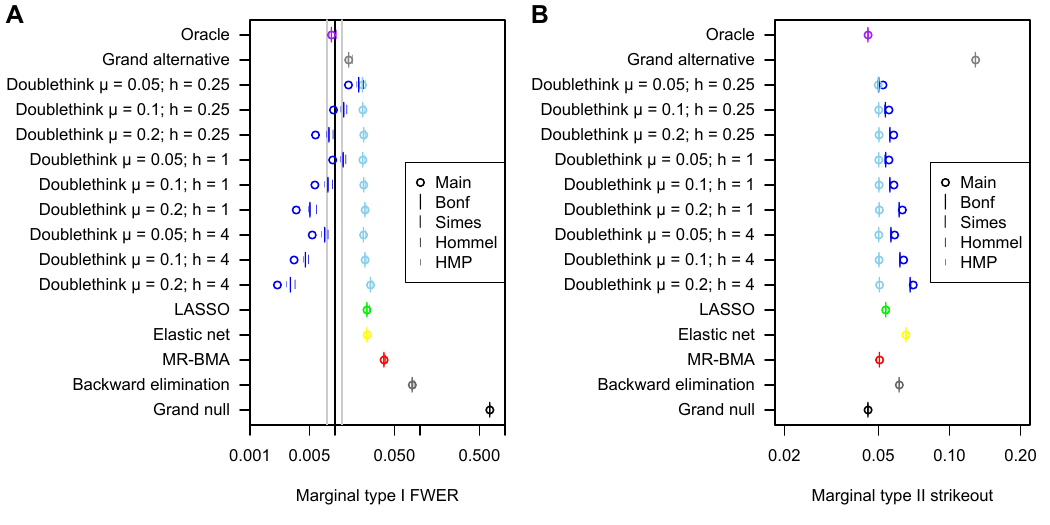}
    \vspace*{-5mm}
    \caption{\smaller Type I Bayes FWER (\textbf{A}) and type II strikeout rate (\textbf{B}) for marginal tests of the significance of individual variables in $10~000$ simulations from the Doublethink prior with $\mu=0.01$ and $h=1$. Expected type I BFWER (black line) is shown, with allowance for Monte Carlo error (grey lines; 95\% confidence interval).}
    \label{fig:sims-marginal-bfwer}
    \centering
    \includegraphics[width=0.9\linewidth]{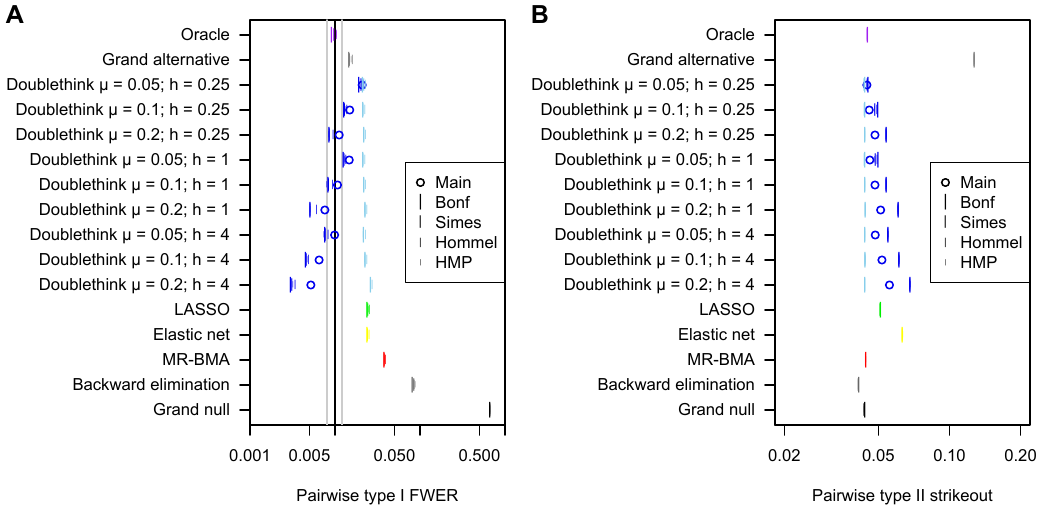}
    \vspace*{-5mm}
    \caption{\smaller Type I Bayes FWER (\textbf{A}) and type II strikeout rate (\textbf{B}) for pairwise tests.}
    \label{fig:sims-pairwise-bfwer}
    \centering
    \includegraphics[width=0.9\linewidth]{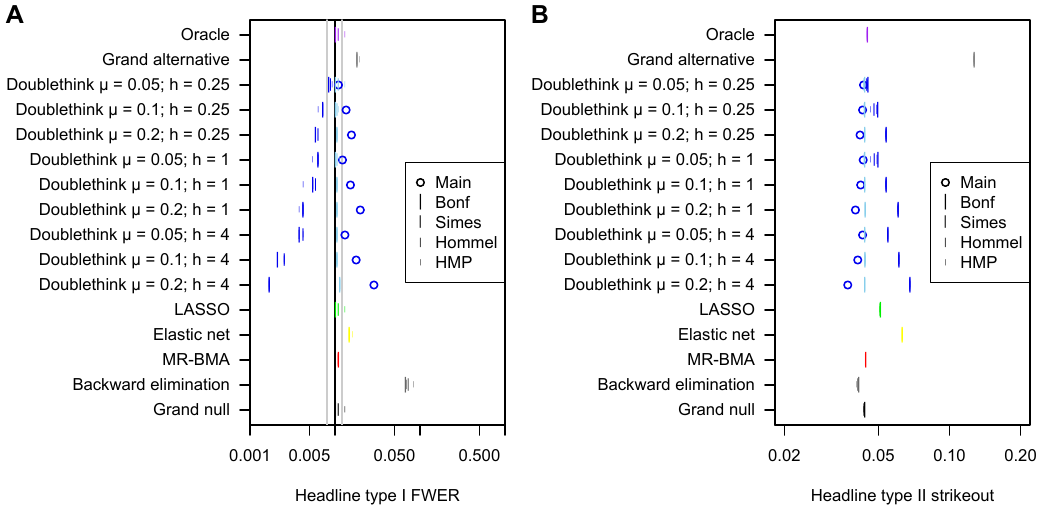}
    \vspace*{-5mm}
    \caption{\smaller Type I Bayes FWER (\textbf{A}) and type II strikeout rate (\textbf{B}) for headline tests.}
    \label{fig:sims-headline-bfwer}
\end{figure}
To compare the false positive (type I error) rates across methods, we began by focusing only on marginal tests for the significance of individual variables. Figure \ref{fig:sims-marginal-bfwer}A, circles, shows that again the oracle method was close to optimal in terms of marginal type I Bayes FWER, nearing the target of $\alpha=0.01$ by achieving 0.0091. The grand alternative model was slightly anti-conservative (0.015). Doublethink model averaging showed a range of errors, from the conservative, 0.002 (when assuming $\mu=0.2$, $h=4$), to the anti-conservative, 0.015 (assuming $\mu=0.05$, $h=0.25$). Doublethink, LASSO and elastic net model selection produced anti-conservative errors ranging from 0.021--0.0026, followed by MR-BMA model selection (0.038), backward elimination (0.081) and the grand null model (0.664), which was strongly anti-conservative, as expected.

For every method capable of producing $p$-values, we were able to apply $p$-value combination methods (Bonferroni, Simes, Hommel and HMP procedures). Results were usually similar across procedures, so Figure \ref{fig:sims-marginal-bfwer}A shows these as vertical lines of differing heights that often overlap. For Doublethink model averaging, they combine the unadjusted $p$-values from Corollary \ref{corollary_po2punadj}, instead of the adjusted $p$-values from Corollary \ref{corollary_po2padj}. They differed from the latter in the anti-conservative direction. For all other methods, the main result utilized Bonferroni correction, so these results coincided, and the other combination tests differed very little.

\emph{Marginal strikeout rates.}
To compare power, we defined the type II strikeout rate as the probability of failing to detect all true signals, which corresponds, for an individual dataset, to 100\% false negative rate. We restrict attention to methods that did not exhibit appreciably anti-conservative BFWER. For marginal tests of the significance of individual variables, the oracle model showed a strikeout rate of 0.045, providing a reference point for other methods (Figure \ref{fig:sims-marginal-bfwer}B). The grand alternative model exhibited the worst strikeout rate of 0.129, demonstrating that it is slightly anti-conservative and strongly under-powered, an undesirable combination. Doublethink model-averaging exhibited strikeout rates of 0.052-0.070. This range of performance across different prior assumptions reflected the corresponding degree of conservatism of the tests (Figure \ref{fig:sims-marginal-bfwer}A).

\emph{Pairwise Bayes FWER.}
Pairwise tests of the intersection null hypothesis that neither of a pair of variables has an effect are expected to be more powerful than marginal tests alone, but the additional multiple testing burden increases the risk of false positives. Figure \ref{fig:sims-pairwise-bfwer}A shows the BFWER for pairwise tests. Only Doublethink model averaging performed these tests natively (circles). For all other methods, including Doublethink model selection, combined tests were performed using the $p$-value combination methods (vertical bars). The pairwise BFWER was barely higher than the marginal BFWER for these combination tests, which is unsurprising since (a) no new information is introduced and (b) the closed testing procedures on which they are based are designed to control the FWER. For Doublethink model averaging, the pairwise BFWERs were all higher than the marginal BFWERs (Figure \ref{fig:sims-marginal-bfwer}A), ranging from the conservative, 0.005, to the anti-conservative, 0.021. This reflects the method's exploitation of the full data for pairwise tests, which is potentially more powerful, and the limitation of the asymptotic approximation for controlling the FWER.

\emph{Pairwise strikeout rates.}
The greater power of the Doublethink approach to pairwise testing can be seen in Figure \ref{fig:sims-pairwise-bfwer}B. Pairwise strikeout rates were lower than marginal strikeout rates (Figure \ref{fig:sims-marginal-bfwer}B), ranging from 0.045--0.056, and rivalling the oracle strikeout rate of 0.045 (albeit for the anti-conservative prior $\mu=0.05$, $h=0.25$). The pairwise strikeout rate for the grand alternative model was nearly unchanged at 0.127.

\emph{Headline Bayes FWER.}
The headline FWER (Figure \ref{fig:sims-headline-bfwer}A) is calculated assuming the grand null hypothesis is true. As before, the oracle model was well calibrated (0.010--0.013) which coincided, in this case, with the grand null hypothesis. Methods that were not generally well calibrated (LASSO, elastic net, MR-BMA and Doublethink model selection) showed headline BFWER in the range 0.010--0.016. Backward elimination remained anti-conservative (0.067--0.084), as did the grand alternative model (0.018--0.019). Combination tests based on Doublethink Corollary \ref{corollary_po2punadj} were conservative (0.002--0.009). Doublethink Corollary \ref{corollary_po2padj} was anti-conservative under the grand null hypothesis (0.011--0.029). While undesirable, earlier simulations suggested this source of inflation should abate for larger sample sizes (Figure \ref{fig:simulations_amd_example}A versus \ref{fig:simulations_amd_example}B). 

\emph{Headline strikeout rates.}
In a closed testing procedure, when there are true signals to be found, the headline or `global' test is the most powerful. The headline strikeout rates (Figure \ref{fig:sims-headline-bfwer}B) are calculated assuming the grand null hypothesis is false. The oracle model again provided a benchmark, 0.045. The grand alternative remained under-powered at 0.127. Combination tests based on Doublethink Corollary \ref{corollary_po2punadj} were under-powered. Headline strikeout error rates for Doublethink model averaging outperformed the oracle model, ranging from 0.037--0.043, but this reflected their small-sample anti-conservatism (Figure \ref{fig:sims-headline-bfwer}A).

\emph{Overall Bayes FWER and FWER$_\rho$.}
\begin{figure}
    \centering
    \includegraphics[width=0.9\linewidth]{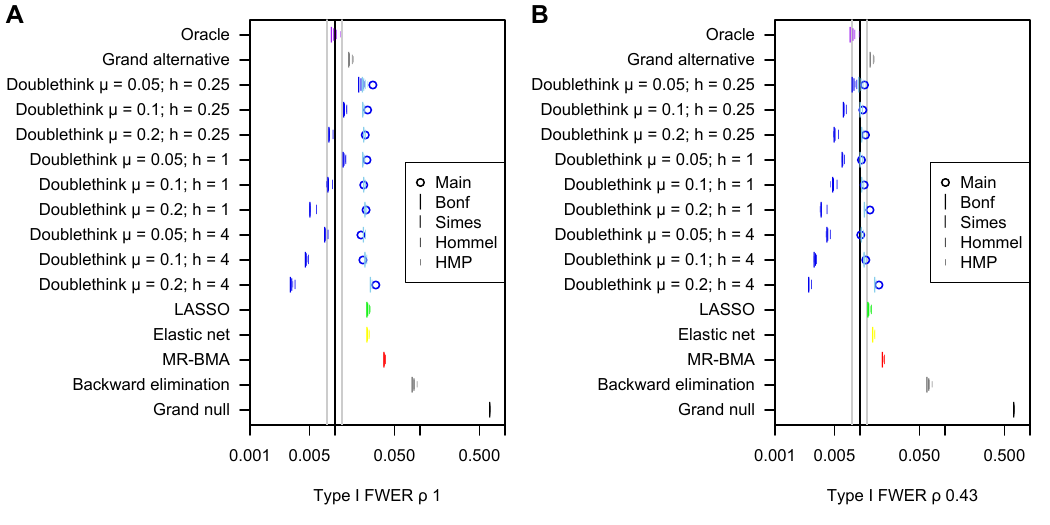}
    \caption{\smaller Type I Bayes FWER for $\rho=1$ (\textbf{A}) and $\rho=0.43$ (\textbf{B}) for all tests of the significance of \emph{any} group of variables in $10~000$ simulations from the Doublethink prior with $\mu=0.01$ and $h=1$. Expected type I BFWER (black line) is shown, with allowance for Monte Carlo error (grey lines; 95\% confidence interval).}
    \label{fig:sims-bfwer-rho}
\end{figure}
We calculated the overall Bayes FWER and the corresponding Bayes FWER$_\rho$, in which variables with absolute correlation coefficients exceeding $\rho$ are grouped together indivisibly, so that sub-groups cannot be tested (Definition \ref{define_fwer_rho}). BFWER$_\rho$ was inflated for Doublethink model averaging at $\rho=1$ (Figure \ref{fig:sims-bfwer-rho}A), but this was mitigated at $\rho=0.43$ (0.010--0.017; Figure \ref{fig:sims-bfwer-rho}B). $\mathrm{BFWER}_{0.43} \leq \mathrm{BFWER}_1$ for all methods, but some remained inflated, including backward elimination (0.061--0.071) and the grand null model (0.644--0.650). The oracle model was well-calibrated at both BFWER$_1$ and BFWER$_{0.43}$ (0.008--0.011).

Importantly, the headline strikeout rate applies irrespective of choice of $\rho$, so power can be achieved at this headline level, while inflation in the FWER is mitigated by defining indivisible groups of null hypotheses. This allowed Doublethink model averaging to compete with the oracle model in terms of both Bayes FWER (Figure \ref{fig:sims-bfwer-rho}B) and headline power (Figure \ref{fig:sims-headline-bfwer}B). 


\section{Discussion}

Often, the goal of hypothesis testing is to identify a few influential variables among many candidates, reflecting the scientific principal of explaining complex observed patterns by simple unobserved processes. For accurately estimating parameters, quantifying their uncertainty, and controlling false positives, model choice is vital, as the simulations showed. Bayesian model averaging is especially useful when the model is uncertain, but for decades frequentists and Bayesians have debated hypothesis testing, seemingly irreconcilably (e.g.\ \citet{edwards1963bayesian, berger1987testing, bayarri2004interplay, gelman2008objections}).

Perhaps it is surprising, therefore, that Bayesian hypothesis testing can be formulated as a closed testing procedure \citep{marcus1976closed} that simultaneously controls the Bayesian FDR and the strong-sense frequentist FWER (Theorem \ref{theorem_bayesCTP}). Particularly surprising because FWER, in contrast to FDR, is often characterized as “an extremely strict criterion which is not always appropriate” \citep{storey2003positive}. Theorem \ref{theorem_bayesCTP} is general, independent of prior or modelling assumptions, complementing known results like Bayesian consistency \citep{doob1949application, schwartz1965onbayes} and the frequentist admissibility of Bayesian procedures \citep{wald1950statistical, berger1985bayesian}. It addresses both the broad question of when Bayesian and frequentist philosophies converge, and the practical question of how to choose significance thresholds. Whereas it may often be convenient to pre-determine one or the other, reporting both the FWER and FDR thresholds offers an opportunity to bridge Bayesian and frequentist approaches to hypothesis testing

In big data regression settings, Theorems \ref{theorem_fpr} and \ref{theorem_fwer} extend work by \citet{johnson2005bayes, johnson2008properties} to interconvert FPR, FWER and FDR thresholds. Corollaries \ref{corollary_po2punadj} and \ref{corollary_po2padj} interconvert the unadjusted and adjusted $p$-values and posterior odds. These results are less general than Theorem \ref{theorem_bayesCTP}, comprising large-sample limits that require assumptions including an appropriate family of prior distributions and independent observations. Clearly their utility depends on the relevance of the Johnson model (Definition \ref{define_johnson}) to the problem at hand. In general the convergence results are pointwise, rather than uniform, with respect to the parameters, subjecting their use to finite-sample inflation. Keeping in mind such caveats, these results nevertheless afford insights into the relationship between the frequentist approach to multiple testing and Bayesian prior assumptions. Corollary \ref{corollary_scaling_fdr_fwer} states that the FWER threshold is asymptotically proportional to (a) the number of variables, $\nu$; (b) the prior odds, $\mu$; and (c) the square root of the prior precision $h$; and asymptotically inversely proportional to (d) the Bayesian threshold, $\tau$; and (e) the square root of the sample size, $n$. The FWER and FDR bounds -- which are not necessarily tight -- are approximately inversely proportional.

Corollary \ref{corollary_scaling_fdr_fwer} recapitulates the Jeffreys-Lindley paradox \citep{jeffreys1939theory, lindley1957statistical}: Bayesian inference is consistent for a fixed FDR threshold, $\tau$, \citep{doob1949application}, whereas the practice of fixing the FWER threshold, $\alpha$, irrespective of sample size, $n$ (e.g.\ at 0.05 or 0.005 \citep{benjaminwagenmakers}), is not consistent because there exists a tangible probability of falsely rejecting a null hypothesis no matter how large the sample \citep{ohagan1995fractional}. This can be solved by inversely scaling $\alpha$ with $\sqrt{n}$, originally advocated by Jeffreys \citep{wagenmakers2023history}, from which point one could report both the FWER and FDR bound.

Interconverting model-averaged posterior odds and adjusted $p$-values using Doublethink (Corollary \ref{corollary_po2padj}) has several limitations that stem from the underlying assumptions. These assumptions include (a) a large sample size, which implies that a Normal approximation to the likelihood is reasonable locally; (b) data that comprise independent realizations of a process satisfying technical conditions which imply locally unique, consistent and asymptotically Normal maximum likelihood estimators \citep{cox1974theoretical}; (c) a specific family of conjugate priors parameterized by $\mu$ and $h$; (d) local alternatives, which implies that the Normal approximation holds in the vicinity of null and alternative hypotheses; and (e) technical assumptions like joint convergence to the limiting distributions. The local alternatives assumption is convenient for deriving an asymptotic distribution for the maximized likelihood ratio \citep{cox1974theoretical, johnson2005bayes, johnson2008properties}, yet it implies that non-zero parameter values scale with $\sqrt{n}$, contrary to consistency and our prior, though this may be a side-issue for large $n$, because the evidence quickly becomes decisive \citep{kass1995reference}. The conjugate prior is convenient for pursuing joint Bayesian-frequentist inference (e.g.\ \citet{zellner1986assessing, liang2008mixtures}), but its covariance structure, based on the Fisher information, is otherwise hard to justify. \citet{kass1995reference} investigated other priors for which the BIC (and by implication the Johnson model) offer a reasonable approximation. The independent data assumption restricts the generality of the approach. It is not recommended to apply Doublethink with $\alpha>0.025$; $p$-values larger than $0.025$ should be interpreted as 1 (Remark 1), but this is hardly limiting if $\alpha$ is to scale inversely with $\sqrt{n}$.

Of principal concern, the large sample assumption means the asymptotic results may be inaccurate for finite samples, as evidenced by the simulations. Two sources of inflation were found: (a) the contribution of higher-order tests involving more than one-degree-of-freedom to the posterior odds, which vanishes asymptotically, and (b) the lack of convergence of the posterior odds to the asymptotically equivalent test statistic (Definition \ref{define_asymptotically_equivalent_test}). The latter is caused by correlation between variables that have zero and non-zero effects. The simplified two-variable model (Definition \ref{define_twovariablemodel}) attributed this inflation to specific parameter values that scale inversely with $\sqrt{n}$, and therefore threaten inflation regardless of $n$. In simulations, not only Doublethink but all methods were affected by such inflation (Figure \ref{fig:sims-bfwer-rho}), which invalidates the fundamental assumption that test statistics are well calibrated under the null. To address it, we formalized the practice of making inferences about indivisible groups of correlated variables, rather than the individual variables concerned, through the notion of FWER$_\rho$ (Definition \ref{define_fwer_rho}). In simulations, this mitigated, but did not eradicate, inflation at $\rho=0.43$.


Hypothesis testing, particularly multiple testing, has been considered a long and difficult-to-reconcile conflict between Bayesian and frequentist approaches \citep{bayarri2004interplay}, in constrast to parameter estimation (e.g. \citet{doob1949application, borwanker1971bernstein}). Doublethink is connected to other approaches that bridge this gap, notably $e$-values \citep{vovk2021values}. Application of the $e$-value approach to the Doublethink model (Definition 7) allows interconversion of the Bayesian FDR and the Bayes FWER (Definition \ref{definition_bayes_fwer}), assuming the prior is correctly specified. The $e$-value bound on the Bayes FWER (Definition \ref{define_evalue_doublethink}) is then approximately proportional to (a) the number of variables, $\nu$; and (b) the prior odds, $\mu$; but not (c) the square root of the prior precision $h$; and inversely proportional to (d) the Bayesian threshold, $\tau$; but not (e) the square root of the sample size, $n$. Therefore the $e$-value bound on the Bayes FWER is conservative, in the large-sample limit, by a factor of approximately $\sqrt{h/n}$, but conversely, the Doublethink threshold on the Bayes FWER could be anti-conservative, in a finite sample, by the corresponding amount.


Another related approach, which extends FWER, is the false discovery proportion (FDP) approach of \citet{goeman2011multiple, goeman2019simultaneous}. The FDP approach comprises a `multilevel' closed testing procedure \citep{meinshausen2008hierarchical} in which all possible combinations of hypotheses are simultaneously controlled via pre-determined thresholds. This means arbitrary combinations of hypotheses can be tested and significant groups of variables freely identified post-hoc. The approach is inherent to Bayesian control of both the FDR and, by Theorem \ref{theorem_bayesCTP}, the FWER. Multilevel testing challenges concepts like fishing for significance, data dredging, and $p$-hacking  \citep{andrade2021harking}.

The Bayesian FDR considered here motivated, but remains distinct from, frequentist FDR concepts. Doublethink controls both the `local' and `global' Bayesian FDR \citep{efron2001empirical}, rather than a frequentist FDR  controlled by Benjamini and Hochberg’s \citeyearpar{benjamini1995controlling} or Storey’s \citeyearpar{storey2002direct, storey2003positive} procedures. One connection to Theorem \ref{theorem_bayesCTP} is that controlling the frequentist FDR controls the FWER at the same level, but only in the weak sense \citep{benjamini1995controlling}. Whereas Corollary \ref{corollary_scaling_fdr_fwer} indicates the FWER bound is $O(\sqrt{n})$ smaller than the Bayesian FDR bound.

Doublethink belongs to a class of combination tests that exploit heavy-tailed distributions, including the Cauchy combination test \citep{liu2020cauchy} and the HMP procedure (\citet{wilson2019harmonic}), through its reliance on the theory of regular variation \citep{karamata1933mode, nagaev1965limit}. The HMP offers model averaging, starting with $p$-values, whereas Doublethink pursues joint Bayesian-frequentist model-averaging, beginning with nested maximized likelihood ratios. A theoretical advance over the HMP is the ability of Doublethink to consider model uncertainty in the null hypothesis, as well as in the alternative hypothesis. Another is the clarification that the relevant limit is $n\rightarrow\infty$, rather than $\nu\rightarrow\infty$. By Corollary \ref{corollary_mlr_limiting_dist}, the model-averaged deviance asymptotically follows a chi-squared distribution with one degree of freedom. This mirroring of the null distribution of the classical likelihood ratio test statistic \citep{wilks1938large, wald1943tests} emerges from the self-similarity or fractal property of sums of heavy-tailed random variables \citep{mandelbrot1968fractional, taqqu1978representation}. A practical advance, demonstrated by the simulations, is the greater power of testing groups of correlated variables using a coherent Bayesian approach, rather than by combining $p$-values derived from tests of the individual variables.

This paper focuses on the frequentist properties of Bayesian model averaging. It implies a limited defence of $p$-values, specifically those based on likelihood ratio tests. Despite their flaws, here $p$-values offer a useful complement to Bayesian reasoning because their distribution under the null hypothesis is defined independently of a prior. The ability to interconvert model-averaged $p$-values with Bayesian sufficient statistics like the model-averaged deviance or posterior odds implies a reprieve from the central criticisms of $p$-values -- their lack of evidentiary value and violation of the likelihood principal (see, e.g.\ \citet{birnbaum1962foundations}).  Simultaneous Bayesian-frequentist hypothesis testing addresses another major flaw of $p$-values -- their indifference towards power -- because the prior determines the performance of the test under the alternative hypothesis. By characterizing the asymptotic properties of Bayesian tests under the null hypothesis, even when the prior is wrong, we have aimed to exploit the advantages of Bayesian model-averaged hypothesis testing while mitigating the subjective influence of the prior by controlling both the FDR and the FWER at a quantifiable level.

\section{Data Availability Statement}
R \citep{Rcore2024} and Nextflow \citep{di2017nextflow} code for figures, tables and simulations has been deposited at \url{https://github.com/danny-wilson/dblthk_amd_sims}. Data files for simulation output have been deposited at \url{https://doi.org/10.5281/zenodo.15360270}.

\section{Acknowledgements and Funding}
The authors wish to thank Jelle Goeman and Peter Gr\"{u}nwald for comments and advice on the paper, and the Mathematisches Forschungsinstitut Oberwolfach, organizers and participants of workshop 2308 Design and Analysis of Infectious Disease Studies. This work was funded by the Robertson Foundation, the Wellcome Trust, and the Royal Society (grant no.\ 101237/Z/13/B).



\bibliographystyle{chicago}
\bibliography{bib.bib}

%% file: DTsupp.tex
\appendix
\counterwithin{equation}{section} 
\renewcommand{\theequation}{\thesection\arabic{equation}}
\setcounter{page}{1}

\begin{center}
    {\large Supplementary information for\\
    \MakeUppercase{\bf Doublethink: simultaneous Bayesian-frequentist\\
    model-averaged hypothesis testing}}
\end{center}
\begin{center}
    {\large Helen R. Fryer, Nicolas Arning, Daniel J. Wilson}
\end{center}
\begin{center}
    {\large 9 May 2025}
\end{center}
\vspace{-1cm}

\section{Regularity conditions}
\label{regCon}
 The following regularity conditions, previously listed by \cite{davidson1970limiting} and Johnson and colleagues (\cite{johnson2008properties,hu2009bayesian}), are assumed. A more detailed explanation of each condition is provided in \cite{davidson1970limiting}.

For almost all $y \in \mathcal{Y}$, $\bs x \in \mathcal{X}$  and  all $\theta \in \Theta$ and $j,k,l=1,...,\nu+\zeta$.

(a) $\frac{\partial \log p(y;\bs x, \theta)}{\partial\theta_j}$, $\frac{\partial^2 \log p(y;\bs x, \theta)}{{\partial\theta_j}{\partial\theta_k}}$ and $\frac{\partial^3\log p(y;\bs x, \theta)}{{\partial\theta_j}{\partial\theta_k}{\partial\theta_l}}$ exist for 

(b) There exists arbitrary functions, $F_{j}(y)$ and $F_{jk}(y)$, that are integrable over $\mathcal{Y}$, such that $\left|\frac{\partial \log p(y;\bs x, \theta)}{\partial\theta_j}\right|<F_{j}(y)$ and  $\left|\frac{\partial^2 \log p(y; \bs x, \theta)}{{\partial\theta_j}{\partial\theta_k}}\right|<F_{jk}(y)$.

(c) The matrix $ {\mathcal{I}}(\theta)$ with elements  \begin{eqnarray}
{\mathcal{I}}(\theta)_{jk}=\mathbb{E}_{\theta}\left[\left({{{\frac{\partial \log p(y;\bs x, \theta)}{\partial\theta_j}}{\bigg|}}_{\theta}}\right)\left({{{\frac{\partial \log p(y;\bs x, \theta)}{\partial\theta_k}}{\bigg|}}_{\theta}}\right) \right]
\end{eqnarray}
is positive definite with a finite determinant.

(d)
\begin{eqnarray}
   {\left|{{{\frac{\partial^3 \log p(y;\bs x,\theta)}{\partial\theta_j\partial\theta_k\partial\theta_l}}{\bigg|}}_{\theta}}\right|}<H_{jkl}(y|\bs x)
   \end{eqnarray}
where there exists an $M>0$ such that $\mathbb{E}_{\theta}\left[H_{jkl}(y;\bs x)\right]<M<\infty$, and $\kappa$, $L>0$, such that 
\begin{eqnarray}
E_{\theta}\left[|H_{jkl}(y;\bs x)-\mathbb{E}_{\theta}[{H_{jkl}(y;\bs x)]}|^{1+\kappa}\right]<L<\infty
\end{eqnarray}

 (e) There exists $\psi,T>0$ such that, whenever $\parallel\theta^{**}-\theta^{*}\parallel\equiv\sum_{j=1}^{\nu+\zeta} \mid\theta^{**}_j-\theta^{*}_j\mid<\psi, \theta^{**}, \theta^{*}\in\Theta$
 \begin{eqnarray}
   \mathbb{E}_{\theta}{\left[\left\{{{{\frac{\partial^3 \log p(y;\bs x, \theta)}{\partial\theta_j\partial\theta_k\partial\theta_l}}{\bigg|}}_{\theta}}\right\}^2\right]}<T<\infty
\end{eqnarray}

(f) There exists $\xi,K>0$, such that
\begin{eqnarray}
   \mathbb{E}_{\theta}{\left[{\left|{{{\frac{\partial \log p(y;\bs x, \theta)}{\partial\theta_j}} {\bigg|}}_{\theta}}\right|}^{2+\eta}\right]}<K<\infty
\end{eqnarray}
\pagebreak
\section{Background theory} \label{secBackThe}

\subsection{Closed testing procedures control the familywise error rate in the strong sense} \label{background_on_ctps}
\citet*{marcus1976closed} introduced the closed testing procedure (CTP) to control the frequentist family-wise error rate (FWER) in the strong sense. Suppose that random element $\bs y$ has a probability mass or density function $p({\bs y} ; {\bs x}, \theta)$ that depends on auxiliary data $\bs x$ and parameters $\theta \in \Theta$.
The aim is to test a set of hypotheses
defined by $\Omega = \{ \omega_{\bs s} \}$, where $\omega_{\bs s} \subset \Theta$ and $\bs s$ is an index. 
$\Omega$ must be closed under intersection, meaning $\omega_{\bs s}, \omega_{\bs s^\prime} \in \Omega$ implies $\omega_{\bs s} \cap \omega_{\bs s^\prime} \in \Omega$. A `local' test controls the false positive rate (FPR) at level $\alpha_{\bs s}$ by rejecting $\theta \in \omega_{\bs s}$ when $\psi_{\bs s}({\bs y})=1$ (rather than 0) such that:
\begin{eqnarray} \label{eq_ctp_fpr}
    \alpha_{\bs s} &:=& \sup_{\theta \in \omega_{\bs s}} \Pr\left(\psi_{\bs s}({\bs y}) = 1 ; {\bs x}, \theta\right).
\end{eqnarray}
A CTP rejects $\theta \in \omega_{\bs s}$ when $\phi_{\bs s}({\bs y}) = 1$, which indicates rejection of all intersection hypotheses in $\Omega$:
\begin{eqnarray}
    \phi_{\bs s}({\bs y}) &=& \min_{\omega_{\bs r} \subseteq \omega_{\bs s} } \psi_{\bs r}({\bs y}). \label{A_CTP_rejects_null_when}
\end{eqnarray}
By construction, rejection of the null hypothesis $\omega_{\bs s}$ by the CTP implies rejection of any and all intersection hypotheses too:
\begin{eqnarray} \label{eq_ctp_logic}
    \phi_{\bs s}({\bs y})=1 &\implies& \phi_{\bs r}({\bs y})=1 \qquad \forall\ \omega_{\bs r} \subseteq \omega_{\bs s} .
\end{eqnarray}
If this is an inherent property of a set of local tests (achieved, for example, by setting the local significance thresholds appropriately), then we have a procedure
\begin{eqnarray} \label{eq_ctp_shortcut}
    \phi_{\bs s}({\bs y}) &=& \psi_{\bs s}({\bs y})
\end{eqnarray}
known as a shortcut CTP. Shortcut procedures are more computationally efficient (c.f. Equation \ref{A_CTP_rejects_null_when}).

The purpose of the CTP is to control the FWER, at a level determined by the local tests. The FPR of test $\phi$ is related to the FPR of local test $\psi$ by
\begin{eqnarray} \label{eq_ctp_fpr_phi}
    \sup_{\theta \in \omega_{\bs s}} \Pr\left( \phi_{\bs s}(\bs y)=1 ; \theta \right) &=& \sup_{\theta \in \omega_{\bs s}} \Pr\left( \bigcap_{\omega_{\bs r}\subseteq\omega_{\bs s}} \left\{ \psi_{\bs r}(\bs y)=1 \right\} ; \theta \right) \nonumber\\
    &\leq& \sup_{\theta \in \omega_{\bs s}} \Pr\left( \psi_{\bs s}(\bs y)=1 ; \theta \right) \ =\ \alpha_{\bs s}.
\end{eqnarray}

Let $\omega_{\tilde{\bs s}}$ denote the intersection of all true null hypotheses in $\Omega$. The CTP controls the FWER given $\theta \in \omega_{\tilde{\bs s}}$ at level
\begin{eqnarray}
\mathrm{FWER}_{\tilde{\bs s}} &:=& 
    \sup_{\theta \in \omega_{\tilde{\bs s}}} \Pr\left( \bigcup_{\omega_{\tilde{\bs s}} \subseteq \omega_{{\bs s}}} \left\{\phi_{\bs s}({\bs y}) = 1\right\} ; {\bs x}, \theta \right) \nonumber\\
    &=& \sup_{\theta \in \omega_{\tilde{\bs s}}} \Pr\left( \phi_{\tilde{\bs s}}({\bs y}) = 1 ; {\bs x}, \theta \right) \nonumber \\
    &\leq& \alpha_{\tilde{\bs s}}
\end{eqnarray}
by the logic of the CTP (Equation \ref{eq_ctp_logic}) and the FPR of test $\phi$ (Equation \ref{eq_ctp_fpr_phi}). In other words, for a familywise error to occur, it is necessary and sufficient that $\theta\in\omega_{\tilde{\bs s}}$ be rejected by test $\phi_{\tilde{\bs s}}(\bs y)$. Since $\omega_{\tilde{\bs s}}$ is unknown, sometimes the weak-sense FWER is used, defined as $\mathrm{FWER}_{\bs 0}$, but this may be anti-conservative when some null hypotheses are false. More often the strong-sense FWER is used, defined as
\begin{eqnarray}
\mathrm{FWER} 
    &:=& \max_{\omega_{\bs s} \in \Omega} \ \mathrm{FWER}_{\bs s} \nonumber \\
    &\leq& \max_{\omega_{\bs s} \in \Omega} \ \alpha_{\bs s}.
\end{eqnarray}

\subsection{Likelihood assumptions for simultaneous Bayesian-frequentist hypothesis testing}

\subsubsection{Likelihood assumptions statement}
We assume that the sample size, $n$, is sufficiently large that the likelihood of $\theta$ constrained by null hypothesis $\bs s$ (i.e., $\theta\in{\omega}_{\bs s}$) approximates that of a Multivariate Normal distribution centered around the maximum likelihood estimate (MLE), i.e., $\hat{\theta}^{\bs s}=\textrm{arg max}_{\theta \in \omega_{\bs s}}\, p(\bs y ; \bs x,\theta)$.
\begin{eqnarray}
\label{normallikform}
p(\bs y ; \bs x,\theta) & \approx & p(\bs y ; \bs x,\hat\theta^{\bs s}) \, \exp\left\{
- \frac{n}{2} \left( \theta^{\phantom{()}}_{\F_{\bs s}} - \hat\theta^{\bs s}_{\F_{\bs s}} \right)^T {\mathcal{I}^{\phantom{()}}_{\F_{\bs s},\F_{\bs s}}}(\tilde\theta) \left( \theta^{\phantom{()}}_{\F_{\bs s}} - \hat\theta^{\bs s}_{\F_{\bs s}} \right)
 \right\}\nonumber\\
&\propto&f_{\mathcal{N}_{|\bs s|+\zeta}}\left(\hat\theta^{\bs s}_{\F_{\bs s}}\,;\,\theta_{\F_{\bs s}}, n^{-1}  \left[\mathcal{I}(\tilde\theta)_{\F_{\bs s},\F_{\bs s}}\right]^{-1} \right)\qquad  (\theta \in {\omega}_{\bs s})\label{NormalLik2}
\end{eqnarray}
Here, $f_{\mathcal{N}_{|\bs s|+\zeta}}$ represents the joint probability density function for a $({|\bs s|+\zeta})$-variate Normal distribution. Furthermore, $\mathcal{I}(\theta)_{\F_{\bs s},\F_{\bs s}}$ is the unit (or per-observation) Fisher information matrix, (defined in Equation \ref{fisher}) relating to the index set of free parameters, $\F_{\bs s}$, evaluated at the true parameter value, $\tilde\theta$. Under the regularity conditions given in Appendix \ref{regCon}, the unit Fisher information matrix can, for discrete data $\bs y$, be expressed by SI Equation \ref{fisher}; for continuous $\bs y$ the summation is replaced by an integral. 
\begin{eqnarray}
\left\{ {\mathcal{I}(\theta)} \right\}_{jk} &=& \dfrac{-1}{n} \sum_{\bs y}  \left[ \dfrac{\partial^2}{\partial \theta_j \, \partial \theta_k} \log p(\bs y ; \bs x, \theta) \right] \, p(\bs y ; \bs x, \theta)\qquad  \{j,k=1\dots \nu+\zeta\}.\label{fisher}
\end{eqnarray}

\subsubsection{Motivation for the Normal approximation to the likelihood}
\label{MotForNorm}
To understand the motivation for the approximate form of the likelihood, we highlight that the theory was developed with large data sets in mind. We refer the reader to asymptotic theory (see for example \cite{davison2003statistical} and Chapter 9 of \cite{cox1974theoretical}), which provides a framework for assessing properties of estimators and statistical tests in the limit as sample sizes tend to infinity. A key theorem from this work demonstrates the asymptotic normality of maximum likelihood estimators for independent data, assuming standard asymptotic theory regularity conditions (covered by those listed in Section \ref{regCon}). Specifically, the theorem says that the difference between the maximum likelihood estimator of a multidimensional parameter and the true value of that parameter converges in distribution to a Multivariate Normal distribution, with the zero vector as the mean, as the number of samples in the data set tends to infinity. Furthermore, the covariance matrix scales with the inverse of the unit information matrix evaluated at the true value of the parameter. In the context of our framework, this gives the following,
\begin{eqnarray}
{\hat\theta^{\bs s}_{\F_{\bs s}}}\overset{d}{\mathop{\to }}\,{\mathcal{N}_{|\bs s|+\zeta }}\left( {\tilde\theta_{\F_{\bs s}}},\tfrac{1}{n}\left[\mathcal{I}({\tilde\theta})_{\F_{\bs s},\F_{\bs s}}\right]^{-1}\right),\qquad n\to \infty\ ,
\label{AsNormhf}
\end{eqnarray}
where $\tilde\theta_{\F_{\bs s}}$ is the unknown true value of $\theta_{\F_{\bs s}}$. The Multivariate Normal form of the likelihood assumed in this work (Equation \ref{NormalLik}) is linked to this result and can be derived through a Taylor expansion of the log-likelihood function around the MLE in a similar manner to proof of this result (\cite{cox1974theoretical}, \cite{davison2003statistical}). Key steps are detailed below. 

First, for $\theta \in{\omega_{\bs s}}$,  define $l(\theta)=l(\theta ;\bs y,\bs x)=\log p(\bs y ;\bs x,\theta)$ to be the log-likelihood function, given $\bs y$ and $\bs x$.  Provided that $\parallel{\theta}-{\hat\theta^{\bs s}}\parallel=
O_p(n^{-1/2})$, a Taylor expansion of the log-likelihood function around the MLE gives
\begin{eqnarray}
\label{Taylorexpansion}
l({\theta})=l({\hat\theta^{\bs s}})+  ({\theta}_{\F_{\bs s}}-{{\hat\theta_{\F_{\bs s}}^{\bs s}}}){{{U}}(\hat{\theta}^{\bs s})_{\F_{\bs s},\F_{\bs s}}} - \frac{1}{2}{{({\theta }_{{{\F}_{s}}}-\hat\theta^{\bs s}_{\F_{\bs s}})}^{T}}{J(\hat\theta^{\bs s})_{\F_{\bs s},\F_{\bs s}}}({\theta }_{{{\F}_{s}}}-\hat\theta^{\bs s}_{\F_{\bs s}})+o_p(1).\
\end{eqnarray}

Where ${U(\theta)}$ is a vector of first-order partial derivatives,
\begin{eqnarray}
{\left\{{U}(\theta) \right\}}_{ij}=\frac{{\partial }l  }{\partial {\theta_{i}} }\
\end{eqnarray}
and ${J(\theta)}$, which is known as the observed information matrix, is a vector that incorporates the second-order partial derivatives.
\begin{eqnarray}
{{\left\{{J}(\theta) \right\}}_{ij}}={-\frac{{{\partial }^{2}}l}{\partial {\theta_{i}}\partial {\theta_{j}}}}
\end{eqnarray}
Crucially, by the definition of the MLE, 
\begin{eqnarray}
{{{U}}(\hat{\theta}^{\bs s})_{\F_{\bs s},\F_{\bs s}}}={\mathbf{0}}\
\end{eqnarray}
Furthermore, in the context of the regularity conditions assumed here (Section \ref{regCon}), the unit Fisher information matrix, ${I}({\theta})$, is a consistent estimator of ${J}({\theta})/n$ \citep[p.\ 302]{cox1974theoretical} and the two can be used interchangeably. In much of the theory we discuss hereon in, $\mathcal{I}({\theta})$ is used as this form is the typical form used in the theorem describing the normality of the MLE (Equation \ref{AsNormhf}), which we refer to later. However, in the case of estimating confidence intervals for the model parameters (as discussed in Sections \ref{classEST} and \ref{bayePE}) the observed information is a quantity that can be calculated more easily, as it does not require expectations to be derived, and is used instead. For $\theta \in {\omega}_{\bs s}$, Equation \ref{Taylorexpansion} can therefore be simplified as follows,
\begin{eqnarray}
\label{logliktayloylor}
l(\mathbf{\theta })=l(\hat{\theta }^{\bs s})-\frac{n}{2}{{({\theta }_{{\F}_{s}}-\hat\theta^{\bs s}_{\F_{\bs s}})}^{T}}{\mathcal{I}(\tilde\theta)_{\F_{\bs s},\F_{\bs s}}}({\theta }_{{{\F}_{s}}}-\hat\theta^{\bs s}_{\F_{\bs s}})+o_p(1),\qquad (\theta \in {\omega}_{\bs s})
\end{eqnarray}

By exponentiating both sides of Equation \ref{logliktayloylor}, the expression for the Multivariate Normal form of the likelihood assumed in this theory is revealed.
\begin{eqnarray}
p(\bs y ;\bs x, \theta) &=& p(\bs y ;\bs x,\hat\theta^{\bs s}) \, \exp\left\{
- \frac{n}{2} ( \theta^{\phantom{()}}_{\F_{\bs s}} - \hat\theta^{\bs s}_{\F_{\bs s}})^T {\mathcal{I}(\tilde\theta)_{\F_{\bs s},\F_{\bs s}}}( \theta^{\phantom{()}}_{\F_{\bs s}} - \hat\theta^{\bs s}_{\F_{\bs s}})
 +o_p(1) \right\},\qquad(\theta \in {\omega}_{\bs s})\nonumber\\ \label{NormalLik}
 \end{eqnarray}

\subsection{Classical results for the regression model} 
\subsubsection{Classical point estimate and variance}
\label{classEST}
In the regression problem specified by Definition \ref{iidregressionprob}, the classical point estimate and confidence region of the parameter $\theta^{\bs s}_{\F_{\bs s}}$ (of null hypothesis ${\omega}_{\bs s}$) are derived from the Normal approximation to the likelihood described above (Equation \ref{NormalLik}). Thus, the classical point estimate is precisely the MLE 
\begin{eqnarray}
\label{classmean}
\hat\theta^{\bs s}_{\F_{\bs s}}.
\end{eqnarray}

A classical variance can also be derived from the Normal approximation to the likelihood and is given by $[n\,{{\mathcal{I}}}(\tilde\theta)_{\F_{\bs s},\F_{\bs s}}]^{-1}$ or its consistent estimator $[{{{J}}}(\tilde\theta)_{\F_{\bs s},\F_{\bs s}}]^{-1}$. In the usual case, where $\tilde\theta$ is unknown, this quantity is also unknown. When it is required that the variance is estimated in practice (e.g.\ in the calculation of confidence regions), then it is typical to replace this variable with another that is a consistent estimator of it. In the large $n$ scenario that we are interested in, $\hat\theta^{\bs s}$ is a consistent estimate of $\tilde\theta$ \citep{cox1974theoretical} and therefore a more useful expression for the classical variance is given by,
\begin{eqnarray}
\label{classvar}\left[{{{J}}}(\hat\theta^{\bs s})_{\F_{\bs s},\F_{\bs s}}\right]^{-1}.
\end{eqnarray}

\subsubsection{Classical confidence region}
Using the Normal approximation to the likelihood, it is possible to provide a  $({\nu }_{s}+\zeta)$-dimensional $100(1-\alpha)\%$ confidence region for the parameter  $\theta^{\phantom{()}}_{\F_{\bs s}}$ 
 \citep[p.\ 137]{stuart1998kendall}, given by,
\begin{eqnarray}
\left\{ \theta^{\phantom{()}}_{\F_{\bs s}} : \left( \theta^{\phantom{()}}_{\F_{\bs s}} - \hat\theta^{\bs s}_{\F_{\bs s}} \right)^T {{J}}(\hat\theta^{\bs s})_{\F_{\bs s},\F_{\bs s}} \left( \theta^{\phantom{()}}_{\F_{\bs s}} - \hat\theta^{\bs s}_{\F_{\bs s}} \right)  \leq Q_{\chi^2_{|\bs s|+\zeta}}(1-\alpha) \right\},\label{ClassicalConfidenceRegion}
\end{eqnarray}

where $Q_{\chi^2_\nu}(p)$ is the quantile function of the chi-squared distribution with $\nu$ degrees of freedom. For an individual parameter, the $100(1-\alpha)\%$ confidence interval simplifies to
\begin{eqnarray}
\left\{ \hat\theta_{\F_{\bs s}} \right\}_j &\pm& \sqrt{\left( \left[ {{J}(\hat\theta^{\bs s})_{\F_{\bs s},\F_{\bs s}}} \right]^{-1} \right)_{jj}  }  \, Q_z\left(1-\frac{\alpha}{2}\right) \label{ClassicalConfidenceInterval}
\end{eqnarray}
where $Q_z(p)$ is the quantile function of the standard Normal distribution. Note the importance of model choice for determining the standard error (the square root term): the index set $\F_{\bs s}$ specifies which terms are fixed (and therefore contribute no uncertainty) and which are estimated.

\subsubsection{Classical hypothesis tests: $p$-values}\label{chtp}
To test whether there is evidence that a particular set of covariates, specified by model $\bs s$, should be included in the regression model, a hypothesis test comparing the maximized likelihood ratio for the test between model $\bs s$ and the grand null, model $\bs 0$, can be performed. When the null hypothesis is true, this test statistic can be used to generate a $p$-value for the test. The maximized likelihood ratio for the test of the grand null hypothesis $\omega_{\bs 0}$ versus the alternative hypothesis $\omega_{\bs s}$, $\omega_{\bs 0} \subset \omega_{\bs s}$ is represented by
\begin{eqnarray}
    \mlr_{{\bs s}} \ =\ \mlr_{{\bs s}:{\bs 0}} &=&
        \frac{
            \sup_{\theta \in \omega_{\bs s}} p({\bs y};{\bs x}, \theta)
        }{
            \sup_{\theta \in \omega_{\bs 0}} p({\bs y};{\bs x}, \theta)
        }\nonumber\\
        &=& 
        \frac{
            \sup_{\theta \in {\Theta_{\bs s}}} p({\bs y};{\bs x}, \theta)
        }{
            \sup_{\theta \in \Theta_{\bs 0}} p({\bs y};{\bs x}, \theta)
        } .
        \label{Rsdef}
\end{eqnarray}
The second line arises from the use of point null hypotheses: see MT Definition \ref{define_regression_problem}. For practical purposes we can express the maximized likelihood ratio in terms of the maximum likelihood estimators:
\begin{eqnarray}
    \mlr_{{\bs s}} &=&
        \dfrac{p(\bs y ;\bs x, \hat\theta^{\bs s})}{p(\bs y ;\bs x, \hat\theta^{\bs 0})}.
\end{eqnarray}
This expression can be evaluated by replacing the denominator with the Normal approximation to the likelihood (MT Equation \ref{NormalLik}) evaluated at the constrained maximum likelihood $(\hat\theta^{\bs 0}_{\F_{\bs s}})$ 
\begin{eqnarray}
\mlr_{\bs s}&=&\dfrac{p(\bs y ;\bs x, \hat\theta^{\bs s})}{p(\bs y ; \bs x, \hat\theta^{\bs s}) \, \exp\left\{-\frac{n}{2}
\left( \hat\theta^{\bs 0}_{\F_{\bs s}} - \hat\theta^{\bs s}_{\F_{\bs s}} \right)^T {\mathcal{I}(\tilde\theta)_{\F_{\bs s},\F_{\bs s}}} \left(\hat \theta^{\bs 0}_{\F_{\bs s}} - \hat\theta^{\bs s}_{\F_{\bs s}} \right)+o_p(1)
\right\}}
\end{eqnarray}
Therefore
\begin{eqnarray}
\mlr_{\bs s}&=& \exp\left\{\frac{n}{2}
\left( \hat\theta^{\bs 0}_{\F_{\bs s}} - \hat\theta^{\bs s}_{\F_{\bs s}} \right)^T {\mathcal{I}(\tilde\theta)_{\F_{\bs s},\F_{\bs s}}} \left(\hat \theta^{\bs 0}_{\F_{\bs s}} - \hat\theta^{\bs s}_{\F_{\bs s}} \right)+o_p(1)
\right\}
\label{MLRfull}
\end{eqnarray}

This expression can be further simplified to derive one that is independent of the nuisance parameters. Although the derivation has been described previously (\cite{cox1974theoretical}, \cite{davison2003statistical}), the key steps are outlined here. First, a partition, compatible with the partition of the parameter vector $(\tilde\theta)$, is defined on the per unit information matrix. To do so, we denote $\B_{\bs s} = \F_{\bs s} \setminus \F_{\bs 0}$ to be the index set of parameters that are free in model $\bs s$ but not in model $\bs 0$.
\begin{eqnarray}\mathcal{I}(\tilde\theta)_{\F_s,\F_s}=\begin{pmatrix}
\mathcal{I}(\tilde\theta)_{\B_{\bs s},\B_{\bs s}}  & \mathcal{I}(\tilde\theta)_{\B_{\bs s},\F_0} \\
\mathcal{I}(\tilde\theta)_{\F_0,\B_{\bs s}}  & \mathcal{I}(\tilde\theta)_{\F_0,\F_0} \\
\end{pmatrix}
\end{eqnarray}
and its inverse.
\begin{eqnarray}\left[\mathcal{I}(\tilde\theta)_{\F_s,\F_s}\right]^{-1}=\begin{pmatrix}
\left( [\mathcal{I}(\tilde\theta)_{\F_{\bs s},\F_{\bs s}}]^{-1} \right)_{\B_{\bs s},\B_{\bs s}}  & \left( [\mathcal{I}(\tilde\theta)_{\F_{\bs s},\F_{\bs s}}]^{-1} \right)_{\B_{\bs s},\F_{\bs 0}}  \\
\left( [\mathcal{I}(\tilde\theta)_{\F_{\bs s},\F_{\bs s}}]^{-1} \right)_{\F_{\bs 0},\B_{\bs s}}   & \left( [\mathcal{I}(\tilde\theta)_{\F_{\bs s},\F_{\bs s}}]^{-1} \right)_{\F_{\bs 0},\F_{\bs 0}}  \\
\end{pmatrix}
\end{eqnarray}

where the dimensions of the partition, are, in each case
\begin{eqnarray}
\begin{bmatrix} 
 |\bs s| \times |\bs s| & |\bs s| \times \zeta\\
\zeta \times |\bs s| & \zeta \times \zeta
\end{bmatrix}
\end{eqnarray}
Then
\begin{eqnarray}
\mlr_{\bs s}&=& \exp\left\{\frac{n}{2}
\begin{pmatrix} 
 - \hat\beta^{\bs s}_{\B_{\bs s}} \\
\hat\gamma^{\bs 0} - \hat\gamma^{\bs s}
\end{pmatrix}^T \begin{pmatrix}
\mathcal{I}(\tilde\theta)_{\B_{\bs s},\B_{\bs s}}  & \mathcal{I}(\tilde\theta)_{\B_{\bs s},\F_0} \\
\mathcal{I}(\tilde\theta)_{\F_0,\B_{\bs s}}  & \mathcal{I}(\tilde\theta)_{\F_0,\F_0} \\
\end{pmatrix} 
\begin{pmatrix} 
- \hat\beta^{\bs s}_{\B_{\bs s}} \\
\hat\gamma^{\bs 0} - \hat\gamma^{\bs s}
\end{pmatrix}+o_p(1)
\right\}
\label{MLRfull_1}
\end{eqnarray}

This can be simplified because, under the large $n$ approximation, the MLEs of the parameters common to model $\bs 0$ and model $\bs s$ are related through the following expression, derived from a Taylor expansion (see p.308 of  \cite{cox1974theoretical} or p.138 of  \cite{davison2003statistical} for proof),
\begin{eqnarray}
\hat{\bs\gamma}^{\bs 0} - \hat{\bs\gamma}^{\bs s} &=& \left[{\mathcal{I}(\tilde\theta)_{\F_0,\F_0}} \right]^{-1}\,{{\mathcal{I}}(\tilde\theta)_{\F_0,\B_{\bs s}}} \; \hat{\beta}^{\bs s}_{\B_{\bs s}}+o_p(n^{-1/2})\label{shorttaylor}
\end{eqnarray}
giving,
\begin{eqnarray}
\mlr_{\bs s}&=& \exp\left\{\frac{n}{2}
\left\{\hat{\bs{\beta}}^{\bs s}_{\B_{\bs s}}\right\}^T \left\{\mathcal{I}(\tilde\theta)_{\B_{\bs s},\B_{\bs s}}-\mathcal{I}(\tilde\theta)_{\B_{\bs s},\F_{\bs 0}}\left[\mathcal{I}(\tilde\theta)_{\F_{\bs 0},\F_{\bs 0}}\right]^{-1}\mathcal{I}(\tilde\theta)_{\F_{\bs 0},\B_{\bs s}}\right\} \hat{\bs{\beta}}^{\bs s}_{\B_{\bs s}}+o_p(1)\right\}\quad
\label{MLRflat}
\end{eqnarray}

Using a standard linear algebra formula (see, for example, \cite{horn2012matrix}) for the inverse of a $2\times 2$ block matrix
\begin{eqnarray}\left([\mathcal{I}(\tilde\theta)_{\F_{\bs s},\F_{\bs s}}]^{-1}\right)_{\B_{\bs s},\B_{\bs s}}=\left\{\mathcal{I}(\tilde\theta)_{\B_{\bs s},\B_{\bs s}}-\mathcal{I}(\tilde\theta)_{\B_{\bs s},\F_{\bs 0}}\left[\mathcal{I}(\tilde\theta)_{\F_{\bs 0},\F_{\bs 0}}\right]^{-1}\mathcal{I}(\tilde\theta)_{\F_{\bs 0},\B_{\bs s}}\right\}^{-1}
\end{eqnarray}

Equation \ref{MLRflat} can be rewritten in a form that is independent of the nuisance parameters. 
\begin{eqnarray}
\mlr_{\bs s}&=& \exp\left\{\frac{n}{2}
\left\{\hat{\bs{\beta}}^{\bs s}_{\B_{\bs s}}\right\}^T \left[\left( \left[ \mathcal{I}(\tilde\theta)_{\F_{\bs s},\F_{\bs s}}  \right]^{-1} \right)_{\B_{\bs s},\B_{\bs s}}\right]^{-1} \hat{\bs{\beta}}^{\bs s}_{\B_{\bs s}}
+o_p(1)\right\}
\label{MLRsimplehf}
\end{eqnarray}

By the large $n$ approximation (\ref{AsNormhf}), the distribution of the MLE for model $\bs s$ is
\begin{eqnarray}
\begin{pmatrix} {\hat\beta^{\bs s}_{\B_{\bs s}}}\\ {\hat\gamma^{\bs s}}\end{pmatrix}\overset{d}{\mathop{\to }}\,{{N}_{{|\bs s|}+\zeta }}\left(\begin{pmatrix} {\tilde\beta_{\B_{\bs s}}}\\ {\tilde\gamma}\end{pmatrix},\frac{1}{n}\begin{pmatrix}
\left( [\mathcal{I}(\tilde\theta)_{\F_{\bs s},\F_{\bs s}}]^{-1} \right)_{\B_{\bs s},\B_{\bs s}}  & \left([\mathcal{I}(\tilde\theta)_{\F_{\bs s},\F_{\bs s}}]^{-1}\right)_{\B_{\bs s},\F_{\bs 0}}  \\
\left([\mathcal{I}(\tilde\theta)_{\F_{\bs s},\F_{\bs s}}]^{-1}\right)_{\F_{\bs 0},\B_{\bs s}}   & \left([\mathcal{I}(\tilde\theta)_{\F_{\bs s},\F_{\bs s}}]^{-1}\right)_{\F_{\bs 0},\F_{\bs 0}}\\
\end{pmatrix} \right),\qquad n\to \infty\nonumber
\\ \label{largeform}
\end{eqnarray}

The marginal distribution of the MLE of ${\bs{\beta}}^{\bs s}_{\B_{\bs s}}$ is then multivariate Normal with mean and covariance revealed by rearrangement of Equation \ref{largeform} and selection of the upper left-hand element of the multivariate covariance matrix.
\begin{eqnarray}\hat{\bs{\beta}}^{\bs s}_{\B_{\bs s}}\overset{d}{\mathop{\to }}\,{\mathcal{N}_{|\bs s|}}\left({\tilde{\bs\beta}}_{\B_{\bs s}},\tfrac{1}{n}\left([\mathcal{I}(\tilde\theta)_{\F_{\bs s},\F_{\bs s}}]^{-1}\right)_{\B_{\bs s},\B_{\bs s}} \right),\qquad n\to \infty
\label{marginalMLEhf}
\end{eqnarray}

In hypothesis testing, we are interested in evaluating the evidence against the null.  When the null is true, $\tilde\beta_{\B_{\bs s}}=0$. Since $\tilde{\gamma}$ can always be assumed to equal $\bs 0$ following reparameterization (provided $\bs 0$ is not at the edge of parameter space), the null can also be specified in full as $\tilde\theta=\bs0$. Substituting these expressions into Equations \ref{marginalMLEhf} gives
 \begin{eqnarray}\hat{\bs{\beta}}^{\bs s}_{\B_{\bs s}}\overset{d}{\mathop{\to }}\,{\mathcal{N}_{|\bs s|}}\left(\bs 0,
\tfrac{1}{n}\left(\left[ \mathcal{I}(\bs 0)_{\F_{\bs s},\F_{\bs s}}  \right]^{-1}\right)_{\B_{\bs s},\B_{\bs s}} \right),\qquad n\to \infty
\label{MLEnullhf}
\end{eqnarray}

and into Equation \ref{MLRsimplehf}, gives
\begin{eqnarray}
\mlr_{\bs s}&=& \exp\left\{\frac{n}{2}
\left\{\hat{\bs{\beta}}^{\bs s}_{\B_{\bs s}}\right\}^T \left[\left( \left[ \mathcal{I}(\bs 0)_{\F_{\bs s},\F_{\bs s}}  \right]^{-1} \right)_{\B_{\bs s},\B_{\bs s}}\right]^{-1} \hat{\bs{\beta}}^{\bs s}_{\B_{\bs s}}
+o_p(1)\right\}.
\label{MLR5464}
\end{eqnarray}

Equations \ref{MLEnullhf} and \ref{MLR5464} lead to Wilks' Theorem (\cite{wilks1938large}); namely, that under standard asymptotic regularity conditions (covered by those in Section \ref{regCon}), under the null, the limiting distribution of $2 \, \log R_{\bs s}$ (as $n$ tends to infinity) is central chi-squared with degrees of freedom equal to the difference in the number of unconstrained parameters between the alternative and the null, $|\bs s|$.
\begin{eqnarray}
2 \, \log \mlr_{\bs s} \ \overset{d}{\mathop{\to }}\   \chi^2_{|\bs s|} \label{d2logR_given_theta0},\qquad \theta\in{\omega}_{\bs 0},\qquad n\to \infty.
\end{eqnarray}
 For a given test statistic, the $p$-value is the probability of obtaining a value of the test statistic at least as extreme as the observed result, under the assumption that the null hypothesis is correct. This leads to the standard result that for large $n$, the $p$-value for the nested likelihood ratio test of model $\bs s$ versus model $\bs 0$ is given by
 \begin{eqnarray}
p_{\bs s} &\to& \Pr\left( \chi^2_{|\bs s|} \geq  2 \, \log \mlr_{\bs s} \right),\qquad \theta\in{\omega}_{\bs 0},\qquad n\to \infty. \label{pvalue}
\end{eqnarray}
Equivalently, the significance threshold (for $\mlr_{\bs s}$) for a false positive rate $\alpha$ is given by
\begin{eqnarray}
\label{classsigTh}
\mlr_{\bs s} \geq e^{\frac{1}{2} Q^{-1}_{\chi^2_{|\bs s|}}(1-\alpha)} .
\end{eqnarray}

\subsubsection{Classical hypothesis tests: power}
To evaluate the power of the likelihood ratio test -- the probability of rejecting the null hypothesis (model ${\omega}_{\bs 0}$) when, in fact, it is false -- it is necessary to consider the likelihood ratio test statistic when the alternative model (model $\bs s$) is true. Following the approach of earlier works (\cite{davidson1970limiting}, \cite{johnson2008properties} and \cite{hu2009bayesian}), a local alternatives assumption can be followed, whereby,   
\begin{eqnarray}
\label{LAfor0}
\textrm{under model  } \bs 0\textrm{:}\quad \theta^{\bs 0}_{\F_{\bs s}}=\begin{pmatrix}{\bs 0}\\{\gamma}\end{pmatrix}
\end{eqnarray}
\begin{eqnarray}
\label{LAfors}
\textrm{under model  } \bs s\textrm{:}\quad  \theta^{\bs s, n}_{\F_{\bs s}}=\begin{pmatrix}{\bs\beta^{\bs s,n}_{\B_{\bs s}}}\\{\gamma}\end{pmatrix}
\end{eqnarray}
Where
\begin{eqnarray}
\beta^{\bs s,n}_{i}={q_i^n}/n^{1/2},  \textrm{ with} \;\lim_{n \to \infty }{q_i^n}=q_{i},\; i=1,2...,|\bs s| .
\label{LAsqrt}
\end{eqnarray}
In simple terms, therefore, the local alternatives hypothesis assumes that ${\bs\beta}^{\bs s, n}_{\B_{\bs s}}$ is a sequence of alternatives that converge to the null vector at rate $n^{1/2}$ and that interest focuses on small departures from the null. The fact that larger departures from the null become self-evident in large sample sizes motivates this scaling assumption; the focus of the statistical analysis is necessarily on the `grey area' of small departures from the null. More importantly, the assumption is convenient because it implies the tail of the distribution of the MLE is still approximately normal near the null value, which greatly simplifies the theory.

Equation \ref{MLRsimplehf} gives an expression for the maximized likelihood ratio.
\begin{eqnarray}
\mlr_{\bs s}&=& \exp\left\{\frac{n}{2}
\left\{\hat{\bs{\beta}}^{\bs s}_{\B_{\bs s}}\right\}^T \left[\left( \left[ \mathcal{I}(\tilde\theta)_{\F_{\bs s},\F_{\bs s}}  \right]^{-1} \right)_{\B_{\bs s},\B_{\bs s}}\right]^{-1} \hat{\bs{\beta}}^{\bs s}_{\B_{\bs s}}
+o_p(1)\right\}\label{iron33}
\end{eqnarray}

When the alternative is true $\tilde{\bs{\beta}}_{\B_{\bs s}}=\bs\beta^{\bs s,n}_{\B_{\bs s}}$ and $\tilde{\theta}=\theta^{\bs s, n}$. Therefore, Equation \ref{iron33} is given by 
\begin{eqnarray}
\mlr_{\bs s}&=& \exp\left\{\frac{n}{2}
\left\{\hat{\bs{\beta}}^{\bs s}_{\B_{\bs s}}\right\}^T \left[\left( \left[ \mathcal{I}(\theta^{\bs s,n})_{\F_{\bs s},\F_{\bs s}}  \right]^{-1} \right)_{\B_{\bs s},\B_{\bs s}}\right]^{-1} \hat{\bs{\beta}}^{\bs s}_{\B_{\bs s}}
+o_p(1)\right\}\label{bp2}
\end{eqnarray}

and Equation \ref{marginalMLEhf} gives the marginal distribution of $\hat{\bs{\beta}}^{\bs s}_{\B_{\bs s}}$.
\begin{eqnarray}
\hat{\bs{\beta}}^{\bs s}_{\B_{\bs s}}\  \overset{d}{\mathop{\to }}\,\  {\mathcal{N}_{|\bs s|}}\left(\bs\beta^{\bs s,n}_{\B_{\bs s}},\tfrac{1}{n}\left([\mathcal{I}(\theta^{\bs s,n})_{\F_{\bs s},\F_{\bs s}}]^{-1}\right)_{\B_{\bs s},\B_{\bs s}} \right),\qquad n\to \infty\label{bp1}
\end{eqnarray}

It follows from Equations \ref{bp2} and \ref{bp1} that the distribution of $2\log \mlr_{\bs s}$ under the sequence of local alternative models specified above converges in distribution to a non-central chi-squared distribution (see  \cite{wald1943tests,davidson1970limiting}) with ${|\bs s|}$ degrees of freedom 
\begin{eqnarray}
2 \, \log \mlr_{\bs s} \ \overset{d}{\mathop{\to }}\   \chi^2_{|\bs s|, \, \delta} \label{d2logR_given_theta},\qquad n\to\infty \end{eqnarray}

where the non-centrality parameter of the chi-squared distribution is
\begin{eqnarray}
\delta &=& n \, \left\{ \bs\beta^{\bs s,n}_{\B_{\bs s}} \right\}^{T} \left[\left( \left[ \mathcal{I}(\theta^{\bs s,n})_{\F_{\bs s},\F_{\bs s}}  \right]^{-1} \right)_{\B_{\bs s},\B_{\bs s}}\right]^{-1} \bs\beta^{\bs s,n}_{\B_{\bs s}}\nonumber\\
&=& \left\{ {\bs q}^{n}_{\B_{\bs s}} \right\}^{T} \left[\left( \left[ \mathcal{I}(\theta^{\bs s,n})_{\F_{\bs s},\F_{\bs s}}  \right]^{-1} \right)_{\B_{\bs s},\B_{\bs s}}\right]^{-1} {\bs q}^{n}_{\B_{\bs s}} \label{thisisdelta}
\end{eqnarray}

Since null hypothesis $\omega_{\bs 0}$ is rejected in favour of model $\bs s$ when $p_{\bs s} \leq \alpha$, the statistical power of the test for fixed $\tilde\theta \in \Theta_{\bs s}$ and false positive rate $\alpha$ would be
\begin{eqnarray}
\Pr\left( \chi^2_{|\bs s|, \delta} \geq  Q^{-1}_{\chi^2_{|\bs s|}}(1-\alpha) \right) . 
\end{eqnarray}


Note however, that $\delta$ is not fixed under $\bs s$, otherwise it would be optimal by the Neyman-Pearson lemma (\citet{neyman1933ix}) to use that value in the likelihood ratio test, instead of estimating it. If, instead of being fixed, the non-centrality parameter was in fact encountered with frequency approximated by
\begin{eqnarray}
\delta \,;\, \bs s &\sim& \textrm{Gamma}\left( \frac{|\bs s|}{2}, \, \frac{|\bs s|}{2 \, \bar{\delta}_{\bs s}} \right),
\end{eqnarray}
where $\mathbb{E}(\delta; \bs s) = \bar{\delta}_{\bs s}$, then
\begin{eqnarray}
2 \, \log R_{\bs s} \,;\, \bs s \overset{d}{\to} \left(1 + \tfrac{\bar{\delta}_{\bs s}}{|\bs s|} \right) \, \chi^2_{|\bs s|}
\label{LogGammaAlternative}
\end{eqnarray}

and the statistical power of a test with variable $\tilde\theta$ would be
\begin{eqnarray}
\Pr\left( \chi^2_{|\bs s|} \geq  \left(1 + \tfrac{\bar{\delta}_{\bs s}}{|\bs s|} \right)^{-1} \, Q^{-1}_{\chi^2_{|\bs s|}}(1-\alpha) \right) . \label{Power2}
\end{eqnarray}

\subsection{Regression model: Bayesian results}
\subsubsection{Bayesian point estimate and variance}
\label{bayePE}
Following Bayes rule, the posterior 
distribution for the free parameters for model $\bs s$, ${f}(\theta_{\F_{\bs s}} \,\big|\, \bs y,\bs x,\bs s)$, is proportional to the product of the likelihood, $p(\bs y;\bs x,\theta)$ with $\theta\in\Theta_{\bs s}$ (Equation \ref{NormalLik2}), and the prior density, $\pi(\theta|{\bs s})$ (specified by MT Equation \ref{ConjugatePrior}). By considering the local alternatives hypothesis, as above, we write $\theta^{\bs s,n}$ in place of $\tilde\theta$ within the likelihood to give
\begin{eqnarray}
{{f}}(\theta_{\F_{\bs s}} \,\big|\, \bs y,\bs x,\bs s)&\propto&p(\bs y\bf;\bs x,\theta)\pi(\theta_{\F_{\bs s}}|\bs s),\qquad (\theta \in \Theta_{\bs s})\nonumber\\
&\propto&\ {f_{\mathcal{N}_{|\bs s|+\zeta}}\left( \hat\theta^{\bs s}_{\F_{\bs s}} \,\bigg|\, \theta^{\phantom{()}}_{\F_{\bs s}},\, n^{-1}  \left[\mathcal{I}(\theta^{\bs s,n})_{\F_{\bs s},\F_{\bs s}}\right]^{-1} \right)} f_{\mathcal{N}_{|\bs s|+\zeta}}\left(\theta_{\F_{\bs s}} \,\bigg|\, \bs{0}_{|\bs s|+\zeta}, {h^{-1} \left[ \mathcal{I}({\bs 0})_{\F_{\bs s},\F_{\bs s}}\right]^{-1}}\right)\nonumber. \\ \label{postasnormals}
\end{eqnarray}

By the definition of the sequence of local alternatives (Equations \ref{LAfors} and \ref{LAsqrt}), $\theta^{\bs s, n}_{\F_{\bs s}}$ is a consistent estimator of $\theta^{0}_{\F_{\bs s}}=\left(\bs 0,\gamma\right)^T$. Since, through transformation, $\gamma$ can be assumed to equal the zero vector, $\theta^{\bs s, n}$ is a consistent estimator of $\bs 0$. Furthermore, since $\mathcal{I}$ is a continuous function, 
\begin{eqnarray}
\mathcal{I}(\theta^{\bs s, n})_{\F_{\bs s},\F_{\bs s}}\to \mathcal{I}({\bs 0})_{\F_{\bs s},\F_{\bs s}}, \qquad n \to  \infty.
\end{eqnarray}

Considering Equation \ref{postasnormals} at the limit as $n$ tends to infinity, substitution of $\mathcal{I}({\bs 0})_{\F_{\bs s},\F_{\bs s}}$ with $\mathcal{I}(\theta^{\bs s, n})_{\F_{\bs s},\F_{\bs s}}$ allows the posterior density to be written as,
\begin{eqnarray}
{f}(\theta_{\F_{\bs s}} \,\big|\, \bs y,\bs x,\bs s)&\propto&\ f_{\mathcal{N}_{|\bs s|+\zeta}}\left( \theta_{\F_{\bs s}} \;\big|\;\tfrac{n}{n+h} \, \hat\theta^{\bs s}_{\F_{\bs s}},\,\tfrac{n}{n+h} \left[ n\,\mathcal{I}(\theta^{\bs s,n})_{\F_{\bs s},\F_{\bs s}}\right]^{-1}\right)\qquad n \to  \infty\label{iron}
\end{eqnarray}

i.e., the posterior distribution for the free parameters tends in distribution to a $|\bs s|+\zeta$-variate Normal distribution.
\begin{eqnarray}
\label{postdistr}
\theta_{\F_{\bs s}} \,\big|\, \bs y,\bs x, \bs s \;\overset{d}{\to}\;  \mathcal{N}_{|\bs s|+\zeta}\left(
 \tfrac{n}{n+h} \, \hat\theta^{\bs s}_{\F_{\bs s}}, \tfrac{n}{n+h} \left[ n\,\mathcal{I}(\theta^{\bs s,n})_{\F_{\bs s},\F_{\bs s}}\right]^{-1} \right), \qquad n \to \infty
\end{eqnarray}

 For the purposes of deriving a posterior mean and variance that can be calculated in practice, the substitution of ${{\mathcal{I}}}(\theta^{\bs s, n})_{\F_{\bs s},\F_{\bs s}}$ with ${{{J}}}(\hat\theta^{\bs s})_{\F_{\bs s},\F_{\bs s}}/n$, gives
\begin{eqnarray}
\label{postdistr_1}
\theta_{\F_{\bs s}} \,\big|\, \bs y,\bs x, \bs s \;\overset{d}{\to}\; \mathcal{N}_{|\bs s|+\zeta}\left(
 \tfrac{n}{n+h} \, \hat\theta^{\bs s}_{\F_{\bs s}}, \tfrac{n}{n+h} \left[{J}(\hat\theta^{\bs s})_{\F_{\bs s},\F_{\bs s}}\right]^{-1} \right), \qquad n \to \infty
\end{eqnarray}

Thus, the posterior mean
\begin{eqnarray}
\label{postmean}
\tfrac{n}{n+h} \, \hat\theta^{\bs s}_{\F_{\bs s}}
\end{eqnarray}
and the posterior variance 
\begin{eqnarray}
\label{postvar}
\tfrac{n}{n+h} \left[ {J}(\hat\theta^{\bs s})_{\F_{\bs s},\F_{\bs s}}\right]^{-1}
\end{eqnarray}
 serve as the Bayesian point estimate and variance, respectively, and each differs from its classical counterpart (Equation \ref{classmean} and \ref{classvar}) by a factor $n/(n+h)$.

For the fixed parameters, the posterior variances and covariances are zero and the posterior means equal the prior means of zero: $\upstar{\bs{V}}^{\bs s}_{ij: i\notin \F_{\bs s}}=0$, $\upstar{\bs{V}}^{\bs s}_{ij : j\notin \F_{\bs s}}=0$, $\upstar{\bs{m}}^{\bs s}_{i: i\notin \F_{\bs s}}=0$.

\subsubsection{Bayesian credibility region}
From the posterior distribution derived in Equation \ref{postdistr_1}, the equal-tailed $100(1-1/(1+\tau))\%$ credibility region is given by 
\begin{eqnarray}
\left\{ \theta^{\phantom{()}}_{\F_{\bs s}} : \tfrac{(n+h)}{n} \left( \theta^{\phantom{()}}_{\F_{\bs s}} - \tfrac{n}{n+h}\hat\theta^{\bs s}_{\F_{\bs s}} \right)^T {{J}(\hat\theta^{\bs s})_{\F_{\bs s},\F_{\bs s}}}\left( \theta^{\phantom{()}}_{\F_{\bs s}} - \tfrac{n}{n+h}\hat\theta^{\bs s}_{\F_{\bs s}} \right)  \leq Q_{\chi^2_{|\bs s|+\zeta}}\left(1-\tfrac{1}{1+\tau}\right) \right\} \label{BayesianConfidenceRegion}
\end{eqnarray}

and the single-parameter credibility interval is given by,
\begin{eqnarray}
\tfrac{n}{n+h} \left\{ \hat\theta_{\F_{\bs s}} \right\}_j &\pm& \sqrt{ \tfrac{n}{n+h} \left(\left[{{{J}}}(\hat\theta^{\bs s})_{\F_{\bs s},\F_{\bs s}}\right]^{-1} \right)_{jj}  }  \, Q_z\left(1-\tfrac{1}{2(1+\tau)}\right) \label{BayesianConfidenceInterval}
\end{eqnarray}

\subsubsection{Bayesian hypothesis testing: the Bayes Factor}
To derive an analytical expression for the Bayes factor, an expression for the marginal probability density function, $m(\bs y |\bs x, \bs s)$, is first derived by integrating the model likelihood (Equation \ref{NormalLik}) over the prior on $\theta$, given model $\bs s$ (MT Equation \ref{ConjugatePrior}). As above, we assume a local alternatives hypothesis under model $\bs s$, as specified by Equations {\ref{LAfor0} to \ref{LAsqrt}}.
\begin{eqnarray}
{m(\bs y |\, \bs x, \bs s)}
&\propto&\int_{\Theta_{\bs s}} p(\bs y ;\bs x,\theta){\pi(\theta | {\bs s)}}\,d \theta\nonumber\\
&\propto&\int_{\Theta_{\bs s}}
\ {f_{\mathcal{N}_{|\bs s|+\zeta}}\left( \hat\theta^{\bs s}_{\F_{\bs s}} \,\big|\, \theta,\, n^{-1}  \left[\mathcal{I}(\theta^{\bs s, n})_{\F_{\bs s},\F_{\bs s}}\right]^{-1} \right)}{ f_{\mathcal{N}_{|\bs s|+\zeta}}\left(\theta  \,\big|\, \bs{0}_{|\bs s|+\zeta}, {h^{-1} \left[ \mathcal{I}({\bs 0})_{\F_{\bs s},\F_{\bs s}}\right]^{-1}}\right)}\,d \theta\nonumber\\
\end{eqnarray}

 Next, though replacement of $\mathcal{I}_{\F_{\bs s},\F_{\bs s}}(\theta^{\bs s, n})$ with its consistent estimator, $\mathcal{I}_{\F_{\bs s},\F_{\bs s}}({\bs 0})$, the  marginal probability mass function can be written as a Multivariate Normal density
\begin{eqnarray}
{m(\bs y |\, \bs x, \bs s)}&\propto&\int_{\Theta_{\bs s}}\ f_{\mathcal{N}_{|\bs s|+\zeta}}\left( \theta \,\big|\tfrac{n}{n+h} \, \hat\theta^{\bs s}_{\F_{\bs s}},\,\tfrac{n}{n+h} \left[ n\,\mathcal{I}({\bs 0})_{\F_{\bs s},\F_{\bs s}}\right]^{-1}\right)\,d \theta\nonumber\\
&\propto& f_{\mathcal{N}_{|\bs s|+\zeta}}\left( \hat\theta^{\bs s}_{\F_{\bs s}} \,\big|\, \bs{0}, \bs{\Sigma}\right)
\label{mgyxs}
\end{eqnarray}

Where 
\begin{eqnarray}\bs\Sigma=
\left(\tfrac{1}{h}+\tfrac{1}{n}\right) \left[ \mathcal{I}({\bs 0})_{\F_{\bs s},\F_{\bs s}}\right]^{-1}
\end{eqnarray}

Equation \ref{mgyxs} can be partitioned into two Multivariate Normal distributions: one defining the marginal distribution $\hat{\bs{\beta}}^{\bs s}_{\B_{\bs s}}$ and the other defining the conditional distribution of $\hat{\bs{\gamma}}^{\bs s}_{\B_{\bs s}}$ given $\hat{\bs{\beta}}^{\bs s}_{\B_{\bs s}}$
\begin{eqnarray}
\label{PYgivenXs}
{m(\bs y |\,\bs x, \bs s)}
&\approx& c\,{f_{\mathcal{N}_{|\bs s|+\zeta}}\left( \hat{\bs{\beta}}^{\bs s}_{\B_{\bs s}} \,\bigg|\, \bs{0}, \bs{\Sigma}_{\B_{\bs s},\B_{\bs s}} \right)}\times \nonumber \\
&& f_{\mathcal{N}_{|\bs s|+\zeta}}\left( \hat{\bs{\gamma}}^{\bs s} \,\bigg|\, \bs{\Sigma}_{\F_{0},\B_{\bs s}} \{ \bs{\Sigma}_{\B_{\bs s},\B_{\bs s}} \}^{-1} \hat{\bs{\beta}}^{\bs s}_{\B_{\bs s}}\left[ \left\{\bs\Sigma^{-1}\right\}_{\F_0,\F_0} \right]^{-1} \right)
\end{eqnarray}

Here $c$ is a scale constant. An expression for $m(\bs y | \bs x, \bs 0)$ can be similarly derived by integrating the model likelihood over the prior, given model $\bs 0$.
\begin{eqnarray}
\label{PYgivenX0}
m(\bs y |\, \bs x, \bs 0) &\approx&c\,{f_{\mathcal{N}_{|\bs s|+\zeta}}\left( \hat{\bs{\beta}}^{\bs s}_{\B_{\bs s}} \,\bigg|\, \bs{0}, \tfrac{h}{n+h} \,  \bs{\Sigma}_{\B_{\bs s},\B_{\bs s}} \right)} \times \nonumber \\&& {f_{\mathcal{N}_{|\bs s|+\zeta}}\left( \hat{\bs{\gamma}}^{\bs s} \,\bigg|\, \bs{\Sigma}_{\F_{0},\B_{\bs s}} \{ \bs{\Sigma}_{\B_{\bs s},\B_{\bs s}} \}^{-1} \hat{\bs{\beta}}^{\bs s}_{\B_{\bs s}}, \left[ \left\{{\bs\Sigma}^{-1}\right\}_{\F_0,\F_0} \right]^{-1} \right)}
\end{eqnarray}

Thus the Bayes factor for the test between model ${\bs s}$ and model $\bs 0$, defined as 
\begin{eqnarray}
    \bF_{\bs s} &=& \dfrac{m(\bs y|\,\bs x, \bs s)}{m(\bs y|\,\bs x, \bs 0)}
\end{eqnarray}
can be simplified (via Equations \ref{PYgivenXs} and \ref{PYgivenX0}) as follows 
\begin{eqnarray}
\bF_{\bs s} &\approx& \frac{ f_{\mathcal{N}}\left( \hat{\bs{\beta}}^{\bs s}_{\B_{\bs s}} \,\bigg|\, \bs{0}, \bs{\Sigma}_{\B_{\bs s},\B_{\bs s}} \right) }
{ f_{\mathcal{N}}\left( \hat{\bs{\beta}}^{\bs s}_{\B_{\bs s}} \,\bigg|\, \bs{0}, \frac{h}{n+h} \,  \bs{\Sigma}_{\B_{\bs s},\B_{\bs s}} \right) }\nonumber \\
&=& \frac{(\textrm{det}(\frac{h}{n+h} \,  \bs{\Sigma}_{\B_{\bs s},\B_{\bs s}}))^{1/2}}{(\textrm{det}({\bs{\Sigma}_{\B_{\bs s},\B_{\bs s}}})^{1/2}} \exp\left\{ -\tfrac{1}{2}\left(1-\tfrac{n+h}{h}\right)
\left\{\hat{\bs{\beta}}^{\bs s}_{\B_{\bs s}}\right\}^T \{ \bs{\Sigma}_{\B_{\bs s},\B_{\bs s}} \}^{-1} \hat{\bs{\beta}}^{\bs s}_{\B_{\bs s}}
\right\}
\end{eqnarray}

Since ${\bs{\Sigma}_{\B_{\bs s},\B_{\bs s}}}$ has dimension $|\bs s| \times |\bs s|$,   ${\textrm{det}\left(\frac{h}{n+h} \,  \bs{\Sigma}_{\B_{\bs s},\B_{\bs s}}\right)}=\left(\frac{h}{n+h}\right)^{|\bs s|} {\textrm{det}(\,  \bs{\Sigma}_{\B_{\bs s},\B_{\bs s}})}$ and therefore,
\begin{eqnarray}
\label{BFnearend}
\bF_{\bs s} &\approx&\left(\tfrac{h}{n+h}\right)^{|\bs s|/2}\exp\left\{ \tfrac{n}{2h}\left\{\hat{\bs{\beta}}^{\bs s}_{\B_{\bs s}}\right\}^T \{ \bs{\Sigma}_{\B_{\bs s},\B_{\bs s}} \}^{-1} \hat{\bs{\beta}}^{\bs s}_{\B_{\bs s}}
\right\}\nonumber\\
&=&\left(\tfrac{h}{n+h}\right)^{|\bs s|/2}\exp\left\{ \tfrac{n}{2h}{\left(\tfrac{1}{h}+\tfrac{1}{n}\right)}^{-1}\left\{\hat{\bs{\beta}}^{\bs s}_{\B_{\bs s}}\right\}^T \left[\left( \left[ \mathcal{I}(\bs 0)_{\F_{\bs s},\F_{\bs s}}  \right]^{-1} \right)_{\B_{\bs s},\B_{\bs s}}\right]^{-1} \hat{\bs{\beta}}^{\bs s}_{\B_{\bs s}}
\right\}
\end{eqnarray}

Through comparison of Equation \ref{BFnearend} with Equation \ref{MLR5464}, it is possible to write the Bayes factor as a function of the MLR, ${\mlr_{\bs s}}$, under the null hypothesis. This expression has been termed the likelihood ratio test-based Bayes factor (\cite{johnson2008properties}, \cite{hu2009bayesian}).
\begin{eqnarray}
\bF_{\bs s}
&\approx& \left(\tfrac{h}{n+h}\right)^{|\bs s|/2}\,({\mlr_{\bs s}})^{n/(n+h)}\label{MLR0}
\end{eqnarray}

It is noteworthy that the local alternatives assumption employed above implies that the hyperparameter, $h$, be proportional to the sample size ($h \propto n$), such that $\beta^{\bs s,n}_{i}$ shrinks with $n$, as given by definition (Equation \ref{LAsqrt}). 

Following the methods of Johnson and colleagues, however (\cite{johnson2008properties,hu2009bayesian}), this local alternatives assumption is dropped from this point forward. Instead, $h$ is assumed to be fixed at a constant to remove the dependence of the prior on $n$. Furthermore, as derived previously (see Lemma 1 of \cite{johnson2008properties}) and repeated in Section \ref{LBFC} for completeness, when $h$ is a constant, the logarithm of the Bayes factor is consistent, allowing for consistency of hypothesis testing. 

Notice further that because the Bayes Factor depends only on a single hyper-parameter, $h$, is it easy to calculate. For $n\gg h$, the main role of the hyper-parameter $h$ is to modify  
the penalty on the number of free parameters in the Bayesian interpretation of the maximized likelihood ratio, $\mlr_{\bs s}$.

The posterior odds for the test of model $\bs s$ to $\bs 0$ is defined to be the product of model prior odds, ${\mu}_{\bs s : \bs 0}={\mu}^{|\bs s|}$ (see MT Equation \ref{doubletpriorodds}) and the Bayes Factor (Equation \ref{MLR0}) and can be written as follows,
\begin{eqnarray}
\po_{\bs s}&\approx&{{\mu}^{|\bs s|}}\left(\tfrac{h}{n+h}\right)^{|\bs s|/{2}}\,(\mlr_{\bs s})^{n/(n+h)}\label{POswithns}
\end{eqnarray}

In the Bayesian framework, the null hypothesis, model $\bs 0$, is rejected in favor of the alternative hypothesis, model $\bs s$, in the sense that the smallest $100\left(1-1/(1+\tau)\right)\%$ posterior model credibility set excludes model $\bs 0$, when $\po_{\bs s} \geq \tau$. Model $\bs 0$ is therefore rejected when the following inequality holds
\begin{eqnarray}
\label{baysigTH}
\mu^{|\bs s|} \left(\tfrac{h}{n+h}\right)^{{|\bs s|}/{2}}\,\mlr_{\bs s}^{n/(n+h)} \geq {\tau}
\end{eqnarray}
\subsubsection{Classical power of the Bayesian test}
 Since the Bayes factor is an increasing function of the MLR, the prior implies that $\bar\delta_{\bs s} = |\bs s| \, n / h$, and Equation \ref{Power2} can be used to calculate the classical power of the Bayesian hypothesis test.
  \begin{eqnarray}
\Pr\left( \chi^2_{|\bs s|} \geq  \left(1 + \tfrac{n}{h} \right)^{-1} \, Q^{-1}_{\chi^2_{|\bs s|}}\left(1-\tfrac{1}{1+\tau}\right) \right) . \label{Power3}
\end{eqnarray}
\subsubsection{The logarithm of the Bayes factor is consistent}\label{LBFC}
As demonstrated by \cite{johnson2008properties}, and repeated here for completeness, the logarithm of the Bayes factor (Equation \ref{MLR0}), as given below, is consistent.
\begin{eqnarray}
\label{logBFeqn}
\log(\bF_{\bs s})&\approx& -{\frac{|\bs s|}{2}}\textrm{log}\left(\tfrac{n}{h}+1\right)+\left(\tfrac{n}{n+h}\right)\textrm{log}\left({\mlr_{\bs s}}\right)
\end{eqnarray}

This means that under model $\bs 0$
\begin{eqnarray}
{\log(\bF}_{\bs s})\overset{p}{\mathop{\to }}-\infty,\qquad n\to \infty
\end{eqnarray}
and under model $\bs s$
\begin{eqnarray}
{\log(\bF}_{\bs s})\overset{p}{\mathop{\to }}\infty,\qquad n\to \infty
\end{eqnarray}

This result follows from noticing that, when the null hypothesis is true, $2 \log \mlr_{\bs s} \ \overset{d}{\mathop{\to }}\   \chi^2_{|\bs s|} $ as $n\to \infty$ (Equation \ref{d2logR_given_theta0}), i.e. can be written as the sum of squares of $|\bs s|$ independent standard Normal distributions. The second term on the right-hand side of Equation \ref{logBFeqn} is thus bounded in probability, whereas the first tends to -$\infty$ as $n\to\infty$. Therefore, $\log(\BF_{\bs s})\overset{p}{\mathop{\to }}-\infty$.

When the alternative hypothesis is true, however, $2\log \mlr_{\bs s} \ \overset{d}{\mathop{\to }}\   \chi^2_{|\bs s|, \, \delta}$ as $n\to\infty$ (Equation \ref{d2logR_given_theta}) i.e.\ represents the sum of independent and identically distributed random variables that each has an expectation that is a linear function of $n$ (Equation \ref{thisisdelta} and MT Equation \ref{fisher}) and finite variance. The second term on the right-hand side of Equation \ref{logBFeqn} is therefore $O_p(n)$. As the first term is $O\left(-\log{(n)}\right)$, the term on the right-hand side dominates the term on the left-hand side, and $\log(\BF_{\bs s})\overset{p}{\mathop{\to }}\infty$. 


Based upon the consistency result demonstrated here, we argue that, whereas the local alternative hypothesis is useful for motivating the asymptotic theory on which the approximate methods are based, in practice $h$ should be fixed, for example, at $h = 1$, as in the unit information prior \citep{kass1995reference}, and further that in the classical significance threshold, $\alpha$, should scale with $n^{-1/2}$, to also make the classical approach consistent.

\section{Proof of Lemma \ref{lemma_scaled_posterior_odds}} \label{proof_lemma_scaled_posterior_odds}

\begin{proof}[Step 1. Define the scaled posterior odds]
To study the large-sample behaviour of the model-averaged posterior odds (MT Eqn. \ref{MPO1}), we can define the scaled odds as
\begin{eqnarray}
\kappa_n \ \po_{\mathcal{A}_\V:\mathcal{O}_\V} &=& \kappa_n \ \dfrac{\sum_{\bs s \in \mathcal{A}_\V} {\po}_{\bs s}}{\sum_{\bs s \in \mathcal{O}_\V} {\po}_{\bs s}}, \qquad \kappa_n = \left( \mu \ \sqrt{\tfrac{h}{n+h}} \right)^{-1} .
\end{eqnarray}
As a reminder of notation, we are interested in testing the null hypothesis $\omega_{\bs v} = \{ \theta : \beta_j = 0 \ \forall\  v_j = 0 \}$. The null hypothesis encompasses all the models
$\mathcal{O}_\V = \{ {\bs s} \in \mathcal{S} : \Theta_{\bs s} \subseteq \omega_{\bs v} \}$ where an individual model $\bs s$ is defined by its parameter space $\Theta_{\bs s} = \{ \theta : \beta_j = 0 \ \forall\  s_j = 0;\ \beta_j \neq 0 \ \forall\  s_j = 1 \}$. $\mathcal{S} = \{0, 1\}^\nu$ is the state space of all models and the alternative hypothesis encompasses the remaining models $\mathcal{A}_\V = \{ \mathcal{S} \setminus \mathcal{O}_\V \}$. The models are disjoint.

It will be convenient to define the set of `basic' alternative models,
\begin{eqnarray}
{\mathcal{T}_\V} &=& \{\bs t : \bs t \in \mathcal{A}_\V, \; t_j=0 \  \forall\ v_j=1\} ,
\end{eqnarray}
the index set of variables-of-interest, to which the null hypothesis pertains,
\begin{eqnarray}
{\mathcal{V}_\V} &=& \{j \in \{1, \dots, \nu\} : v_j = 0\} 
\end{eqnarray}
and the posterior probability of model $\bs s$ versus the other null models in $\mathcal{O}_\V$,
\begin{eqnarray}
w_{\bs s} = \dfrac{\po_{\bs s} }{ \sum_{\bs s^\prime \in \mathcal{O}_\V} \po_{\bs s^\prime} } ,
\end{eqnarray}
using which we can rewrite the scaled model-averaged posterior odds as
\begin{eqnarray}
\kappa_n \ \po_{\mathcal{A}_\V:\mathcal{O}_\V} &=& \sum_{\bs t \in \mathcal{T}_\V} \ \sum_{\bs s \in \mathcal{O}_\V} w_{{\bs s}} \ \kappa_n \ \po_{\bs s + \bs t : \bs s} . \label{define_pobar_via_t}
\end{eqnarray}

In the Doublethink model (MT Definition \ref{define_bmatest}), the prior odds of variable inclusion are $\mu$, independently for each variable, and the posterior odds are
\begin{eqnarray}
\po_{\bs s} &:=& {\kappa_n}^{-|\bs s|} \ {\mlr_{n, \bs s}}^{n/(n+h)}.
\end{eqnarray}
Notice that, in the proof, we use subscript $n$ to highlight the finite-sample distributional properties of the MLR, and we drop the subscript $n$ for the limiting distribution.

\textit{Step 2. Simplification of the posterior odds.} \\
\textit{Step 2a. Independent case.} If we assume that the likelihood factorizes due to independence such that
\begin{eqnarray}
\mlr_{n,{\bs s}} &\equiv& \prod_{j=1}^\nu {\mlr_{n,\bs e_j}}^{\bs s_j}
\end{eqnarray}
where $\{\bs e_j\}_i = \mathbb{I}(i=j)$, then since the prior factorizes due to independence such that
\begin{eqnarray}
\mu_{\bs s} &=& \mu^{|\bs s|},
\end{eqnarray}
the posterior odds have the property that
\begin{eqnarray}
\po_{{\bs s} + \bs t: {\bs s}}&\equiv&\po_{{\bs s'} + \bs t: {\bs s'}}
\end{eqnarray}
for all $\bs s, \bs s' \in \mathcal{O}_\V$ and all $\bs t \in \mathcal{T}_\V$.

Assuming the null hypothesis $\bs v$ is true in the sense that the true parameters $\tilde{\theta} \in \omega_{\bs v}$, then we can define the true model $\tilde{\bs s} \in \mathcal{O}_\V$ such that $\tilde{\theta} \in \Theta_{\tilde{\bs s}}$, where $\Theta_{\tilde{\bs s}} \subseteq \omega_{\bs v}$. This allows us to simplify the scaled model-averaged posterior odds (Equation \ref{define_pobar_via_t}) to
\begin{eqnarray}
\kappa_n \ \po_{\mathcal{A}_\V:\mathcal{O}_\V} &=& \sum_{\bs t \in \mathcal{T}_\V} \kappa_n \ \po_{\tilde{\bs s} + \bs t : \tilde{\bs s}} .
\end{eqnarray}
By the Johnson model, we have
\begin{eqnarray}
\kappa_n \ \po_{\mathcal{A}_\V:\mathcal{O}_\V} &=& \sum_{\bs t \in \mathcal{T}_\V} {\kappa_n}^{1-|\bs t|} \ {\mlr_{n,\tilde{\bs s} + \bs t : \tilde{\bs s}}}^{n/(n+h)} . \label{def_scale_posterior_odds_independence_johnson}
\end{eqnarray}

By Wilks' theorem \citep{wilks1938large, wald1943tests}, assuming standard regularity conditions, (MT Definition \ref{define_lrt}),
\begin{eqnarray}
2 \log \mlr_{n,\tilde{\bs s}+ \bs t:\tilde{\bs s}}&\overset{d}{\rightarrow}& \chi^2_{|\bs t|}, \qquad  n\rightarrow\infty
\end{eqnarray}
where $\chi^2_{\nu}$ represents a chi-squared distribution with $\nu$ degrees of freedom. Equivalently,
\begin{eqnarray}
\mlr_{n,\tilde{\bs s}+ \bs t:\tilde{\bs s}} &\overset{d}{\rightarrow}& \textrm{LG}\left(\tfrac{|\bs t|}{2},1 \right), \qquad n\rightarrow\infty ,
\end{eqnarray}
where $\mathrm{LG}(a, b)$ represents a log-gamma distribution with shape parameter $a$ and rate parameter $b$. It follows that
\begin{eqnarray}
{\mlr_{n,\tilde{\bs s} + \bs t : \tilde{\bs s}}}^{n/(n+h)} &\overset{d}{\rightarrow}& \textrm{LG}\left(\tfrac{|\bs t|}{2},1 \right), \qquad n\rightarrow\infty .
\end{eqnarray}

Slutsky's \citeyearpar{slutsky1925stochastische} theorem (see e.g.\ \citet{van2000asymptotic}) states that if $A_n \overset{d}{\rightarrow} A$, $B_n \overset{p}{\rightarrow} b$ and $C_n \overset{p}{\rightarrow} c$ as $n\rightarrow\infty$, for random variables $A_n, B_n, C_n$ and $A$, and constants $b$ and $c$,
and if $b\neq0$, then $A_n\ B_n + C_n \overset{d}{\rightarrow} A\ b + c$ as $n\rightarrow\infty$. If $b=0$ and $A_n$ is bounded in probability, 
then $A_n\ B_n + C_n \overset{p}{\rightarrow} c$.

We rewrite the scaled posterior odds (Equation \ref{def_scale_posterior_odds_independence_johnson}) as
\begin{eqnarray}
\kappa_n \ \po_{\mathcal{A}_\V:\mathcal{O}_\V} &=& A_n + B_n,
\end{eqnarray}
where
\begin{eqnarray}
A_n &=& \sum_{j\in\mathcal{V}_\V} {\mlr_{n,\tilde{\bs s} + \bs e_j : \tilde{\bs s}}}^{n/(n+h)},\nonumber\\
B_n &=& \sum_{\stackrel{\phantom{\,:}\bs t \in \mathcal{T}_\V\,:}{|\bs t| > 1}} {\kappa_n}^{1-|\bs t|} \ {\mlr_{n,\tilde{\bs s} + \bs t : \tilde{\bs s}}}^{n/(n+h)} . \nonumber
\end{eqnarray}
Noting that $\kappa_n^{1-|\bs t|} \overset{}{\rightarrow} \mathbb{I}(|\bs t|=1)$, then combining Wilks' and Slutsky's theorems, we have for $|\bs t|>1$
\begin{eqnarray}
{\kappa_n}^{1-|\bs t|} \ {\mlr_{n,\tilde{\bs s} + \bs t : \tilde{\bs s}}}^{n/(n+h)} &\overset{p}{\rightarrow}& 0, \qquad\qquad n\rightarrow\infty,
\end{eqnarray}
assuming ${\mlr_{n,\tilde{\bs s} + \bs t : \tilde{\bs s}}}^{n/(n+h)}$ is bounded in probability. Therefore by Slutsky's theorem, $B_n \overset{p}{\rightarrow} b = 0$. 

By Wilks' theorem, we have for $|\bs t|=1$ 
\begin{eqnarray}
{\mlr_{n, \tilde{\bs s} + \bs t : \tilde{\bs s}}}^{n/(n+h)} &\overset{d}{\rightarrow}& \mlr_{\tilde{\bs s} + \bs t : \tilde{\bs s}}, \qquad n\rightarrow\infty.
\end{eqnarray}
Since the independence assumption implies joint convergence in distribution of the summands in $A_n$ to a limiting distribution $A$, Slutsky's theorem implies $A_n + B_n \overset{d}{\rightarrow} A + b$. This simplifies the scaled model-averaged posterior odds to the sum of maximized likelihood ratios involving one degree-of-freedom tests:
\begin{eqnarray}
\kappa_n \ \po_{\mathcal{A}_\V:\mathcal{O}_\V} &\overset{d}{\rightarrow}& \sum_{j\in\mathcal{V}_\V} \mlr_{\tilde{\bs s} + \bs e_j : \tilde{\bs s}}, \qquad n\rightarrow\infty.
\end{eqnarray}

\textit{Step 2b. General case.} Next we establish an equivalent result in the general case. Assuming the Johnson model, we rewrite the scaled model-averaged posterior odds (Equation \ref{define_pobar_via_t}) as
\begin{eqnarray}
\kappa_n \ \po_{\mathcal{A}_\V:\mathcal{O}_\V} &=& \sum_{\bs t \in \mathcal{T}_\V} \ \sum_{\bs s \in \mathcal{O}_\V} w_{{\bs s}} \ \kappa_n^{1-|\bs t|} \ {\mlr_{n, \bs s + \bs t : \bs s}}^{n/(n+h)} \label{def_scaled_posterior_odds_general_johnson}\\
&=& A_n \ w_{\tilde{\bs s}} + B_n \ w_{\tilde{\bs s}} + C_n,  \nonumber
\end{eqnarray}
where
\begin{eqnarray*}
A_n &=& \sum_{j\in\mathcal{V}_\V} {\mlr_{n,\tilde{\bs s} + \bs e_j : \tilde{\bs s}}}^{n/(n+h)} \\
B_n &=& \sum_{\stackrel{\phantom{\,:}\bs t \in \mathcal{T}_\V\,:}{|\bs t| > 1}} \kappa_n^{1-|\bs t|} \ {\mlr_{n,\tilde{\bs s} + \bs t : \tilde{\bs s}}}^{n/(n+h)} \\
C_n &=& \sum_{\bs t \in \mathcal{T}_\V} \ \sum_{\stackrel{\phantom{\,:} \bs s \in \mathcal{O}_\V\,:}{\bs s \neq \tilde{\bs s}}} w_{{\bs s}} \ \kappa_n^{1-|\bs t|} \ {\mlr_{n,\bs s + \bs t : \bs s}}^{n/(n+h)} .
\end{eqnarray*}
The aim is to show, by assumption, that if $A_n \overset{d}{\rightarrow} A$ then $A_n \ w_{\tilde{\bs s}} + B_n \ w_{\tilde{\bs s}} + C_n \overset{d}{\rightarrow} A$, as $n\rightarrow\infty$.

Assuming that ${\mlr_{n,\tilde{\bs s} + \bs e_j : \tilde{\bs s}}}^{n/(n+h)}$, $j\in\mathcal{V}_\V$, converge jointly in distribution to identical but not necessarily independent $\mathrm{LG}(1/2,1)$ random variables, we have
\begin{eqnarray}
A_n &\overset{d}{\rightarrow}& \sum_{j\in\mathcal{V}_\V} \mlr_{\tilde{\bs s} + \bs e_j : \tilde{\bs s}}, \qquad n\rightarrow\infty. \label{def_Slutsky_An_d_A}
\end{eqnarray}

Assuming that ${\mlr_{n, \tilde{\bs s} + \bs t : \tilde{\bs s}}}^{n/(n+h)}$, $t\in\mathcal{T}_\V$, are individually bounded in probability, and noting that $\kappa_n^{1-|\bs t|} \overset{}{\rightarrow} 0$ for $|\bs t|>1$, Slutsky's theorem gives
\begin{eqnarray}
B_n &\overset{p}{\rightarrow}& 0, \qquad n\rightarrow\infty. \label{def_Slutsky_Bn_p_0}
\end{eqnarray}

By Doob's \citeyearpar{doob1949application} theorem, the posterior probability of the true null model, $\tilde{\bs s}$, versus the other null models in $\mathcal{O}_\V$, $w_{\tilde{\bs s}} \overset{a.s.}{\rightarrow} 1$, which implies
\begin{eqnarray}
w_{\tilde{\bs s}} &\overset{p}{\rightarrow}& 1, \qquad n\rightarrow\infty. \label{def_Slutsky_wstilde_p_0}
\end{eqnarray}

Rewriting $C_n$, we have
\begin{eqnarray}
C_n &=& \sum_{\bs t \in \mathcal{T}_\V} \kappa_n^{1-|\bs t|} \ {\mlr_{n, \tilde{\bs s} + \bs t : \tilde{\bs s}}}^{n/(n+h)} \ w_{\tilde{\bs s}} \ \left(\dfrac{\sum_{\bs s \in \mathcal{O}_\V} \po_{\bs s + \bs t}}{\po_{\tilde{\bs s} + \bs t}}  - 1 \right).
\end{eqnarray}
As above, $\kappa_n^{1-|\bs t|} \overset{}{\rightarrow} \mathbb{I}(|t|=1)$, and $w_{\tilde{\bs s}} \overset{a.s.}{\rightarrow} 1$. Doob's theorem is not applicable to the quantity
\begin{eqnarray*}
\dfrac{\po_{\tilde{\bs s} + \bs t}}{\sum_{\bs s \in \mathcal{O}_\V} \po_{\bs s + \bs t}},
\end{eqnarray*}
because $\tilde\theta \notin \Theta_{{\bs s} + \bs t}$ for any $\bs s \in \mathcal{O}_\V$, $\bs t \in \mathcal{T}_\V$. However, $\tilde\theta$ is arbitrarily close to $\Theta_{\tilde{\bs s} + \bs t}$ for all $\bs t \in \mathcal{T}_\V$, by the nature of the point null hypotheses. Therefore, by Schwartz \citeyearpar{schwartz1965onbayes},
\begin{eqnarray}
\dfrac{\po_{\tilde{\bs s} + \bs t}}{\sum_{\bs s \in \mathcal{O}_\V} \po_{\bs s + \bs t}} &\overset{a.s.}{\rightarrow}& 1, \qquad n\rightarrow\infty.
\end{eqnarray}
Assuming that ${\mlr_{n, \tilde{\bs s} + \bs t : \tilde{\bs s}}}^{n/(n+h)}$, $\bs t \in \mathcal{T}_\V$, are individually bounded in probability, Slutsky's theorem gives
\begin{eqnarray}
C_n &\overset{p}{\rightarrow} 0, \qquad n\rightarrow\infty. \label{def_Slutsky_Cn_p_0}
\end{eqnarray}

Consequently, combining Equations \ref{def_Slutsky_An_d_A}, \ref{def_Slutsky_Bn_p_0}, \ref{def_Slutsky_wstilde_p_0} and \ref{def_Slutsky_Cn_p_0} using Slutsky's theorem, Equation \ref{def_scaled_posterior_odds_general_johnson} simplifies to
\begin{eqnarray}
\kappa_n \ \po_{\mathcal{A}_\V:\mathcal{O}_\V} &\overset{d}{\rightarrow}& \sum_{j\,:\,v_j=0} \mlr_{\tilde{\bs s} + \bs e_j : \tilde{\bs s}}, \qquad n\rightarrow\infty,
\end{eqnarray}
as per the independent case.
\end{proof}

\section{Proof of Theorem \ref{theorem_fpr}} \label{proof_theorem_fpr}\label{proofofT2}

\begin{proof}[Step 1. Uniform convergence.]
By Lemma \ref{lemma_scaled_posterior_odds}, continuing the notation in the proof of Lemma \ref{lemma_scaled_posterior_odds} (Appendix \ref{proof_lemma_scaled_posterior_odds}), we have
\begin{eqnarray}
f_n(x) &=& \Pr\left( \kappa_n \ \po_{\mathcal{A}_\V:\mathcal{O}_\V} > x; \ \tilde\theta \right) \nonumber \\
g(x) &=& \Pr\left( \sum_{j\,:\,v_j=0} \mlr_{\tilde{\bs s} + \bs e_j : \tilde{\bs s}} > x; \right) \nonumber \\
\lim_{n\rightarrow\infty} f_n(x) &=& g(x) \qquad \forall\ x.
\end{eqnarray}
To further characterize the large-sample distribution of $\kappa_n \ \po_{\mathcal{A}_\V:\mathcal{O}_\V}$, we will use a uniform convergence result. By \citet[Lemma 2.11]{van2000asymptotic}, since $f_n(x)$ and $g(x)$ are tail distribution functions, the convergence in Lemma \ref{lemma_scaled_posterior_odds} is uniform with respect to $x$: for every $\epsilon>0$ there exists $N_\epsilon \geq 0$ such that
\begin{eqnarray}
|f_n(x) - g(x)| &<& \epsilon \qquad \forall\ n\geq N_\epsilon,\ \forall\ x.
\end{eqnarray}
This result will allow us to interchange the following diagonal limit and iterated limit to give:
\begin{equation}
\lim_{n\rightarrow\infty} \dfrac{f_n(x_n)}{g(x_n)} \ =\ \lim_{m\rightarrow\infty} \ \lim_{n\rightarrow\infty} \dfrac{f_n(x_m)}{g(x_m)} \ =\ \lim_{m\rightarrow\infty} 1 \ =\ 1, \label{def_interchange_diagonal_iterated_limit}
\end{equation}
for $x_n$ increasing in $n$.

\textit{Step 2. Simplification of the convolution.} \\
\textit{Step 2a. Independent case.} A function $f$ satisfying
\begin{eqnarray}
\lim_{x\rightarrow\infty} \dfrac{f(c\ x)}{f(x)} = c^{-\lambda} \qquad \forall\ c>0, \label{def_regvar_step4}
\end{eqnarray}
is said to be regularly varying (at infinity) if $\lambda \neq 0$ and slowly varying (at infinity) if $\lambda = 0$ (\citet{karamata1933mode}, \citet{mikosch1999regular} Definition 1.1.1). A random variable with regularly varying distribution function is called a regularly varying random variable with tail index $\lambda > 0$. The $\mathrm{LG}(a,b)$ distribution is regularly varying with tail index $b$ (see e.g.\ SI Equation 16 of \citet{wilson2019harmonic}).

Regularly varying random variables have `fractal-like' properties: they are said to be closed under convolution. \citet{nagaev1965limit} (see e.g.\ Corollary 1.3.6 of \cite{mikosch1999regular}) states that if $X_1,...,X_k$ are non-negative i.i.d.\ regularly varying random variables, then 
\begin{eqnarray}
 \Pr(X_1+...+X_k>x)\,\sim\, k \ \Pr(X_1>x),\qquad x \to \infty. \label{Mikosch1.3.6}
\end{eqnarray}
Strictly
\begin{eqnarray}
\lim_{x \to \infty}\;\frac{\Pr(X_1+...+X_k>x)}{k\Pr\left(X_1>x\right)}=1 .
\end{eqnarray}

Since $\mlr_{\tilde{\bs s} + \bs e_j : \tilde{\bs s}} \overset{d}{=} \mathrm{LG}(1/2, 1)$,
\begin{eqnarray}
\lim_{x \to \infty} \ 
\dfrac{ \Pr\left( \sum_{j\,:\,v_j=0} \mlr_{\tilde{\bs s} + \bs e_j : \tilde{\bs s}} > x \right) }{ (\sizeV) \ \Pr\left( \mlr_{\tilde{\bs s} + \bs e_k : \tilde{\bs s}} > x \right) }
&=& 1, \qquad \forall\ k\,:\,v_k=0 . \label{def_step4a_result}
\end{eqnarray}

\textit{Step 2b. General case.} \cite{davis1996limit} Lemma 2.1 states that if $X_1,...,X_k$ are non-negative, regularly varying random variables with dependence structure such that
\begin{eqnarray}
    \dfrac{\Pr(X_i>x, X_j>x)}{\Pr(X_1>x)} &\rightarrow& 0, \qquad x\rightarrow\infty, \qquad i\neq j
\label{DavisResnickCondition_djw}
\end{eqnarray}
then
\begin{eqnarray}
 \Pr(X_1+...+X_k>x)\,\sim\, k \ \Pr(X_1>x),\qquad x \to \infty. \label{corr4444_supp}
\end{eqnarray}

This dependence structure is known as asymptotic independence. Below we prove asymptotic independence for pairs of maximized likelihood ratios, under the null hypothesis, in the non-independent case, leading to the same result as in the independent case (Equation \ref{def_step4a_result}):

To characterize the tail behaviour of the bivariate distribution of $( \mlr_{n, \tilde{\bs s} + \bs e_i : \tilde{\bs s}}, \ \mlr_{n, \tilde{\bs s} + \bs e_j : \tilde{\bs s}} )$,  $i, j \in \mathcal{V}_\V$, $i\neq j$, we express the MLRs in terms of score statistics. From Equation \ref{MLRsimplehf} of the background theory (\citet{fisher1925statistical, cox1974theoretical}), we have
\begin{eqnarray}
2\log(\mlr_{n, \tilde{\bs s}+\bs t:\tilde{\bs s}}) &=& U^T_{\tilde{\bs s}+\bs t:\tilde{\bs s}} \ U^{\phantom{T}}_{\tilde{\bs s}+\bs t:\tilde{\bs s}} + o_p(1) \label{2logrsisus}
\end{eqnarray}
where the score vector, of length $|\bs t|$, equals
 \begin{eqnarray}
U_{\tilde{\bs s}+\bs t:\tilde{\bs s}} &=& n^{1/2} \left[\left( \left[ \mathcal{I}(\tilde\theta)_{\F_{\tilde{\bs s}+\bs t},\F_{\tilde{\bs s}+\bs t}}  \right]^{-1} \right)_{\B_{\bs t},\B_{\bs t}}\right]^{-1/2} \hat{\bs{\beta}}^{\tilde{\bs s}+\bs t}_{\B_{\bs t}} . \label{UtoB_djw}
    \end{eqnarray}
(Formally there is a change in the frame of reference: $\tilde{\bs s}$ (above) maps on to model $\bs 0$ (Equation \ref{MLRsimplehf})  while $\tilde{\bs s} + \bs t$ (above) maps on to model $\bs s$ (Equation \ref{MLRsimplehf}).)

Since $\tilde{\bs s}$ is the true model, and therefore $\tilde{\bs \beta}_{\B_{\bs t}} = \bs 0$, by Equation \ref{marginalMLEhf} we have
\begin{eqnarray}
U_{\tilde{\bs s}+\bs t:\tilde{\bs s}} &\overset{d}{\to}& \mathcal{N}_{|{\bs t}|}\left( \bs 0, \bs 1 \right) . \label{Utonorm_djw}
\end{eqnarray}
Using these results, we aim to characterize the bivariate distribution of $(U_{\tilde{\bs s}+\bs e_i:\tilde{\bs s}}, \ U_{\tilde{\bs s}+\bs e_j:\tilde{\bs s}})$, $i, j \in \mathcal{V}_\V$, $i\neq j$, via the distribution of the score vector $U_{\tilde{\bs s}+\bs t:\tilde{\bs s}}$, taking $\bs t = \bs e_i+\bs e_j$.

By defining the scalar coefficient,
\begin{eqnarray*}
B_i &=& n^{1/2} \left[\left( \left[ \mathcal{I}(\tilde\theta)_{\F_{\tilde{\bs s}+\bs e_i},\F_{\tilde{\bs s}+\bs e_i}}  \right]^{-1} \right)_{\B_{\bs e_i},\B_{\bs e_i}}\right]^{-1/2} ,
\end{eqnarray*}
we can use Equation \ref{UtoB_djw} to write
\begin{eqnarray}
    \left( \begin{array}{c}
        U_{\tilde{\bs s}+\bs e_i : \tilde{\bs s}} \\ U_{\tilde{\bs s}+\bs e_j : \tilde{\bs s}}
    \end{array} \right) &=& \left( \begin{array}{cc}
            B_i & 0 \\ 0 & B_j
        \end{array} \right) \left( \begin{array}{c}
        {\hat{\beta}}^{\tilde{\bs s}+\bs e_i}_{\B_{\bs e_i}} \\\hat{\beta}^{\tilde{\bs s}+\bs e_j}_{\B_{\bs e_j}} 
    \end{array} \right) . \label{utob_djw}
\end{eqnarray}
Next, let $\bs w=\tilde{\bs s}+\bs e_i +\bs e_j$ and
\begin{eqnarray*}
\Lambda &=& n \ \left[ \left( \left[ \mathcal{I}(\tilde\theta)_{\F_{\bs w}, \F_{\bs w}} \right]^{-1} \right)_{\B_{\bs e_i + \bs e_j}, \B_{\bs e_i + \bs e_j}} \right]^{-1} .
\end{eqnarray*}
Under the large $n$ approximation, a Taylor expansion (c.f. Equation \ref{shorttaylor} of the background theory) gives
\begin{eqnarray}
\hat{\beta}^{\tilde{\bs s} +\bs e_i}_{\B_{\bs e_i}} &=& \hat{ \beta}^{\bs w}_{\B_{\bs e_i}} +
\left(\Lambda_{\B_{\bs e_i}, \B_{\bs e_i}}\right)^{-1} \ \Lambda_{\B_{\bs e_i}, \B_{\bs e_j}} \ \hat{\bs \beta}^{\bs w}_{\B_{\bs e_j}} +o_p(n^{-1/2}) .\label{thisisD14}
\end{eqnarray}
By defining another scalar coefficient
\begin{eqnarray*}
A_{i,j} &=& \left(\Lambda_{\B_{\bs e_i}, \B_{\bs e_i}}\right)^{-1} \ \Lambda_{\B_{\bs e_i}, \B_{\bs e_j}}
\end{eqnarray*}
element selection gives
 \begin{eqnarray}
 \hat{\beta}^{\tilde{\bs s}+\bs e_i}_{\B_{\bs e_i}} &=& \hat{ \beta}^{\bs w}_{\B_{\bs e_i}} + A_{i, j}
 \ \hat{\bs \beta}^{\bs w}_{\B_{\bs e_j}}+o_p(n^{-1/2})\\ \implies
\left( \begin{array}{c}
        {\hat{\beta}}^{\tilde{\bs s}+\bs e_i}_{\B_{\bs e_i}} \\\hat{\beta}^{\tilde{\bs s}+\bs e_j}_{\B_{\bs e_j}} 
    \end{array} \right)&=&   \left( \begin{array}{cc}
        1 & A_{i, j} \\ A_{j, i} & 1
    \end{array} \right)  \left( \begin{array}{cc}
        {\hat{\beta}}^{\bs w}_{\B_{\bs e_i}} \\ \hat{\beta}^{\bs w}_{\B_{\bs e_j}} 
    \end{array} \right) \\
\implies \left( \begin{array}{c}
        U_{\tilde{\bs s}+\bs e_i : \tilde{\bs s}} \\ U_{\tilde{\bs s}+\bs e_j : \tilde{\bs s}}
    \end{array} \right) &=& \left( \begin{array}{cc}
            B_i & 0 \\ 0 & B_j
        \end{array} \right)  \left( \begin{array}{cc}
        1 & A_{i, j} \\ A_{j, i} & 1
    \end{array} \right) \ 
        {\hat{\beta}}^{\bs w}_{\B_{\bs e_i+\bs e_j}}  
 \qquad  (\textrm{by Eqn. \ref{utob_djw}})\\&=&
    \left( \begin{array}{cc}
            B_i & B_i \ A_{i, j} \\ B_j \ A_{j, i} & B_j
        \end{array} \right) \ \Lambda^{-1/2} \ U_{\bs w:\tilde{\bs s}} \quad\quad\quad  (\textrm{by Eqn. \ref{UtoB_djw}})
 \end{eqnarray}
From Equation \ref{Utonorm_djw}, $U_{\bs w:\tilde{\bs s}}\overset{d}{\to} \mathcal{N}_{2}\left( \bs 0, \bs 1 \right)$. Therefore, the joint distribution of the scores is (in the large $n$ limit) a linear transformation of a standard bivariate normal distribution and, through multivariate normal affine transformation rules, is therefore also a bivariate normal distribution. Thus, for some correlation coefficient ($-1<c<1$), we can write
\begin{eqnarray}
    \left( \begin{array}{c}
        U_{\tilde{\bs s}+\bs e_i : \tilde{\bs s}} \\ U_{\tilde{\bs s}+\bs e_j : \tilde{\bs s}}
    \end{array} \right)
    &\overset{d}{\to}& \mathcal{N}_{2}\left( 
        \left( \begin{array}{c}
            0 \\ 0
        \end{array} \right) 
    ,
        \left( \begin{array}{cc}
            1 & c \\ c & 1
        \end{array} \right) 
    \right) .
 \end{eqnarray}

Therefore, for $i, j\in \mathcal{V}_\V, i\neq j$,
\begin{eqnarray}
    \Pr\left( \mlr_{\tilde{\bs s}+\bs e_i : \tilde{\bs s}} > x, \ \mlr_{\tilde{\bs s}+\bs e_j : \tilde{\bs s}} > x;\ \tilde{\theta}  \right)
    &=&
    4 \Pr\left( U_{\tilde{\bs s}+\bs e_i : \tilde{\bs s}} > \sqrt{2 \log (x)}, \ U_{\tilde{\bs s}+\bs e_j : \tilde{\bs s}} > \sqrt{2 \log (x)} \right) \nonumber \\
    &\sim& \dfrac{1}{\pi} \ \dfrac{(1+c)^2}{\sqrt{1-c^2}} \ \log(x)^{-1} \ x^{-2/(1+c)}, \qquad x\rightarrow\infty.
\end{eqnarray}
Here, the last line is from a standard asymptotic form for the bivariate Normal distribution \citep{zhou2017tail}. By \citet{wilks1938large} we have the marginal tail probability, which can be written as (see e.g. \citet{mikosch1999regular}, \citet{wilson2019harmonic} SI Equation 16):
\begin{eqnarray}
\Pr\left(\mlr_{\tilde{\bs s}+ \bs e_i:\tilde{\bs s}} > x;\, \tilde{\theta} \right) &\sim& \dfrac{1}{\sqrt{\pi}} \ \log(x)^{-1/2} \ x^{-1}, \qquad x\rightarrow\infty.
\end{eqnarray}
The Davis-Resnick condition (Equation \ref{DavisResnickCondition_djw}) is therefore satisfied because
\begin{eqnarray}
    \dfrac{
        \Pr\left( \mlr_{\tilde{\bs s}+\bs e_i : \tilde{\bs s}} > \dfrac{\tau}{\mu\sqrt{\xi_n}}, \ \mlr_{\tilde{\bs s}+\bs e_j : \tilde{\bs s}} > \dfrac{\tau}{\mu\sqrt{\xi_n}}\ {\tilde\theta} \right)
    }{
        \Pr\left(\mlr_{\tilde{\bs s}+ \bs e_k:\tilde{\bs s}} > \dfrac{\tau}{\mu\sqrt{\xi_n}};\ {\tilde\theta} \right)
    }
    &\sim& \dfrac{1}{\sqrt{\pi}} \ \dfrac{(1+c)^2}{\sqrt{1-c^2}} \ \log\left(\dfrac{\tau}{\mu\sqrt{\xi_n}}\right)^{-\frac{1}{2}} \ \left(\dfrac{\tau}{\mu\sqrt{\xi_n}}\right)^{-\frac{(1-c)}{(1+c)}} \nonumber\\&\rightarrow& 0, \qquad n\rightarrow\infty\qquad (i,j,k\in \mathcal{V}_\V, i\neq j).
\end{eqnarray}
Therefore
\begin{eqnarray}
\lim_{x \to \infty}
\dfrac{ \Pr\left( \sum_{j\,:\,v_j=0} \mlr_{\tilde{\bs s} + \bs e_j : \tilde{\bs s}} > x \right) }{ (\sizeV) \ \Pr\left( \mlr_{\tilde{\bs s} + \bs e_k : \tilde{\bs s}} > x \right) }
&=& 1, \qquad \forall\ k\,:\,v_k=0 . \label{def_step4b_result}
\end{eqnarray}

\textit{Step 3. Conclusion.}  By the definition of regular variation (Equation \ref{def_regvar_step4}), with $\lambda = 1$ for a LG$(1/2, 1)$ distribution, we have
\begin{eqnarray}
\lim_{x \to \infty}
\dfrac{ |\mathcal{V}_\V| \Pr\left( \mlr_{\tilde{\bs s} + \bs e_k : \tilde{\bs s}} > x \right) }{ \Pr\left( \mlr_{\tilde{\bs s} + \bs e_k : \tilde{\bs s}} > x/|\mathcal{V}_\V| \right) }
&=& 1, \qquad \forall\ k \in \mathcal{V}_\V .
\end{eqnarray}
Applying the product rule for limits, the result for both the independent (Equation \ref{def_step4a_result}) and general case (Equation \ref{def_step4b_result}) can be expressed as
\begin{eqnarray}
\lim_{x \to \infty}
\dfrac{ \Pr\left( \sum_{j \in \mathcal{V}_\V} \mlr_{\tilde{\bs s} + \bs e_j : \tilde{\bs s}} > x \right) }{ \Pr\left( \mlr_{\tilde{\bs s} + \bs e_k : \tilde{\bs s}} > x/|\mathcal{V}_\V| \right) }
&=& 1, \qquad \forall\ k \in \mathcal{V}_\V .
\end{eqnarray}

Writing $x_n = \kappa_n \ \tau$, the uniform convergence assumption (Equation \ref{def_interchange_diagonal_iterated_limit}) implies
\begin{eqnarray}
\lim_{n \to \infty}
\dfrac{ \Pr\left( \kappa_n \ \po_{\mathcal{A}_\V:\mathcal{O}_\V} > \kappa_n \,\tau; \ \tilde\theta \right) }{ \Pr\left( \mlr_{\tilde{\bs s} + \bs e_k : \tilde{\bs s}} > \dfrac{\kappa_n \ \tau}{|\mathcal{V}_\V|} \right) }
&=& 1, \qquad \forall\ k \in \mathcal{V}_\V ,
\end{eqnarray}
or, equivalently,
\begin{eqnarray}
\lim_{n \to \infty}
\dfrac{ \Pr\left( \po_{\mathcal{A}_\V:\mathcal{O}_\V} > \tau; \ \tilde\theta \right) }{ \Pr\left( \chi^2_1 > 2 \log \dfrac{\tau}{(\sizeV) \ \mu \ \sqrt{\xi_n}} \right) }
&=& 1, \qquad \xi_n=\frac{h}{n+h}. \label{justlimofalpha}
\end{eqnarray}

Finally we have
\begin{eqnarray}
\underset{\theta\in\omega_{\bs v} \ n \to \infty}{\sup \, \ \lim}
\dfrac{ \Pr\left( \po_{\mathcal{A}_\V:\mathcal{O}_\V} > \tau; \ \theta \right) }{ \Pr\left( \chi^2_1 > 2 \log \dfrac{\tau}{(\sizeV) \ \mu \ \sqrt{\xi_n}} \right) }
&=& 1.
\end{eqnarray}
\end{proof}

\section{Proof of Theorem \ref{TH2FWER}} \label{proofT3}
\begin{proof}
By the logic of a CTP (Definition \ref{definitionctp}), the FWER is controlled at or below
\begin{eqnarray}
    \alpha &\leq& \max \{ \alpha_{\bs v} : \omega_{v} \in W \},
\end{eqnarray}
where the FPR for the test of null hypothesis $\omega_{\bs v}$ is
\begin{eqnarray}
   \alpha_{\bs v} &=& \sup_{\theta \in \omega_{\bs v}}\,\Pr\left(
    \psi_{\bs v}(\bs y) = 1 ; \bs x, \theta
   \right),
\end{eqnarray}
and $\psi_{\bs v}(\bs y) = \mathbb{I}(\PO_{\omega_{\bs v}^\complement:\omega_{\bs v}} \geq \tau)$ for a Bayesian test (Definition \ref{definitionbayestest}).

Moreover, by Theorem \ref{theorem_bayesCTP}, the Bayesian test is a shortcut CTP, so $\phi_{\bs v}(\bs y) = \psi_{\bs v}(\bs y)$. 

In the Doublethink model (Definition \ref{define_bmatest}) we specify the null hypotheses as $\omega_{{\bs v}} = \{ \theta : \beta_j=0\ \forall\ {v}_{j}=0 \}$, $\bs v \in \{0, 1\}^\nu$, and we use the notation $\po_{\mathcal{A}_\V:\mathcal{O}_\V} = \PO_{\omega_{\bs v}^\complement:\omega_{\bs v}}$.

By Theorem \ref{theorem_fpr} we have
\begin{eqnarray}
    \sup_{\theta \in \omega_{\bs v}} \lim_{n\rightarrow\infty} \frac{\Pr(\po_{\mathcal{A}_\V:\mathcal{O}_\V} \geq \tau;\, \theta)}{\Pr\left( \chi^2_1
    \geq 2 \log \tfrac{\tau}{(\sizeV) \, \mu \, \sqrt{\xi_n}} \right)} &=& 1.
\end{eqnarray}

Since $(\sizeV)\leq\nu$ and the tail probability of the chi-squared distribution is monotonic decreasing
\begin{eqnarray}
    \Pr\left( \chi^2_1
    \geq 2 \log \dfrac{\tau}{\nu \, \mu \sqrt{\xi_n}} \right) 
    &\geq& \Pr\left( \chi^2_1
    \geq 2 \log \dfrac{\tau}{(\sizeV) \, \mu \sqrt{\xi_n}} \right)\qquad \forall\; \V.
\end{eqnarray}
This implies
\begin{eqnarray}
    \max_{\omega_{\bs v} \in \Omega} \ \sup_{\theta \in \omega_{\bs v}} \lim_{n\rightarrow\infty} \frac{
        \Pr\left( \po_{\mathcal{A}_\V:\mathcal{O}_\V} \geq \tau;\; \theta \right)
    }{
        \Pr\left( \chi^2_1
    \geq 2 \log \tfrac{\tau}{\nu \, \mu \, \sqrt{\xi_n}} \right)
    } &\leq& 1.
\end{eqnarray}
We express this as an asymptotic bound, where convergence is pointwise with respect to $\theta$:
\begin{eqnarray}
    \alpha &:=& \max_{\omega_{\bs v} \in \Omega} \ \sup_{\theta \in \omega_{\bs v}} \Pr\left( \po_{\mathcal{A}_\V:\mathcal{O}_\V} \geq \tau;\; \theta \right)
    \ \overset{\mathrm{pw}}{\lesssim} \  \Pr\left( \chi^2_1
    \geq 2 \log \dfrac{\tau}{\nu \, \mu \, \sqrt{\xi_n}} \right), 
    \quad n\rightarrow\infty \label{alpha_fwer_supp}
\end{eqnarray}

\end{proof}

\section{Proof of Theorem \ref{theorem_uniformconvergence_twovariablemodel}.} \subsection*{Uniform convergence in a simplified two-variable model.}\label{proof_theorem_uniformconvergence_twovariablemodel}
\begin{proof}
Let $\theta=\left( \beta_1, \beta_2, \gamma\right)^T$, where $\gamma$ is the intercept and $\beta_1$ and $\beta_2$ are the regression coefficients for the model 
\begin{eqnarray}
    Y\overset{d}{=}\,{\mathcal{N}_{n}}\left((X_1,X_2,1)\,\theta,\sigma^2 I_n \right),
\end{eqnarray}
Here, $\sigma^2$ is assumed to be known and $I_n$ is the identity matrix of dimension $n$. The unit Fisher information is given by:
\begin{eqnarray} \mathcal{I}(\theta)=\tfrac{1}{n\sigma^2}(X_1,X_2,1)^T(X_1,X_2,1)
=\tfrac{1}{\sigma^2}\left(\begin{smallmatrix}
     1&\rho&0\\
     \rho&1&0\\
     0&0&1
\end{smallmatrix}\right)\label{thefisher}
\end{eqnarray}
And its inverse is given by:
\begin{eqnarray} {\mathcal{I}(\theta)}^{-1}
=\left(\begin{matrix}
     \left(I(\theta)_{\B,\B}\right)^{-1}& \begin{smallmatrix}0\\ 0 \end{smallmatrix}\\
     \begin{smallmatrix}0 & 0 \end{smallmatrix}  &\sigma^2
\end{matrix}\right)\qquad\left(\textrm{where  } I(\theta)_{\B,\B}=\tfrac{1}{\sigma^2} \left(\begin{smallmatrix}
     1&\rho \\
     \rho&1
\end{smallmatrix}\right)\right)
\end{eqnarray}

From Equation \ref{thefisher}
\begin{eqnarray}
    \left[\mathcal{I}(\theta)_{\F_{\bs s},\F_{\bs s}}\right]^{-1}=\begin{cases}        \sigma^2\left(\begin{smallmatrix}
     1&0  \\
     0&1
\end{smallmatrix}\right)& \textrm{if  } \bs s \in \left\{ 
(1,0)^T,(0,1)^T \right\}\label{Ifsfs}\\{\mathcal{I}(\theta)}^{-1}&\textrm{if  } \bs s =(1,1)^T 
    \end{cases}
\end{eqnarray}
Following Step 2b of the proof of Theorem \ref{Corr20}, the score statistics for the models $\bs s=(1,0)^T$ and $\bs s=(0,1)^T$ can be expressed as linear transformations of the grand alternative model, $\bs s=(1,1)^T$ 
\begin{eqnarray}
U_{\bs s} &=& n^{1/2} \left[\left( \left[ \mathcal{I}(\theta)_{\F_{\bs s},\F_{\bs s}} \right]^{-1} \right)_{\B_{\bs s},\B_{\bs s}}\right]^{-1/2} \hat{\bs{\beta}}^{\bs s}_{\B_{\bs s}},\qquad \left({\textrm{by Eqn. \ref{UtoB_djw}}}\right)\label{neweq94}\\
&=& \begin{cases}\tfrac{n^ {1/2}}{\sigma}\hat{\bs{\beta}}^{\bs s}_{\B_{\bs s}} & \textrm{if  }\bs s \in \{(1,0)^T,(0,1)^T\}\\ {n^{1/2}}L\hat{\bs{\beta}}^{\bs s}&\textrm{if  }\bs s= (1,1)^T\label{case1b} \end{cases}\label{Uscases}
\end{eqnarray}
Here $L$ is defined so that $ I(\theta)_{\B,\B}=L^TL$. 
Since the likelihood surface is multivariate normal
\newcommand\scalemath[2]{\scalebox{#1}{\mbox{\ensuremath{\displaystyle #2}}}}
\begin{eqnarray}
\hat{\bs \beta}^{\bs s}_{\B_{\bs s}}&=&\hat{\bs \beta}^{11}_{\B_{\bs s}} + \left[\mathcal{I}(\theta)_{\B_{\bs s},\B_{\bs s}}\right]^{-1}\,\mathcal{I}(\theta)_{\B_{\bs s},\B_{\bs s'}}\,  \hat{\bs \beta}^{11}_{\B_{\bs s'}}\quad\left(\textrm{c.f. Eqn. \ref{thisisD14}}\right)\label{en127now}\nonumber\\
&=&\hat{\bs \beta}^{11}_{\B_{\bs s}} +\rho \ \hat{\bs \beta}^{11}_{\B_{\bs s'}}\qquad \qquad \textrm{for  } (\bs s,\bs s')\in \{\left({\scalemath{0.8}{(1,0)^T}}, {\scalemath{0.8}{(0,1)^T}}\right),\left({\scalemath{0.8}{(0,1)^T}}, {\scalemath{0.8}{(1,0)^T}}\right)\}
\end{eqnarray}
Therefore,
\begin{eqnarray}
\begin{pmatrix}U_{10}\\U_{01}\end{pmatrix} &=&n^{1/2}\sigma\, I(\theta)_{\B,\B}\,\hat{\bs \beta}^{11}\qquad\left(\textrm{by Eqns. \ref{Uscases} and \ref{en127now}}\right)\nonumber\\&=&n^{1/2}\sigma\,\, L^TL\,\hat{\bs \beta}^{11}\nonumber\\&=&\sigma L^T\,U_{11}\qquad\left(\textrm{by Eqn. \ref{Uscases}}\right)\nonumber\nonumber\\
&=&\sigma\,L^T\begin{pmatrix}W\\Z\end{pmatrix}\qquad\left({\textrm{let  }U_{11}=\left(\substack{W\\Z}\right)}\right)\nonumber\\
&=&\begin{pmatrix}\sqrt{1-\rho^2}&\rho\\0&1\end{pmatrix}\begin{pmatrix}W\\Z\end{pmatrix}\qquad\left(\textrm{let } L:=\tfrac{1}{\sigma}\begin{pmatrix}\sqrt{1-\rho^2}&0\\\rho&1\end{pmatrix}\right) \nonumber\\&=&\begin{pmatrix}\sqrt{1-\rho^2}W+\rho Z\\Z\end{pmatrix}\label{U10U01}
\end{eqnarray}

Let $\tilde{\theta}=\left(\substack{\tilde{\beta}\\\tilde{\gamma}}\right)$ be the true parameter vector. The distribution of $U_{11}$ can be derived as follows
\begin{eqnarray}
\hat{\bs \beta}^{11}&\overset{d}{\mathop{=}}&\mathcal{N}_2 \left({\tilde{\bs\beta}},\tfrac{1}{n}[\mathcal{I}(\tilde\theta)^{-1}]_{\B,\B} \right)\qquad\left(\textrm{c.f. Equation \ref{marginalMLEhf}}\right)\nonumber\\
\implies\qquad\quad U_{11}&\overset{d}{\mathop{=}}&\mathcal{N}_2 \left(n^{\frac{1}{2}}L{\tilde{\bs\beta}},{I}_2\right)\qquad\qquad \left(\textrm{Eqn. \ref{Uscases}} \right)\nonumber\\ \implies\qquad\begin{pmatrix}W\\Z\end{pmatrix}&\overset{d}{\mathop{=}}&\mathcal{N}_2 \left(\tfrac{n^{1/2}}{\sigma}\left(\begin{smallmatrix}\sqrt{1-\rho^2}&0\\\rho&1\end{smallmatrix}\right){\tilde{\bs\beta}},{I}_2\right) \nonumber\\&\overset{d}{\mathop{=}}&\mathcal{N}_2 \left(\tfrac{n^{1/2}}{\sigma}\left(\begin{smallmatrix}\sqrt{1-\rho^2}\tilde{\beta}_1\\\rho \tilde{\beta}_1+\tilde{\beta}_2\end{smallmatrix}\right),I_2\right) \label{zwdistrib}
\end{eqnarray}

To consider the limiting false positive rate, define restrictions on the size of the coefficients. We will exclude from the parameter space small non-zero values of $\bs \beta$, i.e.\ 
$\Theta^- = \{\theta \in \Theta : |\beta_j| \notin (0, \beta_\mathrm{min}) \ \forall\ j=1\dots\nu\}$,
so that ${\omega}_{\bs v}^*= \omega_{\bs v} \cap \Theta^-$ for all $\bs v \in \{0, 1\}^\nu$. For this simplified model, there are three possible null hypotheses, $\Omega^*=\left\{\omega^*_\V \right\}$, comprising $\bs v = (0, 0)^T$, $\bs v = (0, 1)^T$ and $\bs v = (1, 0)^T$. Since the third of these mirrors the second theoretically, we do not consider it further. Among the two remaining hypothesis tests, the false positive rate can be calculated for a single true model specified by $\tilde{\bs s}$:
\begin{eqnarray}
\begin{array}{rlrl}
\textrm{Case 1:}& 
 \V = (0,0)^T &\textrm{and}&\tilde{s}= (0,0)^T\\
\textrm{Case 2:}& 
 \V = (0,1)^T &\textrm{and}&\tilde{s}= (0,0)^T\\
\textrm{Case 3:}&
 \V = (0,1)^T &\textrm{and}& \tilde{s}= (0,1)^T
\end{array}
\end{eqnarray}
Since $\tilde{\beta}_1=0$ for each case,\begin{eqnarray}
\begin{pmatrix}W\\Z\end{pmatrix}&\overset{d}{\mathop{=}}&\mathcal{N}_2\left(\scalemath{0.7}{\begin{pmatrix}0\\{\sqrt{n}}\tilde{\beta}_2/\sigma\end{pmatrix}},I_2\right)\qquad\qquad\left(\textrm{by Eqn. \ref{zwdistrib}}\right). \label{zwdistrib_beta1_zero}
\end{eqnarray}

The model-averaged posterior odds is given by
\begin{eqnarray}
\po_{\mathcal{A}_\V:\mathcal{O}_\V}(W,Z)&=&\dfrac{\sum_{\bs s \in \mathcal{A}_\V} {\po}_{\bs s}(W,Z)}{\sum_{\bs s \in \mathcal{O}_\V} {\po}_{\bs s}(W,Z)}\qquad\qquad\left(\textrm{MT Eqn. \ref{MPO1}}\right)\label{bigpooseq}
\end{eqnarray}
where
\begin{eqnarray}
  \po_{\bs s}(W,Z)&=&{{\mu}^{|\bs s|}}\left(\tfrac{h}{n+h}\right)^{|\bs s|/{2}}\,(R_{\bs s}(W,Z))^{n/(n+h)}\qquad(\textrm{Eqn. \ref{POswithns}})\nonumber\\&=&
  (\mu \sqrt{\xi_n})^{|\bs s|}e^{(1-\xi_n){U_{\bs s}}^TU_{\bs s}/2}\qquad\left(\textrm{since}\;\, \xi_n=\tfrac{h}{n+h}\right)\label{tie139}.
\end{eqnarray}
and 
\begin{eqnarray}{U_{\bs s}}^TU_{\bs s}=&\begin{cases}(\sqrt{1- \rho^2}W+\rho Z)^2&s=(1,0)^T\\Z^2&s=(0,1)^T\\W^2+Z^2&s=(1,1)^T\end{cases}\label{UsTUs}
\end{eqnarray}

$\\$
$\\$
\underline{Case 1 or 2}: $\left(\V=(0,0)^T\textrm{  or  }\V=(0,1)^T,\textrm{  and  }\tilde{\bs s}=(0,0)^T \right)$

Since $\tilde{s}=(0,0)^T$, then $\tilde{\beta_2}=0$ and $(W,Z)^T=(W,\check{Z})^T \overset{d}{=}\mathcal{N}_2\begin{pmatrix}\mathbf{0}, I_2\end{pmatrix}$, i.e.\ $(W,Z)^T$ involves no free parameters. Let $\theta^{\dagger}=(0,0,0)^T$
\begin{eqnarray}
\underset{\tilde{\theta} \in \omega_{00}^*}{\sup}\Pr\left(\po_{\mathcal{A}_\V:\mathcal{O}_\V}(W,Z)\geq \tau;\;\tilde{\theta}\right)
&=&\Pr\left(\po_{\mathcal{A}_\V:\mathcal{O}_\V}(W,Z)\geq \tau;\;{\theta}^{\dagger}\right)
\end{eqnarray}
It follows that
\begin{eqnarray}
\underset{ \phantom{ \tilde\theta } n\rightarrow\infty \phantom{ \tilde\theta } }{\lim}\underset{\substack{\tilde{\theta} \in \omega_{00}^*}}{\sup}\, 
\frac{\Pr\left(\po_{\mathcal{A}_\V:\mathcal{O}_\V}\geq \tau;\;\tilde{\theta}\right)}{
\Pr\left( \chi^2_1
\geq 2\log \tfrac{\tau}{(\sizeV) \mu \sqrt{\xi_n}} \right)}
&=& \underset{ \phantom{ \tilde\theta } n\rightarrow\infty \phantom{\theta } }{\lim}\frac{\Pr\left(\po_{\mathcal{A}_\V:\mathcal{O}_\V}\geq \tau;\; {\theta}^{\dagger}\right)}{
\Pr\left( \chi^2_1
\geq 2\log \tfrac{\tau}{(\sizeV) \mu \sqrt{\xi_n}} \right)}\nonumber\\&=& 1 \label{case1result} \qquad\qquad \left({\textrm{by Eqn. \ref{justlimofalpha}}}\right)
\end{eqnarray}

\underline{Case 3}: $\left(\V=(0,1)^T \textrm{  and  }\tilde{\bs s}=(0,1)^T\right)$

\textit{Step 1. Derive an expression for} $\po_{\mathcal{A}_\V:\mathcal{O}_\V}(W,Z)$

Since $\tilde{\bs s}=(0,1)^T$, then $\tilde{\beta}_2\neq0$, (specifically $|\tilde{\beta}_2|\geq \beta_{\textrm{min}}$), and 
the distribution of $Z$ depends on
the unknown $\tilde{\beta}_2$.  From Equations \ref{bigpooseq}, \ref{tie139} and \ref{UsTUs} we have
\begin{eqnarray}
\po_{\mathcal{A}_\V:\mathcal{O}_\V}(W,Z)
&=&\; \frac{\PO_{11}+\PO_{10}}{\PO_{01}+\PO_{00}}\\
&=&\; \frac{(\mu \sqrt{\xi_n})^2\,e^{(1-\xi_n)(W^2+Z^2)/2}+{\mu \sqrt{\xi_n}}{e^{(1-\xi_n)(\sqrt{1-\rho^2}W+\rho Z)^2/2}}}{{\mu \sqrt{\xi_n}}e^{(1-\xi_n)Z^2/2}+1}\\
&=&\mu \sqrt{\xi_n}\,e^{(1-\xi_n) W^2/2}\; \left(\frac{1+\tfrac{1}{\mu \sqrt{\xi_n}}{e^{-(1-\xi_n)(\rho W - Z\sqrt{1-\rho^2})^2/2}}}{1+ \tfrac{1}{\mu \sqrt{\xi_n}}e^{-(1-\xi_n)Z^2/2}}\right)
\end{eqnarray}

\textit{Step 2.  We define a simpler random variable, $\po^{+}(W,Z)$, that stochastically dominates $\po_{\mathcal{A}_\V:\mathcal{O}_\V}(W,Z)$, meaning } 
\begin{eqnarray}
\Pr\left(\po_{\mathcal{A}_\V:\mathcal{O}_\V}(W,Z)\geq \tau;\; \tilde{\theta}\right)
\leq\Pr\left(\po^+\left(W,Z\right)\geq\tau;\; \tilde{\theta}\right), \label{proof_th4_stochastic_bound}
\end{eqnarray}
where
\begin{eqnarray}
\po^{+}\left(W,{Z}\right)=\mu \sqrt{\xi_n}e^{(1-\xi_n)W^2/2}\; \iota_n(Z),
\end{eqnarray}
\begin{eqnarray}
\iota_n(Z)&=& \begin{cases}
1+\tfrac{1}{\mu \sqrt{\xi_n}} & Z^2\leq z_n^2\\ 1+\tfrac{1}{\mu \sqrt{\xi_n}}e^{{-\frac{(1-\xi_n)}{2}(|\rho| w_n - |Z|\sqrt{1-\rho^2})}^2}& Z^2>z_n^2
\end {cases}, \label{justinn}
\end{eqnarray}
\begin{eqnarray}
    {w_n}&=& \sqrt[+]{\tfrac{2}{1-\xi_n}\log \tfrac{\tau}{\mu \sqrt{\xi_n}}} , \label{proof_th4_inflation_factor}
\end{eqnarray}
and
\begin{eqnarray}
    z_n&=&\frac{|\rho| w_n}{\sqrt{1-\rho^2}} .
\end{eqnarray}
Note that $\iota_n(Z)$ acts as an `inflation factor', in the sense that (a) it always equals at least one and (b) it acts to multiply $\po^{+}(W,Z)$ above its asymptotic identity with $\po_{\tilde{\bs s}}(W,Z)$, $\tilde{\bs s}=(0,1)^T$.

To demonstrate stochastic dominance, we consider three different conditions on $(W,Z)$ in turn.

\textbf{Condition 1}: $W^2\geq {w_n}^2$. 
In this case, $\po^{+}\left(W,Z\right)$ already exceeds $\tau$, and therefore its tail probability must dominate the tail probability of $\po_{\mathcal{A}_\V:\mathcal{O}_\V}\left(W,Z\right)$:
\begin{align}
\dfrac{\po^{+}\left(W,Z\right)}{\iota_n(Z)} &= \mu \sqrt{\xi_n}e^{(1-\xi_n)W^2/2} \nonumber \\
&\geq \tau &\quad& \forall\ W^2\geq w_n^2  . \nonumber
\intertext{Since $\iota_n(Z)\geq 1$ this implies}
\po^{+}\left(W,Z\right) &\geq \tau &\quad& \forall\ W^2\geq w_n^2 \nonumber
\end{align}
and therefore
\begin{equation}
\Pr\left(\po_{\mathcal{A}_\V:\mathcal{O}_\V}\left(W,Z\right)\geq\tau \middle| W^2\geq w_n^2;\;\tilde{\theta}\right) \leq \Pr\left(\po^{+}\left(W,Z\right)\geq \tau \middle| W^2\geq w_n^2;\;\tilde{\theta}\right) = 1  .
\end{equation}

\textbf{Condition 2:} $Z^2\leq z_n^2$. In this case, $\po^{+}\left(W,Z\right)$ bounds $\po_{\mathcal{A}_\V:\mathcal{O}_\V}\left(W,Z\right)$, so its tail probability must dominate :
\begin{align}
\dfrac{\po_{\mathcal{A}_\V:\mathcal{O}_\V}\left(W,Z\right)}{\po_{\tilde{\bs s}}\left(W,Z\right)} &\leq \dfrac{\po^+\left(W,Z\right)}{\po_{\tilde{\bs s}}\left(W,Z\right)} &\quad& \forall \ Z^2\leq z_n^2 \nonumber
\intertext{because}
\frac{1+\tfrac{1}{\mu \sqrt{\xi_n}}{e^{-(1-\xi_n)(\rho W - Z\sqrt{1-\rho^2})^2/2}}}{1+ \tfrac{1}{\mu \sqrt{\xi_n}}e^{-(1-\xi_n)Z^2/2}} &\leq 1 +\tfrac{1}{\mu\sqrt{\xi_n}} &\qquad& \forall \ Z^2\leq z_n^2 \nonumber
\end{align}
and therefore
\begin{equation}
\Pr\left(\po_{\mathcal{A}_\V:\mathcal{O}_\V}\left(W,Z\right)\geq\tau\middle|Z^2\leq z_n^2;\;\tilde{\theta}\right) \leq \Pr\left(\po^{+}\left(W,Z\right)\geq \tau \middle| Z^2\leq z_n^2 ;\;\tilde{\theta}\right).
\end{equation}

\textbf{Condition 3}: $W^2<{w_n}^2$ and $Z^2> z_n^2$. In this region $\po^{+}\left(W,Z\right)$ again bounds $\po_{\mathcal{A}_\V:\mathcal{O}_\V}\left(W,Z\right)$, so its tail probability must dominate :
\begin{eqnarray}
s_Z\,\rho\, W\leq|\rho||W|&<&|\rho|\, w_n<|Z|\sqrt{1-\rho^2} \qquad\qquad (\textrm{let  }s_Z:=\textrm{sgn}(Z))\nonumber\\ 
\implies \qquad
\left(s_Z\rho\, W - |Z|\sqrt{1-\rho^2}\right)^2&\geq&\left(|\rho|\, w_n - |Z|\sqrt{1-\rho^2}\right)^2\nonumber\\ 
\implies \qquad\left(\rho\, W - s_Z|Z|\sqrt{1-\rho^2}\right)^2&\geq&\left(|\rho|\, w_n - |Z|\sqrt{1-\rho^2}\right)^2\nonumber\\
\implies \qquad
e^{-(1-\xi){(\rho\, W - Z\sqrt{1-\rho^2})^2}/2}&\leq&e^{-(1-\xi){(|\rho|\, w_n - |Z|\sqrt{1-\rho^2})^2}/2}\nonumber\\
\implies\qquad \po_{\mathcal{A}_\V:\mathcal{O}_\V}\left(W,Z\right)&\leq&\po^{+}\left(W,Z\right) \nonumber
\end{eqnarray}
and therefore
\begin{align}
\Pr\left(\po_{\mathcal{A}_\V:\mathcal{O}_\V}\left(W,Z\right)\geq\tau\middle| W^2<{w_n}^2,\, Z^2>z_n^2;\;\tilde{\theta}\right) \leq \qquad\qquad\qquad\qquad\qquad\qquad \notag \\ 
\qquad\qquad\qquad\qquad\qquad\qquad \Pr\left(\po^{+}\left(W,Z\right)\geq \tau \middle| W^2<{w_n}^2,\, Z^2>z_n^2 ;\;\tilde{\theta}\right).
\end{align}

The three conditions cover all eventualities and therefore imply stochastic dominance:
\begin{eqnarray}
\Pr\left(\po_{\mathcal{A}_\V:\mathcal{O}_\V}\left(W,Z\right)\geq \tau;\;\tilde{\theta}\right)&\leq&\Pr\left(\po^+\left(W,Z\right)\geq\tau;\; \tilde{\theta}\right) . \label{proof_th4_step2_result}
\end{eqnarray}

\textit{Step 3. Finding a simplifying expression for the supremum over} $\tilde{\theta} \in \omega_{01}^*$ \textit{of} ${\Pr\left(\po_{\mathcal{A}_\V:\mathcal{O}_\V}(W,Z)\geq\tau;\tilde{\theta}\right)}$.

Let $f_{\mathcal{N}}$ be the pdf of a standard normal random variable, $a=\tfrac{\sqrt{n}}{\sigma}\tilde{\beta}_2$, $a_{\mathrm{min}}=\tfrac{\sqrt{n}}{\sigma}{\beta}_{\mathrm{min}}$ and $A_n=\left\{a \in \mathbb{R}: |a|\geq a_{\mathrm{min}}\right\}$. Then Equations \ref{zwdistrib_beta1_zero} and \ref{proof_th4_step2_result} imply that
\begin{eqnarray}
\underset{\substack{\tilde{\theta} \in \omega_{01}^*}}{\sup}\;{\Pr\left(\po_{\mathcal{A}_\V:\mathcal{O}_\V}(W,Z)\geq\tau;\tilde{\theta}\right)}& \leq&\underset{\substack{a \in A_n}}{\sup}\;
\int\limits_{-\infty}^{\infty}\;\Pr\left(\po^{+}\left(W,z\right)\geq\tau; a\right)f_{\mathcal{N}}(z-a)dz\nonumber\\
&=&\underset{\substack{a \in A_n}}{\sup}\;
\int\limits_{-\infty}^{\infty}\;\Pr\left(W^2\geq  \tfrac{2}{1-\xi_n} \log \left(\frac{\tau}{\mu \sqrt{\xi_n}\,{\iota}_n(z)}\right)\right)f_{\mathcal{N}}(z-a)d{z}\nonumber \\
&=&\underset{\substack{a \in A_n}}{\sup}\;
\int\limits_{-\infty}^{\infty}\;k_n(z)f_{\mathcal{N}}(z-a)d{z}\nonumber\\&&\qquad\qquad\left(\textrm{Eqn. \ref{zwdistrib_beta1_zero}},\;k_n(z):=\Pr\left(\chi_1^2\geq  \tfrac{2}{1-\xi_n} \log \left(\frac{\tau}{\mu \sqrt{\xi_n}\,{\iota}_n(z)}\right)\right)\right)\nonumber\\
&=&\underset{\substack{a \in A_n}}{\sup}\;
\int\limits_{-\infty}^{\infty}\;k_n(z+a)f_{\mathcal{N}}(z)d{z}\qquad(\textrm{employing } z \mapsto z+a)\nonumber\\
&=&\underset{\substack{a \geq a_{\mathrm{min}}}}{\sup}\;
\int\limits_{0}^{\infty}\;\left(k_n(z-a)+k_n(z+a)\right)f_{\mathcal{N}}(z)d{z}\nonumber\\&&
\qquad\qquad\qquad\left(\textrm{since }f_{\mathcal{N}}(z)=f_{\mathcal{N}}(-z) \textrm{ and } k_n(z)=k_n(-z)\right)\nonumber\\
&=&\underset{\substack{a \geq a_{\mathrm{min}}}}{\sup}\;
\int\limits_{-\infty}^{\infty}\;k_n(z+a)f_{\mathcal{N}}(z)d{z} . \label{supoverpositive}
\end{eqnarray}
\textit{Step 4. Using the Expectation-Maximization (EM) algorithm to further simplify the expression for the supremum over} $\tilde{\theta} \in \omega_{01}^*$ \textit{of} ${\Pr\left(\po_{\mathcal{A}_\V:\mathcal{O}_\V}(W,Z)\geq\tau;\tilde{\theta}\right)}$.

The Expectation-Maximization algorithm \citep{dempster1977maximum} allows the marginal probability of an event $\bbar{Y}$, when it can be expanded as
\begin{equation*}
 \Pr(\bbar{Y}|a)=\int\limits_{-\infty}^{\infty}p(\bbar{Y},z\,|\,a)dz=\int\limits_{-\infty}^{\infty}
\Pr(\bbar{Y}\,|\,z,a)\, p(z\,|a)dz,
\end{equation*}
to be maximized with respect to $a$ via the recursion
\begin{eqnarray}
a_{m+1}&=&\underset{\substack{a \geq a_{\mathrm{min}}}}{\mathrm{argmax}}\;
\mathbb{E}_{\bbar{Y},a_m}\left[\log\,p(\bbar{Y},z\,|a)\, \right]\nonumber\\
&=&\underset{\substack{a \geq a_{\mathrm{min}}}}{\mathrm{argmax}}\int\limits_{-\infty}^{\infty}\;
\log\,p(\bbar{Y},z\,|a)\, p(z\,|\,\bbar{Y} ,a_m)dz\nonumber\\
&=&\underset{\substack{a\geq a_{\mathrm{min}}}}{\mathrm{argmax}}\int\limits_{-\infty}^{\infty}\;
\left(\log\,\Pr(\bbar{Y}\,|\,z,a)+\log\,p(z\,|\,a)\right)\,p(z\,|\,\bbar{Y} ,a_m) dz .\nonumber\\
\end{eqnarray}
Let $\bbar{Y}$ be the event that $\po_{\mathcal{A}_\V:\mathcal{O}_\V}(W,Z)\geq\tau$, then $\Pr(\bbar{Y}|z,a)=\Pr(\bbar{Y}|z)=k_n(z)$. Furthermore, $p(z\,|\,a)=f_{\mathcal{N}}({z}-{a})$  giving
\begin{eqnarray}a_{m+1}&=&\underset{\substack{a \geq a_{\mathrm{min}}}}{\mathrm{argmax}}\;
\int\limits_{-\infty}^{\infty}\;\left(\log k_n(z)+\log f_{\mathcal{N}}(z-a)\right)p({z}\,|\,\bbar{Y} ,a_m)dz\nonumber\\
&=&\underset{\substack{a \geq a_{\mathrm{min}}}}{\mathrm{argmax}}\;
\int\limits_{-\infty}^{\infty}\;-\tfrac{1}{2}(z-a)^2\,p({z}\,|\,\bbar{Y} ,{a_m})d{z}\nonumber\\
&=&\underset{\substack{a \geq a_{\mathrm{min}}}}{\mathrm{argmin}}\;\mathbb{E}_{\bbar{Y},a_m}\left[(Z-a)^2\right] .
\end{eqnarray}
Notice that \begin{eqnarray}
   \frac{\partial}{\partial\, a}\mathbb{E}_{\bbar{Y},a_m}\left[(Z-a)^2\right]&=&2a-2\mathbb{E}_{\bbar{Y},a_m}[Z] 
\end{eqnarray}
and
\begin{eqnarray}
   \frac{\partial^2}{\partial^2\, a}\mathbb{E}_{\bbar{Y},a_m}\left[(Z-a)^2\right]=2 .
\end{eqnarray}

Since the second derivative is nonnegative on its entire domain, it is therefore convex, implying that the minimum is unique.
This implies
\begin{eqnarray}
a_{m+1}=\begin{cases}
    \mathbb{E}_{\bbar{Y},a_m}[Z] &\textrm{if}\;\,\mathbb{E}_{\bbar{Y},a_m}[Z]\geq a_{\textrm{min}}\\a_{\textrm{min}}& \textrm{if}\;\,\mathbb{E}_{\bbar{Y},a_m}[Z]< a_{\textrm{min}}
\end{cases}
\label{amplu1itit}
\end{eqnarray}
Furthermore,
\begin{eqnarray}
\mathbb{E}_{\bbar{Y},a_m}[Z]&=&\int\limits_{-\infty}^{\infty}z\,p({z}\,|\,\bbar{Y} ,{a_m})d{z}\nonumber\\
&=& \int\limits_{-\infty}^{\infty}z\;\frac{\Pr(\bbar{Y}\,|\,{z},{a_m})p({z}\,|\,{a_m})}{\Pr(\bbar{Y}|\,a_m)}d{z}\nonumber\\
&=& \frac{1}{\Pr(\bbar{Y}|\,a_m)}\int\limits_{-\infty}^{\infty}z\,k_n(z)f_{\mathcal{N}}(z-a_m)d{z}\nonumber\\
&=& \frac{1}{\Pr(\bbar{Y}|\,a_m)}\int\limits_{-\infty}^{\infty}(z+a_m)\,k_n(z+a_m)f_{\mathcal{N}}(z)d{z}\nonumber\\
&=&a_m+\frac{1}{\Pr(\bbar{Y}|\,a_m)}\int\limits_{-\infty}^{\infty}z\,k_n(z+a_m)f_{\mathcal{N}}(z)d{z}\nonumber\\
&=&a_m-{\Psi_m}\qquad\left(\textrm{let}\;\Psi_m:=-\frac{1}{\Pr(\bbar{Y}|\,a_m)}\int\limits_{-\infty}^{\infty}z\,k_n(z+a_m)f_{\mathcal{N}}(z)d{z}\right)
\end{eqnarray}
allowing Equation \ref{amplu1itit} to be rewritten as
\begin{eqnarray}
a_{m+1}=\begin{cases}
    a_m- \Psi_m &\textrm{if}\;\,a_m- \Psi_m\geq a_{\textrm{min}}\\a_{\textrm{min}}& \textrm{if}\;\,a_m-\Psi_m< a_{\textrm{min}}
\end{cases}
\end{eqnarray}

Let $a_m=0$.
\begin{eqnarray}
    \int\limits_{-\infty}^{\infty}z\,k_n(z+a_m)f_{\mathcal{N}}(z)d{z}&=&\int\limits_{0}^{\infty}zf_{\mathcal{N}}(z)\left(k_n(z+a_m)-k_n(z-a_m)\right)dz=0\\
    \implies\qquad \Psi_m&=&0\nonumber\\ \implies
a_{m+1}&=&\begin{cases}
    0 &\textrm{if}\;\,0= a_{\textrm{min}}\\a_{\textrm{min}}& \textrm{if}\;\,0< a_{\textrm{min}}\nonumber\\
\end{cases}\\
&=&a_{\textrm{min}}
\end{eqnarray}
In general (i.e., for $a_m\geq0$), therefore
\begin{eqnarray}
a_{m+1}=\begin{cases}
    a_m- \Psi_m &\textrm{if}\;\,0<a_m\geq a_{\textrm{min}}+\Psi_m\\a_{\textrm{min}}& \textrm{if}\;\,0<a_m< a_{\textrm{min}}+\Psi_m\\a_{\textrm{min}}& \textrm{if}\;\,a_m=0
\end{cases}\label{amplus1cases}
\end{eqnarray}
Let $a_m>0$.   
\begin{eqnarray}
\int\limits_{-\infty}^{\infty}z\,k_n(z+a_m)f_{\mathcal{N}}(z)d{z}&=& \int\limits_{-\infty}^{0}z\,k_n(z+a_m)f_{\mathcal{N}}(z)d{z}+\int\limits_{0}^{\infty}z\,k_n(z+a_m)f_{\mathcal{N}}(z)d{z}\nonumber\\ &=& -\int\limits_{0}^{\infty}z\,k_n(-z+a_m)f_{\mathcal{N}}(z)d{z}+\int\limits_{0}^{\infty}z\,k_n(z+a_m)f_{\mathcal{N}}(z)d{z}\qquad\left(f_{\mathcal{N}}(x)=f_{\mathcal{N}}(-x)\right)\nonumber\\ &=& \int\limits_{0}^{\infty}zf_{\mathcal{N}}(z)\left(k_n(z+a_m)-k_n(z-a_m)\right)dz\qquad\left(k_n(x)=k_n(-x)\right)
\end{eqnarray}
Notice that $zf_{\mathcal{N}}(z)\geq0\ \forall\ z\geq0$ and   $\,zf_{\mathcal{N}}(z)>0\ \forall\ z>0$. Furthermore, $k_n(z+a_m)\leq k_n(z-a_m)\ \forall\ z \geq0$ and $k_n(z+a_m)< k_n(z-a_m)\ \forall\ z> a_m+z_n$. Therefore 
\begin{eqnarray}
\int\limits_{-\infty}^{\infty}z\,k_n(z+a_m)f_{\mathcal{N}}(z)d{z}&<&0\nonumber\\
\implies\qquad \Psi_m&>&0
\end{eqnarray}
In general (i.e., for $a_m\geq0$), the first case in Equation \ref{amplus1cases} therefore cannot be satisfied at the stationary point, since $a \neq a-\Psi$ for $0<a$. Since the EM algorithm is guaranteed to reach a stationary point, $a=a_{\textrm{min}}$  is the only remaining solution. In summary, $a=a_{\textrm{min}}$ (${\beta}_2=\beta_{\textrm{min}}$) allowing Equation \ref{supoverpositive} to be rewritten as
\begin{eqnarray}
\underset{\substack{\tilde{\theta} \in \omega_{01}^*}}{\sup}\;{\Pr\left(\po_{\mathcal{A}_\V:\mathcal{O}_\V}(W,Z)\geq\tau;\tilde{\theta}\right)}&\leq&
\int\limits_{-\infty}^{\infty}\;k_n(z+a_{\mathrm{min}})f_{\mathcal{N}}(z)d{z}\nonumber\\&=&
\int\limits_{-\infty}^{\infty}{\Pr\left(\chi_1^2\geq  \tfrac{2}{1-\xi_n} \log \frac{\tau}{\mu \sqrt{\xi_n}\,\iota_n(z+a_{\mathrm{min}})}\right)}f_{\mathcal{N}}(z)dz\nonumber\\&=&\int\limits_{-\infty}^{\infty}{\Pr\left(\chi_1^2\geq {2} \log \left(\frac{\tau}{\mu \sqrt{\xi_n}\,\iota_n(z+a_{\mathrm{min}})}\right)^{\tfrac{1}{1-\xi_n}}\right)}f_{\mathcal{N}}(z)dz\nonumber\\&=&\int\limits_{-\infty}^{\infty} H\left(\left(\frac{\tau}{\mu \sqrt{\xi_n}\,\iota_n(z+a_{\mathrm{min}})}\right)^{\tfrac{1}{1-\xi_n}}\right)f_{\mathcal{N}}(z)dz
\end{eqnarray}

where $H(x)=\Pr(\chi_1^2\geq {2}\log x)=\Pr\left(\textrm{LG}(\tfrac{1}{2},1)>x\right)$. 

\textit{Step 5. Setting up an expression for the asymptotic limit.}

\begin{eqnarray}
\underset{ \phantom{ \tilde\theta } n\rightarrow\infty \phantom{ \tilde\theta } }{\lim}\; 
\frac{ \underset{\substack{\theta \in {\omega}_{01}^*}}{\sup}\;\Pr\left(\po_{\mathcal{A}_\V:\mathcal{O}_\V}\geq \tau\; ;\; \theta\right)}{
\Pr\left( \chi^2_1
\geq \tfrac{2}{1-\xi_n}\log \frac{\tau}{\mu \sqrt{\xi_n}} \right)}&\leq&\underset{ \phantom{ \tilde\theta } n\rightarrow\infty \phantom{ \tilde\theta } }{\lim}\int\limits_{-\infty}^{\infty}{\frac{H\left(\left({\tau}/{\mu \sqrt{\xi_n}\,\iota_n(z+a_{\mathrm{min}})}\right)^{\tfrac{1}{1-\xi_n}}\right)}{H \left(\left({\tau}/{\mu \sqrt{\xi_n}} \right)^{\tfrac{1}{1-\xi_n}}\right)}f_{\mathcal{N}}(z)dz}\nonumber\\
&=&\underset{ \phantom{ \tilde\theta } n\rightarrow\infty \phantom{ \tilde\theta } }{\lim}\int\limits_{-\infty}^{\infty}f_n(z)dz\label{wherewegoto}\\
\
&& \left(\textrm{let}\;
f_n(z):=\frac{H\left(\left({\tau}\big/{\mu \sqrt{\xi_n}\,\iota_n(z+a_{\mathrm{min}})}\right)^{\tfrac{1}{1-\xi_n}}\right)}{H \left(\left({\tau}\big/{\mu \sqrt{\xi_n}} \right)^{\tfrac{1}{1-\xi_n}}\right)}f_{\mathcal{N}}(z)\right)\label{theleadup}\nonumber
\end{eqnarray}

\textit{Step 6. Use of Potter's bounds to find an expression that bounds the function} $f_n(z)$

Theory on Potter's Bounds (see for example, Proposition 1.4.2 of \cite{kulik2020heavy}) implies that since $\textrm{LG}(\tfrac{1}{2},1)$ is a non-negative regularly random variable with tail index $1$, then for each $\varepsilon>0$ there exists a finite positive constant $A_{\varepsilon}$, such that for all $x\geq1$ and $y\geq1$
\begin{eqnarray}
\frac{{H}(x/y)}{{H}(x)}
&\leq& A_{\varepsilon}\,y^{1+\varepsilon}\label{keveither}
\end{eqnarray}
Since $\left({\tau}/{\mu \sqrt{\xi_n}}\right)^{\tfrac{1}{1-\xi_n}}\geq1$ and ${\iota_n(z+a_{\mathrm{min}})}^{{\tfrac{1}{1-\xi_n}}}\geq1$, it follows that
\begin{eqnarray}
f_n(z)&=&\frac{H\left(\left({\tau}/{\mu \sqrt{\xi_n}\,\iota_n(z+a_{\mathrm{min}})}\right)^{\tfrac{1}{1-\xi_n}}\right)}{H \left(\left({\tau}/{\mu \sqrt{\xi_n}} \right)^{\tfrac{1}{1-\xi_n}}\right)}f_{\mathcal{N}}(z)\nonumber\\
&\leq& A_{\varepsilon} \;\iota_n(z+a_{\mathrm{min}})^{\tfrac{(1+\varepsilon)}{1-\xi_n}}\,f_{\mathcal{N}}(z)\nonumber\\ &=& A_{\varepsilon} \;\iota_n(z+a_{\mathrm{min}})^{\tfrac{(1+\varepsilon)}{1-\xi_n}}\,\tfrac{1}{\sqrt{2 \pi \sigma^2}} \,e^{{-z^2}/{2\sigma^2}}\nonumber\\\implies
\log f_n(z)&\leq& \log B +\tfrac{(1+\varepsilon)}{(1-\xi_n)} \log\iota_n(z+a_{\mathrm{min}})-{z^2}/{2\sigma^2}\label{logmodffff}
\end{eqnarray}
where $B:={A_{\varepsilon}}/{\sqrt{2\pi\sigma^2}}$ and by Equation \ref{justinn}
\begin{eqnarray}
\iota_n(z+a_{\textrm{min}})&=& \begin{cases}
1+\tfrac{1}{\mu \sqrt{\xi_n}} & |z+a_{\textrm{min}}|\leq z_n\\ 1+\tfrac{1}{\mu \sqrt{\xi_n}}\textrm{exp}\{{-\frac{(1-\xi_n)(1-\rho^2)}{2}(z_n - |z+a_{\textrm{min}}|)}^2\}& |z+a_{\textrm{min}}|>z_n
\end {cases}
\end{eqnarray}

\textit{Step 7. Show that the function} $f_n(z)$ \textit{is bounded by a scaled normal pdf}

In what follows, we consider how the term $\tfrac{(1+\varepsilon)}{(1-\xi_n)} \log\iota_n(z+a_{\mathrm{min}})-{z^2}/{2\sigma^2}$ is bounded across different regions of $z$-space. This involves noting how the following elements scale with $n$ 
\begin{eqnarray}a_{\textrm{min}}&=&\tfrac{\sqrt{n}\beta_{\mathrm{min}}}{\sigma}=O(\sqrt{n})\label{order1}\\z_n&=& \sqrt{{\tfrac{2 \rho^2}{(1-\xi_n)(1-\rho^2)}\log \tfrac{\tau}{\mu \sqrt{\xi_n}}}}=O(\sqrt{\log n})\label{order2}\\\tfrac{(1+\varepsilon)}{(1-\xi_n)}{\log\left(\tfrac{2}{\mu\sqrt{\xi_n}}\right)}&=&O(\log{n}) \label{order3}\\
a_\textrm{min}-z_n&=&O(\sqrt{n}) 
\end{eqnarray}

\textbf{Region i}:  $z<-a_{\textrm{min}}+z_n$
\begin{eqnarray}
\iota_n(z+a_{\textrm{min}})&\leq& 
    1+\tfrac{1}{\mu \sqrt{\xi_n}}\nonumber \\&<&\tfrac{2}{\mu \sqrt{\xi_n}}\qquad \left(\forall n>n_C;\; \textrm{for some } n_C\right)\nonumber\\ \implies
\tfrac{1}{z^2}\tfrac{(1+\varepsilon)}{(1-\xi_n)} \log\iota_n(z+a_{\mathrm{min}})
&<&\tfrac{1}{z^2}\tfrac{(1+\varepsilon)}{(1-\xi_n)}{\log\left(\tfrac{2}{\mu\sqrt{\xi_n}}\right)}\nonumber
\\&<& \tfrac{1}{(z_n-a_\textrm{min})^2}\tfrac{(1+\varepsilon)}{(1-\xi_n)}{\log\left(\tfrac{2}{\mu\sqrt{\xi_n}}\right)}\quad\left(=o(1)\right)\quad(\textrm{Eqns. \ref{order1}-\ref{order3}} )\nonumber\\&<&\delta\qquad(\forall\delta>0,n>n_{\delta} \textrm{ for some } n_{\delta}\geq n_C) \nonumber \\
\implies
\tfrac{(1+\varepsilon)}{(1-\xi_n)} \log\iota_n(z+a_{\mathrm{min}})-{z^2}/{2\sigma^2}
&<&-(\tfrac{1}{2\sigma^2}-\delta){z^2}\quad(\textrm{choose } \delta<\tfrac{1}{2\sigma^2})\nonumber\\&=& -Dz^2\qquad \left(\forall n>n_{\delta}; D:=\tfrac{1}{2\sigma^2}-\delta>0   \right)\nonumber\\ \label{refreg1}
\end{eqnarray}

\textbf{Region ii}:  $|z+a_{\textrm{min}}|> z_n$ and $\tfrac{1}{\mu \sqrt{\xi_n}}\textrm{exp}\left\{-\tfrac{(1-\xi_n)(1-\rho^2)}{2}\left(z_n - |{z}+a_{\textrm{min}}|\right)^2\right\} \leq 1$

\begin{eqnarray}
\iota_n(z+a_{\textrm{min}})&=& 
    1+\tfrac{1}{\mu \sqrt{\xi_n}}\textrm{exp}\left\{-\tfrac{(1-\xi_n)(1-\rho^2)}{2}\left(z_n - |{z}+a_{\textrm{min}}|\right)^2\right\}\nonumber\\&\leq&2\nonumber\\
    \implies \tfrac{(1+\varepsilon)}{(1-\xi_n)} \log\iota_n(z+a_{\mathrm{min}})
&\leq&\tfrac{(1+\varepsilon)}{(1-\xi_n)}{\log 2}\nonumber\\
&\leq&\log E\quad (\forall n>n_E; \textrm{ for some } E>0,\,  n_E \geq 1)\nonumber\\\implies \tfrac{(1+\varepsilon)}{(1-\xi_n)} \log\iota_n(z+a_{\mathrm{min}})-\tfrac{z^2}{2\sigma^2}
&\leq&\log E-\tfrac{z^2}{2\sigma^2}\quad (\forall n>n_E)\label{refreg2}
\end{eqnarray}

\textbf{Region iii}:  $z>0$ and $\tfrac{1}{\mu \sqrt{\xi_n}}\textrm{exp}\left\{-\tfrac{(1-\xi_n)(1-\rho^2)}{2}\left(z_n - |{z}+a_{\textrm{min}}|\right)^2\right\} > 1$

\begin{eqnarray}
 \iota_n(z+a_{\textrm{min}})&=& 
    1+\tfrac{1}{\mu \sqrt{\xi_n}}\textrm{exp}\left\{-\tfrac{(1-\xi_n)(1-\rho^2)}{2}\left(z_n - |{z}+a_{\textrm{min}}|\right)^2\right\}\nonumber\\&<& 
    \tfrac{2}{\mu \sqrt{\xi_n}}\textrm{exp}\left\{-\tfrac{(1-\xi_n)(1-\rho^2)}{2}\left(z_n - |{z}+a_{\textrm{min}}|\right)^2\right\}\nonumber\\
    &<& 
    \tfrac{2}{\mu \sqrt{\xi_n}}\textrm{exp}\left\{-\tfrac{(1-\xi_n)(1-\rho^2)}{2}\left(a_{\textrm{min}}-z_n\right)^2\right\}\qquad \quad(\forall n>n_F;\;\textrm{ for some } n_F)
\end{eqnarray}
The above follows because $z>0$ and $a_{\textrm{min}}-z_n>0\; \forall\; n>n_F$. Therefore
    \begin{eqnarray}
    \tfrac{(1+\varepsilon)}{(1-\xi_n)} \log\iota_n(z+a_{\mathrm{min}})&<&{(1+\varepsilon)}\left\{\tfrac{1}{(1-\xi_n)}\log\left(\tfrac{2}{\mu\sqrt{\xi_n}}\right)-\tfrac{(1-\rho^2)}{2}(a_{\textrm{min}}-z_n)^2\right\}\nonumber\\&\leq&0\quad(\forall n>n_G;\;\textrm{ for some } n_G\geq n_F \textrm{: see Eqns \ref{order1} and \ref{order2}})\nonumber\\ \tfrac{(1+\varepsilon)}{(1-\xi_n)} \log\iota_n(z+a_{\mathrm{min}})-\tfrac{z^2}{2\sigma^2}&\leq&-\tfrac{z^2}{2\sigma^2}\quad(\forall n>n_G)\label{refreg3}
\end{eqnarray}

\textbf{Region iv}:  $0>z>-a_{\textrm{min}}+z_n$ and $\tfrac{1}{\mu \sqrt{\xi_n}}\textrm{exp}\left\{-\tfrac{(1-\xi_n)(1-\rho^2)}{2}\left(z_n - |{z}+a_{\textrm{min}}|\right)^2\right\} > 1$

Let 
\begin{eqnarray}\varrho=\textrm{max}\left(|z_n-z-a_{\textrm{min}}|,|z| \right)\geq \tfrac{1}{2}|z_n-a_{\textrm{min}}|=O(\sqrt{n}) \label{thisisvarrho}
\end{eqnarray}
Then
\begin{eqnarray}
&&\tfrac{1}{\varrho^2}\left(\tfrac{(1+\varepsilon)}{(1-\xi_n)} \log\iota_n(z+a_{\mathrm{min}})-\tfrac{z^2}{2\sigma^2}\right)\\
&&\qquad\qquad\qquad\leq{(1+\varepsilon)}\left\{\tfrac{1}{\varrho^2(1-\xi_n)} \log\left(\tfrac{2}{\mu\sqrt{\xi_n}}\right)-\tfrac{(1-\rho^2)}{2\varrho^2}\left(z_n -{z}-a_{\textrm{min}}\right)^2\right\}-\tfrac{z^2}{2\varrho^2\sigma^2}\nonumber\\&&\qquad\qquad\qquad\leq\eta-H\qquad \left(H:=\textrm{min}\left(\tfrac{(1+\epsilon)(1-\rho^2)}{2},\,\tfrac{1}{2\sigma^2}\right)>0\right) \qquad (\forall \eta>0 \textrm{ for some } n>n_{\eta})\nonumber
\end{eqnarray}
The above line follows because $\tfrac{1+\varepsilon}{\varrho^2(1-\xi_n)} \log\left(\tfrac{2}{\mu\sqrt{\xi_n}}\right)=o(1)$ (Eqns. \ref{order3} and  \ref{thisisvarrho}) and because either $|z_n-z-a_{\textrm{min}}|/\varrho=1$ or $|z|/\varrho=1$ (Eqn. \ref{thisisvarrho}). Choosing $\eta<H$, it therefore follows that
\begin{eqnarray}\tfrac{(1+\varepsilon)}{(1-\xi_n)} \log\iota_n(z+a_{\mathrm{min}})-\tfrac{z^2}{2\sigma^2}
&\leq&-K\varrho^2\qquad\left(K:=\eta-H>0\right)\nonumber\\&\leq& -Kz^2 \qquad (\forall n>n_{\eta}; \textrm{ since } \varrho^2 \geq z^2)\label{refreg4}
\end{eqnarray}

Equations \ref{logmodffff}, and \ref{refreg1}-\ref{refreg4} therefore imply that there exists an $L>0$, $M>0$, and $n_0 \geq1$ such that
\begin{eqnarray}
\log f_n(z)&\leq& \log L-{Mz^2}\qquad(\forall n>n_0)\nonumber\\ \implies\qquad
f_n(z)&\leq&L\,\textrm{exp}\left(-{Mz^2}\right)\qquad (\forall n>n_0)\label{newgeez}
\end{eqnarray}

\textit{Step 8 Application of the Lebesgue dominated convergence theorem to demonstrate `uniform' (lim sup) convergence.} 

The Lebesgue dominated convergence theorem \citep{van1968weak} states that if $h_n:\mathbb{R} \mapsto [-\infty,\infty]$ are (Lebesgue) measurable functions such that the pointwise limit $h(z): = \lim_{n\to\infty} h_n(z)$ exists and there is an integrable $g:\mathbb{R} \mapsto [0,\infty]$ with $|h_n(z)|\leq g(z)$ for all $n$ and all $z \in \mathbb{R}$, then 
\begin{eqnarray}
\underset{ \phantom{ \tilde\theta } n\rightarrow\infty \phantom{ \tilde\theta } }{\lim}\int\limits_{-\infty}^{\infty}h_n(z)\;dz=\int\limits_{-\infty}^{\infty}h(z)\;dz
\end{eqnarray}

Let $h_n(z):=f_{n+n_0}(z)$, then Equation \ref{newgeez} gives
\begin{eqnarray}
\left|h_n(z)\right|&\leq&L\,\textrm{exp}\left(-{Mz^2}\right):=g(z)\qquad (\forall n)\label{newgeez2}
\end{eqnarray}

Notice first that since $g(z)$ (defined in Eqn. \ref{newgeez2}) is proportional to a Normal pdf, it is integrable; and second that 
\begin{eqnarray}
    h(z): = \lim_{n\to\infty} h_n(z)= \lim_{n\to\infty} f_n(z)=f_{\mathcal{N}}(z)
\end{eqnarray}

Equation 
\ref{wherewegoto} then gives\begin{eqnarray}
\underset{ \phantom{ \tilde\theta } n\rightarrow\infty \phantom{ \tilde\theta } }{\lim} 
\frac{\underset{\substack{\tilde{\theta} \in {\omega}_{01}^*}}{\sup}\;\Pr\left(\po_{\mathcal{A}_\V:\mathcal{O}_\V}\geq \tau\; ;\; \tilde{\theta}\right)}{
\Pr\left( \chi^2_1
\geq \tfrac{2}{1-\xi_n}\log \tfrac{\tau}{\mu \sqrt{\xi_n}} \right)}\leq\int\limits_{-\infty}^{\infty}f_{\mathcal{N}}(z)\;dz=1\label{case2result_prelim}
\end{eqnarray}

Since $\xi_n>0$, this implies
\begin{eqnarray}
\underset{ \phantom{ \tilde\theta } n\rightarrow\infty \phantom{ \tilde\theta } }{\lim} 
\frac{\underset{\substack{\tilde{\theta} \in {\omega}_{01}^*}}{\sup}\;\Pr\left(\po_{\mathcal{A}_\V:\mathcal{O}_\V}\geq \tau\; ;\; \tilde{\theta}\right)}{
\Pr\left( \chi^2_1
\geq 2\log \tfrac{\tau}{\mu \sqrt{\xi_n}} \right)}\leq 1\label{case2result}
\end{eqnarray}

Equations \ref{case1result} and \ref{case2result} cover all Cases (1-3). In generality, we can therefore write 
\begin{eqnarray} 
    \lim_{n\rightarrow\infty}\;\underset{\substack{\theta \in\omega_{\bs v}^{*}}}{\sup}\;
    \;
    \frac{\Pr\left( \po_{\mathcal{A}_\V:\mathcal{O}_\V} \geq \tau \ ; \theta \right)}{\Pr\left( \chi^2_1
    \geq 2 \log \tfrac{\tau}{(\sizeV) \, \mu  \sqrt{\xi_n}} \right)} &\leq& 1, \qquad(\forall\ \omega_{\bs v}^*\in \Omega^*) . \label{limsupalphav}
\end{eqnarray}

\end{proof}

\section{The \texorpdfstring{$\alpha \leq 0.025$}{alpha <= 0.025} threshold in Doublethink} \label{si_section_p0.02}
This section explains MT Remark \ref{remark_p0.025_threshold}. We consider the distribution of the test statistic
\begin{eqnarray}
    \tilde{\po}_{\mathcal{A}_\V:\mathcal{O}_\V} &=& \sum_{j \in \mathcal{V}_\V} \po_{\tilde{\bs s}+{\bs e}_j:\tilde{\bs s}},
\end{eqnarray}
which is asymptotically equivalent (see proof of Theorem \ref{theorem_fpr}) to the posterior odds, $\po_{\mathcal{A}_\V:\mathcal{O}_\V}$, as $n\rightarrow\infty$  (Definition \ref{define_asymptotically_equivalent_test}), where $\tilde{\bs s}$ is the `true' model and $\{{\bs e}_j\}_k = \mathbb{I}(j=k)$. It tests the null hypothesis that $\beta_j=0$ for all $j \in \mathcal{V}_\V$, where $\mathcal{V}_\V = \{ j\ :\ v_j=0\}$.

We rewrite it as
\begin{eqnarray}
    \frac{\tilde{\po}_{\mathcal{A}_\V:\mathcal{O}_\V}}{(\sizeV)\, \mu\, \sqrt{\xi_n}} &=& \frac{1}{|\mathcal{V}_\V|} \sum_{j \in \mathcal{V}_\V} \mlr_{\tilde{\bs s}+{\bs e}_j:\tilde{\bs s}}^{1-\xi} \ \sim\ \frac{1}{|\mathcal{V}_\V|} \sum_{j \in \mathcal{V}_\V} \mlr_{\tilde{\bs s}+{\bs e}_j:\tilde{\bs s}},
\end{eqnarray}
as $n\rightarrow\infty$, where $\xi_n=h/(n+h)$ and $\mlr_{\tilde{\bs s}+{\bs e}_j:\tilde{\bs s}} \overset{d}{=} \mathrm{LG}(1/2, 1)$, by \cite{wilks1938large}.

Under independence, assuming $\tilde{\bs s}$ is true, meaning $\theta\in\Theta_{\tilde{\bs s}}$, this problem is equivalent to studying
\begin{eqnarray} \label{eq_mean_of_LG1_RVs}
    \bar{X}_k &=& \frac{1}{k} \sum_{j=1}^k X_j,
\end{eqnarray}
where $2\log X_j$, $j=1\dots k$ are independent and identically distribution chi-squared random variables with one degree of freedom. Equivalently, $X_j$, $j=1\dots k$ are independent and identically distributed log-gamma random variables with shape parameter 0.5 and scale parameter 1.

Theorem \ref{theorem_fpr} and Corollary \ref{corollary_mlr_limiting_dist} are equivalent to the asymptotic approximation
\begin{eqnarray} \label{si_eq_p0.02_target}
    \Pr(\bar{X}_k \geq x) &\sim& \Pr(\chi^2_1 \geq 2 \log x), \qquad x\rightarrow\infty.
\end{eqnarray}
However, this was derived via a step approximating the more direct expression
\begin{eqnarray} \label{si_eq_p0.02_direct}
    \Pr(\bar{X}_k \geq x) &\sim& k \Pr(\chi^2_1 \geq 2 \log k\,x), \qquad x\rightarrow\infty.
\end{eqnarray}

\begin{figure}
    \centering
    \includegraphics[width=0.45\linewidth]{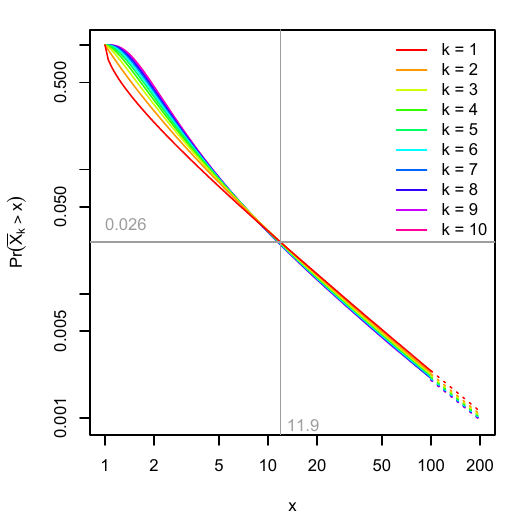}
    \caption{\smaller Tail probabilities for the mean of $k$ LG(1/2,1) random variables for $k=1,\dots,10$ based on 10 million simulations (solid coloured lines) and, for $x>100$, the theory of regular variation (Equation \ref{si_eq_p0.02_direct}; dashed coloured lines). Grey lines indicate the smallest $x$ at which the tail probability $\Pr(\bar{X}_k \geq x)$ is greatest at $k=1$.}
    \label{si_fig_p0.025_threshold}
\end{figure}

We used simulations to investigate whether the approximation (Equation \ref{si_eq_p0.02_target}) is conservative, under independence, for all $k$ at all probabilities less than 0.0259846 in the sense that
\begin{eqnarray}
    \Pr(\bar{X}_1 \geq x) &\geq& \Pr(\bar{X}_k \geq x), \quad \forall \quad k\geq 1,\ x\geq x_\mathrm{crit}.
\end{eqnarray}
This was borne out numerically because the red line ($k=1$) in Figure \ref{si_fig_p0.025_threshold} is above all other coloured lines ($k>1$) when $x\geq x_\mathrm{crit}$.

We solved $x_\mathrm{crit}$ numerically based on the following empirical observation: the solution to $\Pr(\bar{X}_1 \geq x)=\Pr(\bar{X}_k \geq x)$, $x>1$, was solved at increasingly smaller values of $x$ as $k$ increased. This is apparent in the graph because the red line ($k=1$) crosses the pink line ($k=10$) at smaller values of $x$ than the purple line ($k=9$), and so on.

Empirically, therefore, it was sufficient to numerically solve the convolution
\begin{eqnarray}
    \Pr(\bar{X}_1 \geq x_\mathrm{crit}) &=& \Pr(\bar{X}_2 \geq x_\mathrm{crit}), \qquad x_\mathrm{crit}>1 \nonumber \\
    &=& \int_1^\infty p_{\mathrm{LG}(0.5, 1)}(y) \, \Pr(X_1 + y \geq 2 \, x_\mathrm{crit}) \mathrm{d}y ,
\end{eqnarray}
where $p_{\mathrm{LG}(0.5,1)}$ is the density function of a log-gamma random variable with shape parameter 0.5 and scale parameter 1. This yielded
\begin{eqnarray}
    x_\mathrm{crit} &=& 11.92362, \nonumber \\
    \Pr(\bar{X}_1 \geq x_\mathrm{crit}) &=& 0.0259846.
\end{eqnarray}

\bibliographystyle{chicago}
\bibliography{bib.bib}